\DocumentMetadata{}
\documentclass[manuscript,screen,acmsmall]{acmart}
\usepackage{hyperref}
\usepackage{hyperxmp}
\AtBeginDocument{%
  }

\setcopyright{acmlicensed}
\copyrightyear{2025}
\acmYear{2025}
\acmDOI{XXXXXXX.XXXXXXX}
\acmConference[CSCW'26]{}{}{}

\begin{document}

\title{Retweets, Receipts, and Resistance: \\ Discourse, Sentiment, and Credibility in Public Health Crisis Twitter}

\author{Tawfiq Ammari}
\affiliation{%
  \institution{Rutgers University}
  \city{New Brunswick, NJ}
  \country{USA}}
\email{tawfiq.ammari@rutgers.edu}

\author{Anna Gutowska}
\affiliation{%
  \institution{Rutgers University}
  \city{New Brunswick, NJ}
  \country{USA}}
\email{anna.gutowska@rutgers.edu}

\author{Jacob Ziff}
\affiliation{%
  \institution{Rutgers University}
  \city{New Brunswick, NJ}
  \country{USA}}
\email{jacob.ziff@rutgers.edu}

\author{Casey Randazzo}
\affiliation{%
  \institution{Rutgers University}
  \city{New Brunswick, NJ}
  \country{US}}
\email{cer124@rutgers.edu}

\author{Harihan Subramonyam}
\affiliation{%
  \institution{Stanford University}
  \city{Stanford, CA}
  \country{USA}}
\email{harihars@stanford.edu}

\renewcommand{\shortauthors}{Ammari et al.}

\begin{abstract}
As the COVID-19 pandemic evolved, the Centers for Disease Control and Prevention (CDC) used Twitter to disseminate safety guidance and updates, reaching millions of users. This study analyzes two years of tweets from, to, and about the CDC using a mixed-methods approach to examine discourse characteristics, credibility, and user engagement. We found that the CDC’s communication remained largely one-directional and did not foster reciprocal interaction, while discussions around COVID-19 were deeply shaped by political and ideological polarization. Users frequently cited earlier CDC messages to critique new and sometimes contradictory guidance. Our findings highlight the role of sentiment, media richness, and source credibility in shaping the spread of public health messages. We propose design strategies to help the CDC tailor communications to diverse user groups and manage misinformation more effectively during high-stakes health crises.
\end{abstract}

\ccsdesc[500]{Human-centered computing~Human computer interaction (HCI)}

\keywords{Twitter, Sentiment Polarization, Credibility Signals, Public health crisis, Crisis communication, Public policy discourse, Social media affordances, Public Engagement}


\received{24 March 2025}
\received[revised]{XX XX XXXX}
\received[accepted]{XX XX XXXX}

\maketitle

\section{Introduction}
A 1894 smallpox outbreak in Milwaukee highlighted how political tensions and poor public health communication infrastructure can undermine public health efforts, despite technically competent medical leadership \cite{leavitt2003public}. In contrast, a 1947 New York City outbreak shows how coordinated messaging across different media platforms (e.g., radio, print media), and civic partnerships can foster trust and compliance \cite{leavitt2003public}. These cases reveal that effective communication—not just medical intervention—is crucial in managing health crises, a lesson still relevant today amid challenges like vaccine misinformation and health access disparities.

Crisis communication involves delivering factual information to the public during unexpected emergencies that typically fall outside an organization’s control and require an immediate response \citep{zahry2023risk}. In health crises, information becomes “a necessity, not a luxury,” as public health messaging often mirrors political persuasion in its strategic framing to promote specific outcomes \citep{dennison2019rising, entman1993framing}.

To guide such communication, the Crisis and Emergency Risk Communication (CERC) framework offers a multistage model linking message design to audience response during emergencies \citep{miller2021being}. Developed through the interdisciplinary work of Seeger (communication) and Reynolds (public health), CERC functions as an “applied theory,” equipping practitioners with principles to manage crises like COVID-19 effectively \citep[p. 12]{miller2021being}.

A core tenet of CERC is the importance of two-way communication during crises to reduce uncertainty \citep{reynolds2005crisis}. This includes mechanisms for public feedback, supporting stakeholder engagement \citep{renn2015stakeholder}. Public-facing agencies such as the CDC or FEMA are encouraged to adopt symmetrical two-way communication to foster mutual understanding and adapt strategies accordingly \citep{grunig2008excellence, waters2011tweet}. However, studies show that CDC’s online communication has predominantly remained top-down rather than incorporating feedback from the public \citep{gesser2014risk}.

While CERC provides strong guidance for traditional media, it falls short in addressing communication via social media—now a central medium in crisis response \citep{zhang2019social}. Social media platforms are widely used by both communities and crisis organizations, offering opportunities to (1) enhance situational awareness, (2) facilitate community-level support, and (3) better understand public needs during disasters.

Social media might be an important communication medium at times of crisis. However, its use introduces new challenges like the spread of misinformation \citep{hobbs2024low} and the need to focus on shorter communication messages in an increasingly polarized environment \citep{rao2020retweets,reuter2018social}. 

To investigate these challenges, this study analyzed the discourse of the Centers for Disease Control and Prevention (CDC) during the COVID-19 Pandemic by focusing on four dimensions identified by \cite{zhang2019social}: (1) dissemination patterns; (2) message content; (3) the public's experience; and (4) rumor and trust.\footnote{Twitter re-branded to X after the collection of the data used in this study and so, we refer to it as Twitter throughout this paper.} 

\subsection{Dissemination Patterns}
Twitter’s platform-specific affordances—including hashtags, mentions, retweets, likes, replies, and quote-tweets—shape how users construct audiences and propagate messages \citep{devito_platforms_2017, leonardi_social_2017, faraj_materiality_2013}. Hashtags help users build “imagined audiences” by marking participation and signaling affiliation with causes or communities, such as \#BLM or \#stopasianhate \citep{litt2012knock, stewart_et_al_17, freelon2016beyond}. Mentions expand reach by directing messages to visible, often influential targets, especially in contentious contexts \citep{mcgregor2019social, hemsley2018tweeting}. Retweets amplify messages and are often interpreted as endorsements \citep{kim2012role, bonson2019twitter}; likes signal approval and selectively extend visibility \citep{Sekimoto_et_al_2020}; replies enable re-framing of content \citep{zade_et_al_24}; and quote-tweets allow users to reshape narratives within their own networks \citep{Garimella_et_al_16}. Given how different affordances of the platform \citep{faraj_materiality_2013, leonardi_social_2017} can affect the way messages are disseminated, we ask:
\begin{quote}
\textbf{RQ1: What are the properties of users whose messages are most widely propagated through the various affordances of the platform (e.g., retweets, quote-tweets, replies, and mentions)?}
\end{quote}

\subsection{Message Content, Rumor, and Trust}
At its core, the CERC model identifies key features of effective crisis messaging—clarity, consistency, accuracy, credibility, and timeliness—that shape how individuals cognitively and emotionally process information. These components align with research in information science and human-computer interaction, which emphasizes the importance of message quality and source trustworthiness in driving user engagement with crisis-related content  \cite{GULATI2024100495, shah2022source}. For example, credible and actionable content has been shown to improve information uptake in public health contexts \cite{journalmedia5020046} \citep{journalmedia5020046}, while timely and empathetic messaging can enhance users’ emotional regulation and sensemaking during fast-evolving disasters \citep{palen2016crisis}. The model further elaborates how message features influence intermediate audience states—such as risk perception, trust in authorities, and self-efficacy—which in turn mediate key behavioral outcomes, including guideline compliance, rumor rejection, and protective actions. These dynamics are consistent with empirical findings that underscore how emotional tone and perceived credibility shape decision-making in crisis contexts \citep{VANDERMEER2014526,VANDOORN201974}.

Given that earlier work shows how the use of low-credibility information sources \citep{wang_et_al_23,hobbs2024low}, the use of rich media in the message \citep{jin2016novel,gupta2013faking}, and the sentiment portrayed in a message \citep{valdez2020social} all affect the way messages are propagated in a social network, we ask:

\begin{quote}
\textbf{RQ2: How do the properties of a message affect the way they are propagated in the CDC public?}    
\end{quote}

\subsection{Public's experience}
Public audiences rely on “interpretive schemata” to make sense of crisis organization messages. These schemata are shaped by emotional valence (e.g., social identity polarized through “us vs. them” framing) and ideological divides, which emerge through disagreement over how problems are defined (e.g., recognizing a pandemic) and how evaluations are expressed (e.g., slogans like \#maskssavelives; \citealt{bruggemann2023debates}).

Given that earlier work shows how the properties of the users disseminating the messages, specifically their follower-following ratio (signifying their influence in the network) and their verified status \citep{harris2024perceived} affect the way their messages are propagated to the public, we ask:
\begin{quote}
\textbf{RQ3: How are messages from verified users and users with high follower-following ratios in the CDC public different from other users?}
\end{quote}

This study investigates how the CDC communicated during the COVID-19 pandemic on Twitter and how the public responded. Using a mixed-methods approach, we analyzed over 275,000 tweets sent from, to, or about the CDC between January 2020 and January 2022. Our findings reveal that discourse was dominated by highly polarized topics, including vaccines, school masking, and claims about whether the pandemic was a hoax. The network of mentions showed signs of fragmentation and echo chambers, where users mostly interacted within ideologically aligned groups. Accounts sharing low-credibility sources were less likely to be mentioned by others but more likely to initiate mentions, indicating an active strategy to shape discourse. Verified users, though more likely to be retweeted, also frequently shared low-credibility content. Based on these findings, we offer design and policy recommendations to improve public health communication, particularly in managing evolving guidelines and countering misinformation in high-stakes, fast-changing crises.

\section{Related Work}
In this section, we review prior work that informs our study of CDC-related discourse. We begin by examining research on Twitter affordances and audience management in the context of public communication and crisis engagement. Next, we explore how message content—such as rich media and verification status—shapes engagement. We conclude by discussing scholarship on digital communication in public health crises, with a focus on trust, credibility, and institutional messaging.

\subsection{Twitter affordances and audience management} \label{sec:affordances}
Technological affordances are defined as the relationship between users and "their perceptions of environments" \citep[P.361]{pavelko_muderinos_2020}. Affordances can explain the influence of different technologies on human behavior \citep{faraj_materiality_2013,leonardi_social_2017}. Twitter has numerous platform-specific affordances that enable people to communicate with each other, associate with content, and provide or receive direct \citep{devito_platforms_2017} or indirect feedback \citep{randazzo_23} due to the searchability and permanence of the content \citep{leonardi_social_2017}. We will focus on three main areas: (1) hashtags; (2) interactions including mentions, and (3) Twitter affordances including: retweets \citep{kim2012role,boyd2010tweet}, likes \citep{mcdonald2021please,huang_et_al_18,wang2016catching}, replies \citep{kim2012role}, and quote-tweets \citep{Garimella_et_al_16,zade_et_al_24}. 

\subsubsection{Imagined Audiences Built with Hashtags}
Hashtags are keywords that allow users to search for or find a topic of interest. This allows users to build an imagined audience \citep{litt2012knock} whom they believe to be interested in a particular topic, ``to mark participation, assert individual identity, promote group identity, and support or challenge a frame.'' \citep{stewart_et_al_17} For example, hashtags make issues of interest like Black Lives Matter (\#BLM) and \#stopasianhate more searchable and accessible, which in turn allows users to reach wider audiences \citep{tong_et_al_22}. Additionally, \cite{huang_et_al_2010} argue that hashtags are conversational tools that motivate people to share their thoughts on the topic of interest. 

Hashtags like \#BLM act as markers of identity and alignment with broader value-driven communities \citep{zappavigna2018communing, freelon2016beyond}. They help signal users’ intended or imagined audiences—those who share similar values or political commitments. Such hashtags also function as affective anchors, enabling individuals to express their emotions around key issues \citep{papacharissi2016affective}. In times of crisis, users often seek out hashtags to access “situational updates” relevant to their experiences \citep{lachlan2016social}. Crisis-response organizations like FEMA or the CDC can leverage or popularize hashtags to broadcast critical information to large and relevant audiences \citep{dalrymple2016facts}.

\subsubsection{Imagined Audiences Built with Visible Mentions} Another way to build an audience is by using mentions (e.g., @user). When a Tweet mentions a user, only the target user receives a notification, but other users can view and search for the Tweet. In other words, users searching for Tweets made by an account will also see tweets mentioning that account. This means one can expand their audience by targeting an account belonging to a person (e.g., politician or journalist) or organization central to the discourse on a particular topic \citep{mcgregor2019social}. Users will typically write negative Tweets mentioning a specific account to send a hostile message rather than start a discussion with the target user \citep{hemsley2018tweeting} with the hope that a provocative message to an influential node in the network would increase the size of their audience. Addressing specific users with messages with a negative sentiment valence can increase  polarization in online interactions \citep{kountouri2023polarizing}. 

\subsubsection{Twitter Propagation Affordances: Retweets, Likes, Responses, and Quote-tweets}Retweets propagate messages throughout the Twitter network, and whether intended or not, serve as endorsements of the original Tweet message \citep{kim2012role} and are thus associated with the virality of a message \citep{bonson2019twitter}. Earlier work trained models using the number of retweets to better understand more influential messages in the network (c.f.,\citep{jenders_2013,Suh_et_al_2010,bakshy_et_al_2011}). Likes also propagate messages to a more  targeted audience than retweets \citep{Rathnayake_17, Sekimoto_et_al_2020}. Thus, they show the popularity of the message since the user ``liking'' the Tweet is an implicit approval of the message \citep{bonson2019twitter,Sekimoto_et_al_2020}.

Responses to Tweets are usually associated with commitment to an issue \citep{bonson2019twitter} as it provides a chance to re-frame the conversation \citep{zade_et_al_24} by providing a positive or negative view, thus pushing the user's view of the original message \citep{Garimella_et_al_16}. Quote-tweets, much like forwarded messages, not only allow the user to push the original message in a new direction, but also ``broadcasts'' \citep{zade_et_al_24} this new contextualized view to networks in which they are embedded \citep{Garimella_et_al_16}.   

\subsection{Rich Media, Verified Users, and Misinformation on Twitter}
In this section, we will start by reviewing the effects of rich media use by social media users. We then study verified Twitter users, and finally dis/misinformation on Twitter.

\subsubsection{Rich Media on Twitter}
While social media was traditionally more text-heavy, social media sites like Twitter, Instagram, and TikTok afford users the capacity to create more rich content - including images and videos. Given the words limits on Twitter, incorporating rich media allows users to extend their text posts \citep{jin2016novel,gupta2013faking}.

Shi et al. show that rich media increases the likelihood that users will propagate messages (e.g., through likes and retweets[\citeauthor{shi2018determinants},\citeyear{shi2018determinants};\citeauthor{jiang2019diary},\citeyear{jiang2019diary}]) and increase the perceived credibility of the message \citep{yin2020incorporating,LI2022113752}. Richer messages also increase the persuasiveness \citep{zhou2021characterizing} and the reach \citep{wang_et_al_23} of COVID-19 misinformation.

\subsubsection{Verified Twitter accounts} \label{related:verified_acct} Verified Twitter accounts are central to the mechanism by which users search for and propagate information on the platform \citep{gonzalez2021bots,paul2019elites}, especially in the event of a crisis like COVID \citep{zhou_et_al_23}. In fact, Twitter shared that it would increase the number of local verified accounts sharing information about COVID \citep{TwitterSupport2020} to provide more trusted information in a timely manner \citep{Lunden2020}. One example \cite{zhou_et_al_23} provide is Dr. Eric Ding (@DrEricDing), who became a verified account on January 25, 2020. Additionally, self-disclosure by verified accounts about their COVID challenges increased the chance of self-disclosure in the network \citep{du2022contribution}.

However, verified accounts can also become ``perceived experts'' who spread misinformation about vaccines \citep{harris2024perceived}, especially regarding the side-effects of new COVID vaccines like the Oxford/AstraZeneca \citep{hobbs2024low}. This duality highlights the complex role of verified accounts in shaping public understanding during health crises.

\subsubsection{Misinformation and Distrust} This complexity is heightened by widespread misinformation and growing distrust in institutions. While social media offers timely public health updates, it also enables the spread of conspiracy theories—often as responses to fear and uncertainty. During the Zika outbreak, for example, social media functioned both as a news source and a platform for rumor and conspiracy \citep{kou_at_al_17}, partly due to gaps in official communication \citep{gui_et_al_17}. Additionally, COVID-19 saw the rise of medical and scientific populist actors that included health professionals like Dr. Judy Mikovitz who countered the general consensus held by the majority of medical experts on issues like masking and vaccination \cite{premat2024introduction}.

Mistrust is especially pronounced among vulnerable populations already skeptical of public health authorities \citep{abelson2009does}. Historical cases like HIV/AIDS show how misinformation can lead to rejection of testing and treatment \citep{christina_2010}. This resurfaced during COVID, as memes and images spread misinformation \citep{wang_et_al_23}, often infused with hate and discrimination \citep{liao_23,Vishwamitra_24,javed_et_al_23}

\subsection{Digital Communication in Public Health Crises}
 Public health agencies tend to utilize social media as a means of broadcasting crisis management information rather than a two-way communication tool \citep{Graham_14}. By altering this perspective and examining the public’s sentiment through social media interactions, information gaps and overlooked concerns can be addressed sooner, limiting distrust.

\subsubsection{Emerging Infectious Diseases: Rolling uncertainty, risk, and public relations}
\cite{fox2012medical} suggests embracing the ``inevitable uncertainty'' of medicine where scientists and medical professionals have to contend with ``certainty for now'' instead of more objective certainty. This uncertainty is significantly higher for emerging infectious diseases (EIDs) like Ebola and COVID because the provision of medical expertise and the touting of existing protocols provides an ``illusion of certainty.'' \citep{dalrymple2016facts} Earlier work shows that when organizations like the CDC emphasize certainty in a chaotic uncertain times, this amplifies the uncertainty that the general public feels, and increases doubts about the institutions as well \citep{dalrymple2016facts}.

\subsubsection{Sentiment in Digital Public Health Interventions}
Previous research has categorized crisis communication sentiment using positive, negative, and neutral labels \citep{beigi2016overview,unlu2023exploring}. On platforms like Twitter, sentiment polarization often reflects underlying political polarization \citep{haupt2021characterizing, valiavska2022politics}. This is evident in movements like “Liberate,” which opposed public health interventions such as masking, vaccination, and school closures. These groups mobilized around hashtags like \#Liberate to express resistance and amplify their stance \citep{ewing2021navigating, kwok2021tweet, manguri2020twitter}.

\subsubsection{Framing Messages to the Public}
Organizations such as FEMA and the CDC strive to craft their messaging in ways that avoid political entanglement, instead emphasizing their core missions \citep{Murphree29052009}. Previous research highlights the importance of viewing a crisis through multiple lenses—including those of various stakeholders—and tailoring the framing of messages accordingly \citep{keri_et_al_2005}.

\section{Dataset}
 We used the twarc Python library for data mining via the Twitter API to collect the tweets used in this study. This was possible through the use of an individualized bearer token granted by Twitter. The inclusion criteria for the tweets in this study were tweets that were produced by users in the United States, in English, posted between January 2020 and January 2022, and not retweets. Furthermore, we specifically only included tweets that were sent to the official @CDCgov account, from @CDCgov, used the hashtag ‘\#CDC’, tagged @CDCgov, or mentioned the string ‘CDC’. In total, we collected 21,843 tweets sent to @CDCgov, 11,810 from @CDCgov, 20,012 using ‘\#CDC’, 53,104 tagging @CDCgov, and 162,295 mentioning the string ‘CDC’. Finally, we collected any other Tweets using the same ``Conversation ID'' which indicate that they are under the same discussion thread. In total, 275,124 tweets were collected. While we did not collect all retweets, we do have the number of times each tweet is retweeted with a total of 3,331,350 retweets. 

\section{Methods}
\label{sec:meth}
To examine how discourse around the CDC unfolded on Twitter during the COVID-19 pandemic, we used a mixed-methods approach combining topic modeling, sentiment analysis, and credibility classification with regression, classification, and network modeling. We first extracted features from tweets—such as topics, sentiment, and source credibility—and then used these to analyze patterns of engagement, user influence, and message propagation. This section details our the features used in our modeling, our analytical techniques and modeling strategies.

\subsection{Features used in data analysis} \label{subsec:meth_features}
In this section, we present all the features used to train classifiers, regression, network models. We start with a description of topics and themes created using topic modeling. We then describe how we measured sentiment, identified low-credibility sources, and finally used platform signals (e.g., retweets) and user properties (e.g., verified user) as features.

\subsubsection{Topic Modeling with BERTopic (N=71)} \label{sec:topics}
We applied contextualized topic modeling using the BERTopic framework \citep{grootendorst2022bertopic}, which clusters BERT embeddings \citep{reimers_sentence-bert_2019} with HDBSCAN \citep{campello_density-based_2013}. This method is better suited for short, symbol-rich texts like tweets than traditional approaches such as LDA \citep{blei_dynamic_2006,hays_cultural_1998}, which often produce less intelligible topics in such contexts \citep{egger2022topic}. Word embeddings capture semantic relationships by mapping terms into a vector space where distance reflects contextual similarity \citep{mikolov_efficient_2013,griffiths_topics_2007}.

To identify the optimal number of topics, we trained 25 models by varying HDBSCAN’s minimum cluster size from 15 to 255. We evaluated each model using Gensim’s \textit{CoherenceModel} \citep{rehurek_coherencemodel}, which assesses topic interpretability by measuring the degree of semantic similarity among top terms within a topic. Coherence scores have been shown to align well with human judgments of topic quality \citep{roder_exploring_2015}. The final model, with a minimum cluster size of 45, achieved the highest coherence score (0.39) and identified 71 topics.

\begin{figure}[h!]
    \centering
    \includegraphics[width=0.99\linewidth]{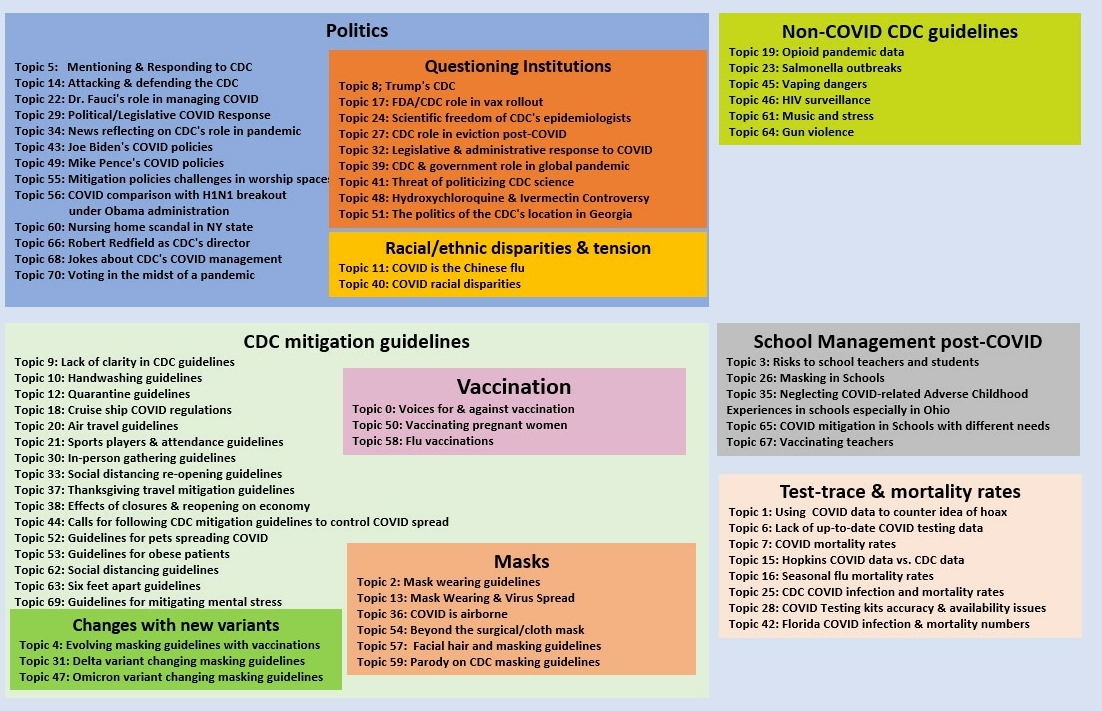}
    \caption{Topic groupings into themes}
    \label{fig:top_themes}
\end{figure}

\paragraph{Qualitative Analysis} \label{sec:qualitative_analysis}
In a virtual public square, discourse constructs social contexts \citep{schiffrin1987discourse, howarth2000discourse} as it reflects communication emphasis as expressed by participants
\citep{potter1987discourse}. When qualitatively analyzing discourse, communication content becomes more than information exchange (i.e., mentions) or diffusion (i.e., retweets; \cite{matassi2023know}]). Discourse can become a resource to structure knowledge and make sense of a situation \citep{hardy2005discourse}. Topic models must be interpreted by humans \citep{chang_reading_2009,dou_paralleltopics_2011} to contextualize them in the social/political environment \citep{baden2020hybrid,yarchi2021political}. The following are the steps used in this qualitative interpretative process:
\begin{enumerate} 
    \item \textbf{Selecting top Positive, Negative, and Neutral Sentiment Comments}: All topics had a sentiment distribution with more extreme scores than a normal distribution \citep{kallner2017laboratory}. Given these distributions, we used Tag2Vec to develop top words describing different sentiments associated with each topic. Tag2Vec, using word embeddings (similar to the ones discussed in \S\ref{sec:topics}), provides a list of the words closest to a tag (positive vs. negative valence of a topic \cite{chen2017doctag2vec}). It was used before to determine the linguistic differences between throwaway and pseudonymous users on Reddit \citep{ammari_et_al_19}. We randomly sampled 100 tweets for each category of \textit{top topic} X \textit{sentiment}. When there was no sample across any of the sentiment criteria, we sampled another 100 tweets from this topic. This gave us a total of 6,000 tweets.  
    \item \textbf{Consensus on Topics}: After an iterative coding process and weekly meetings to discuss individual code books, we reached consensus on the different themes and the polarization within each \citep{McDonald_et_al_19}. We also identified overarching themes as shown in Figure \ref{fig:top_themes}.
\end{enumerate}

\paragraph{A note on quoting tweets and identifying Twitter accounts}
\cite{fiesler_participant_2018} argue that Twitter users do not expect to be quoted verbatim in academic research, even if they are not named. They suggest engaging in \cite{bruckman_studying_2002}'s recommended levels of user disguise when quoting users in a research study. To protect the privacy of Twitter users, we primarily describe the contents of tweets without direct quotations. We only name public organizations such as the CDC or news organizations and organizations.

Figure \ref{fig:top_themes} shows the breakdown of topics into five main themes covering a similar discourse around COVID mitigation and politics.  

\subsubsection{Sentiment analysis (N=1)} \label{sec:meth_sent}
To understand how the sentiments of tweets affected the way that people discussed topics in the CDC public, we used VADER, a sentiment analysis tool that is frequently applied to Twitter text \citep{elbagir2019twitter} which calculates the valence and polarity for each word (positive, negative, or neutral) along with the percentage of said polarity in a composite score. VADER labels text as having positive sentiment if the composite score is above 0.05, negative sentiment if the score is below -0.05, and neutral sentiment if the score is between -0.05 and 0.05 \citep{hutto2014vader}. We used composite score in regression models, and we used positive, negative, and neutral labels for classifiers (see \S\ref{meth:regress_classify}). 

\subsubsection{Credibility of Information Sources (N=1)} \label{sec:credibility}
We used the Iffy Index of Unreliable Sources,\footnote{\url{https://iffy.news/index/}} a public-facing dataset of low-credibility sources\footnote{\url{https://docs.google.com/spreadsheets/d/1ck1_FZC-97uDLIlvRJDTrGqBk0FuDe9yHkluROgpGS8/gviz/tq?tqx=out:csv&sheet=Iffy-news}} reviewed by Media Bias/Fact Check,\footnote{\url{https://mediabiasfactcheck.com/methodology/}} and was used to determine the credibility of medical data about COVID-19 vaccination information on Twitter \citep{pierri_2023} based on the following criteria:
\begin{enumerate}
    \item Domain: Site address without ``https:\/\/'' and ``www.'' \newline
    (e.g. google.com instead of https:\/\/www.google.com)
    \item Site rank: The Alexa rank of a site’s traffic
    \item Year online: The year the site was established
    \item Name: The site name
    \item Factual categorization: Very high, high, mostly factual, mixed, low, very low
    \item Bias (least biased/pro-science, right-center/left-center, left/right, questionable/conspiracy pseudoscience)
\end{enumerate}

The Iffy Index only includes sites with a low-credibility rating and categorizes as either Conspiracy/Pseudoscience (CP) or Questionable Source/Fake News (FN). 

\subsubsection{Twitter Platform Signals (N=9)} \label{subsec:platform_sig} As discussed in \S\ref{sec:affordances}, Twitter provides affordances for propagating messages and expanding your audience. We will use the features identified in the section and listed below:
 \begin{enumerate}
        \item Retweet count (RT)
        \item Quote Tweet count (QT)
        \item Tweet likes (Like)
        \item Direct responses to Tweets (Rep)
        \item Rich Media: Tweet uses rich data (e.g., pictures or videos) 
        \item Verified account
    \end{enumerate}

Additionally, we add the following features to describe the use of Twitter in the CDC public:
\begin{enumerate}
    \item Total number of Tweets by user (count): this feature represents the total number of Tweets for a user over the period of data collection within the CDC public. The higher this number, the higher the commitment of the user to this discourse.  
    \item Tenure in the CDC Twitter public (date of last tweet - date of first tweet). While the Tweet count provides one measure of commitment to the discourse, this is another measure that shows how long the user was engaged in this discussion.
    \item FFR: Following/Follower ratio: the higher this ratio, the higher the number of people interested in their content without the target user following many others. If the value is less than one, then the user can be described as a ``mass follower'' who follows other accounts for the``sole purpose of gaining more users'' for themselves \cite{anger_kittle_2011}. 
\end{enumerate}

\subsection{Understanding platform signals and content analysis} \label{meth:regress_classify}
To better understand platform signals, we trained four regressions to measure what increases the likelihood of retweet counts, quote tweet counts, replies to tweets, and likes for Tweets (see Tables \ref{tab:regression_results_1} and \ref{tab:regression_results_2}). 

Additionally, we trained seven classifiers to better understand what exemplified verified accounts, Tweets responding to the CDC, Tweets using low-credibility links, Tweets using rich media (e.g., pictures and videos), positive, negative, and neutral Tweets based on Vader sentiment analysis (see \S\ref{sec:meth_sent}). The top classifiers, along with the positive and negatively predictive features are shared in Tables \ref{tab:classifier_metrics_part1} and \ref{tab:classifier_metrics_part2}.

\subsubsection{Hyperparameter Tuning}
When building all of the models identified above, we used 82 features as described below:
\begin{enumerate}
    \item Twitter topics as defined in \S\ref{sec:topics} (N=71).
    \item Vader sentiment analysis as defined in \S\ref{sec:meth_sent} (N=1). 
    \item Low credibility source identifier as defined in \S\ref{sec:credibility} (N=1).
    \item Twitter platform signals as defined in \S\ref{subsec:platform_sig} (N=9).
\end{enumerate}

We set an 80/20 random train/test split and used Sklearn's standard scaler to scale data before training \citep{pedregosa2011scikit}.  To find the top R\textsuperscript{2} in our regression models, we used Random Search Cross Validation (with 5-fold cross-validation\cite{vishnu2023recurrent}) to investigate the performance of Ridge and Lasso regressions. We tuned for $\alpha$ and the solver (["auto", "svd", "cholesky", "lsqr", "sparse\_cg", "sag", "saga", "lbfgs"]) in Ridge. We also tuned for $\alpha$ and selection (["cyclic", "random"]) in Lasso regression \citep{wang2011random}.

As for the seven classifiers, we again set an 80/20 random train/test split and used Sklearn's standard scaler to scale data before training. To find the model with the top Area Under the Curve (AUC) - a common metric that evaluates a model's fit \citep{swets2014signal,fawcett2006introduction,lemeshow1982review} - we used Grid Search Cross Validation (with 5-fold cross-validation) to find the model with the best AUC. A model is considered acceptable as long as the AUC value is above 0.7. The accuracy, recall, F1, and AUC metrics are reported in average and for the top model in Tables \ref{tab:reply_to_CDC}, \ref{tab:classifier_metrics_part1}, and \ref{tab:classifier_metrics_part2}. We trained our models on the following classifiers: (1) logistic regression classifier; (2) Random Forest Classifier; (3) Gradient Boosting Classifier; and (4) Gaussian Classifier. 

\subsubsection{Explainabilty}
For the Ridge and Lasso Regressions, we used the Regressors package \citep{haas_regressors} to determine significant coefficients (presented in Tables \ref{tab:regression_results_1} and \ref{tab:regression_results_2}).

However, given that six of our top classifiers were random forest classifiers, whilst the seventh is a Gradient Boosting Classifier, we applied post-hoc analysis to explain the models. We used SHapely Additive exPlanations (SHAP), a post-hoc explainable AI analysis tool to explain each model \citep{huang2023increasing,SHOEMAKER202383}. Shapley values, based on game theory \citep{shaplundberg2017unified}, allow us to analyze the effects of each feature on output of the model. In turn, this allowed us to rank top predictors and to show dependence between predictors.  

\subsubsection{Analyzing mention ties: Exponential Random Graph Models} \label{sec:ERGM_mention}
Exponential Random Graph Models (ERGM) model patterns within a network which can be seen as evidence for ongoing structural processes, which are also influenced by node attributes \citep{robins_lusher_2012}. ERGM models ``extend the logic of multivariate regression to relational data: the presence or absence of a binary tie in a network is an outcome variable predicted by a vector of independent variables called parameters.'' \citep{keegan_et_al_2012}[p.431] A positive parameter increases the likelihood of a tie in the network, while a negative parameter indicates a decreased likelihood of tie formation. The parameter equivalent to the intercept in a regression model is the \textit{edges parameter} which ``reflects the log-likelihood of a network tie appearing entirely by chance'' which in turn describes the ``network density if no other effects were present.'' \citep[p.432]{keegan_et_al_2012}. Another network-specific or ``endogenous'' ERGM parameter is mutuality, or reciprocity which measures the tendency for a tie from one node to another to be reciprocated in the other direction. A higher mutual value increases the value of ``cooperative network ties`` between network nodes \citep{axelrod1981evolution}. Exogenous node-level coefficients indicate whether actors with a specific attribute are more or less likely to be connected to alters in the network \citep{lusher2013exponential}. 

For example, in the medical field, \cite{Srinivasan_Jiang_2023} created a disease co-occurrence network and showed that chronic diseases increase the likelihood of a tie to COVID diagnosis, thus proving a link between the two. 

\paragraph{ERGM parameters}
First, we will describe the network we formed to analyze using ERGM.  Twitter users can explicitly ``mention'' other users by including their usernames in the text of the tweet. A user \textit{u}, can mention another user \textit{v}, by tweeting ``@v, what are you up to today.'' We created a directed mentions network based on user mentions composed of 59,304 nodes and 315,495 edges. We then used the Statnet R package \citep{handcock2019package} to fit the ERGM model on the mentions network using the following parameters: (1) Endogenous parameters: (a) edges; (b) mutuality and (2) exogenous parameters: (a) average theme weights as defined in \S\ref{sec:qualitative_analysis} given the computational cost of incorporating all 71 topics \citep{kim2025basic}; (b) user sentiments broken by positive, negative, and neutral based on the VADER score Positive, negative, and neutral (see \S\ref{sec:meth_sent}); (c) credibility of sources as presented in \ref{sec:credibility}; and (d) user-specific features as presented in \S\ref{subsec:platform_sig}.  

\section{Findings}
We start this section by analyzing how CDC discourse is affected by Twitter platform affordances (e.g., retweets, likes, etc.) in \S\ref{sec:finding_rt}. Next, we analyze how credibility of sources in a Twitter message, the use of rich media, and the sentiment of that message affect COVID discourse in the CDC public in \S\ref{sec:sent_photo_credibility}. Finally, in \S\ref{sec:verified_accts}, we analyze how user properties, like being verified or influential in the network (based on FFR), are associated with user behavior in the CDC public.

\begin{table}[ht]
    \centering
    \caption{This table compares the CDC, responses to it, verified users, and responses to them, tweets using pictures and responses to them, as well as the use of low-credibility sources and responses to these tweets. Specifically, we focus on the echo and average sentiment for each. Echo is the ratio between retweet/tweets. The closer the value to 1, the less the likelihood of their tweets being retweeted. The higher the echo the higher the number of retweets per Tweet. At times, the properties might overlap. For example, we show how often verified users use images.}
    \begin{tabular}{|p{2.5cm}|l|l|l|l|p{1cm}|l|l|}
    \hline
        \textbf{Behavior and Affordances} & \textbf{Total} & \textbf{Unique} & \textbf{Echo} & \textbf{Sent.} & \textbf{Low-cred. (\%)} & \textbf{Rich (\%)} & \textbf{Ver. (\%)} \\ \hline
        \textbf{CDC} & 2,688,505 & 11,810 & 227.65 & 0.17 & 0 & 100 & 100 \\ \hline
        \textbf{Respond to CDC} & 122,813 & 21,843 & 5.623 & -0.01 & 4.16 & 34.50 & 6.12 \\ \hline
        \textbf{Verified} & 2,903,733 & 26,428 & 109.87 & 0.10 & 10.11 & 75.61 & -\\ \hline
        \textbf{Respond to verified} & 139,058& 26,205 & 5.31 & -0.01 & 3.96 & 35.07 & 11.02 \\ \hline
        \textbf{Rich media} & 3,032,625 & 99,004 & 30.63 & 0.01 & 11.54 & - & 20.18 \\ \hline
        \textbf{Respond to pictures} & 136,131 & 27,783 & 4.90 & -0.01 & 4.28 & 43.98 & 8.07 \\ \hline
        \textbf{Low credibility} & 36,008 & 17,433 & 2.07 & -0.16 & - & 65.55 & 5.11 \\ \hline
         \textbf{Respond to low credibility} & 702 & 468 & 1.50 & -0.07 & 62.61 & 61.11 & 2.78 \\ \hline
    \end{tabular}
    \label{tab:behavior_echo}
\end{table}

\subsection{Platform affordances, user properties, sentiment and information credibility }\label{sec:twitter_engagement}
Table \ref{tab:behavior_echo} provides a comparative overview of Twitter discourse involving the CDC during the COVID-19 pandemic. It highlights that CDC tweets had the highest virality (echo ratio of 227.65). In contrast, responses to the CDC and verified users had lower engagement and more neutral or negative sentiment. Tweets containing rich media were more likely to be shared and a notable portion (11\%) cited low-credibility sources. Interestingly, tweets from verified accounts also included comparable portions of low-credibility content at 10\%, and those using such sources often relied heavily on rich media.

In this section, we discuss how Twitter engagements afforded by the platform like retweets, quote-tweets, likes, mentions, and responses are associated with different themes of COVID discussions. We will specifically focus on direct responses to the CDC and, in turn, direct responses to users from the CDC.

\subsubsection{What is in a retweet and what gets more likes?} \label{sec:finding_rt}
As Table \ref{tab:regression_results_1} shows, discussions for or against the vaccine (Topic 0) are more likely to be retweeted and to receive likes. While discussions of lock-downs and re-openings, especially in schools (Topics 38 and 65), were less likely to receive likes, they were retweeted at a higher rate. Similarly, engaging in Topic 1, introducing data showing that COVID is not a hoax, increases the chance of being retweeted while reducing the chance of receiving a like.

The chance of receiving likes and being retweeted is higher when Tweets are posted by verified users and when there is a high number of Tweets in the CDC public. Users with a higher follower-following ratio (FFR) and more direct replies are also more likely to receive likes. However, the more direct replies received by a user, the less likely they are to be retweeted. We discuss direct replies next.

\subsubsection{What gets more direct replies and quote-tweets?} \label{findings:replies_quotes}
Engaging in discourse for and against the vaccine (Topic 0) increases the chances of receiving a response and being quote-tweeted. The same applies to discussing masking in schools (Topic 26). Both of these topics were polarizing as they pit people who believed that each mitigation tool was important to control COVID spread and others who were against their implementation because (1) they thought that the mitigation strategies are not as effective as advertised; or (2) they thought that forcing these mitigation measures were against their freedom. A related topic early in the pandemic focused on the very presence of COVID as a virus. While some were questioning the magnitude of the effects on society, others were sharing data, especially from the Johns Hopkins dataset, to counter data from the CDC and other official channels that were considered not forthcoming about the effects of the virus. However, Topic 38, focusing on the effects of lock-downs on society, received fewer responses and quote-tweets. Interestingly, this topic received fewer likes as well. However, it did increase the number of retweets. 

While being verified is more associated with being retweeted, being a verified user makes it less likely to be quote-tweeted. Furthermore, quote-tweeted users have a lower follower-following ratio (influence in the network) and less total tweets. 

\begin{figure}[ht]
    \centering
    \includegraphics[width=1\linewidth]{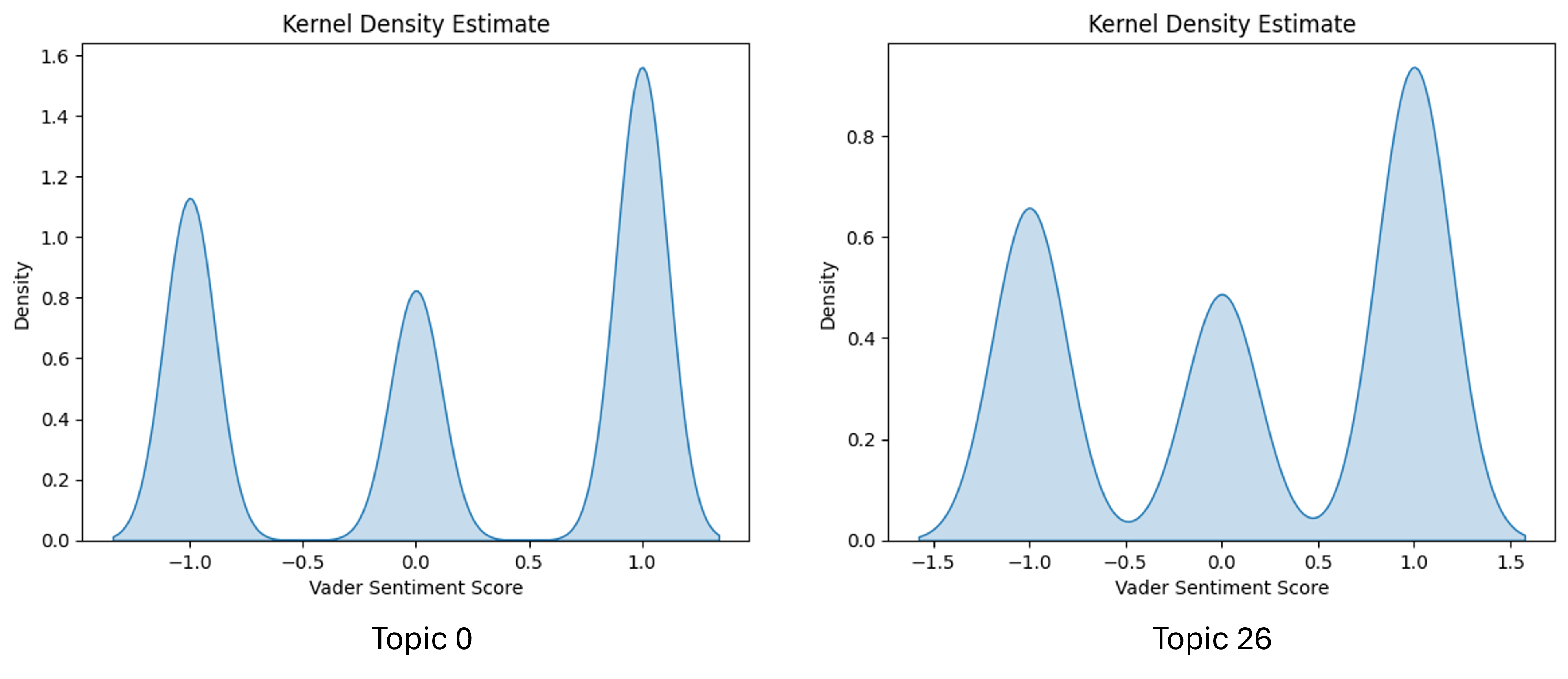}
    \caption{This figure shows the sentiment distribution for two highly propagated topics—Topic 0 (vaccination discourse) and Topic 26 (masking in schools). Both exhibit bimodal sentiment distributions, indicating strong polarization. Positive sentiment clusters align with public health advocacy, neutral clusters contain logistical or institutional information, and negative clusters reflect misinformation, conspiracy theories, and ideological resistance.}
    \label{fig:kde}
\end{figure}

\paragraph{Polarization of Top Propagated Topics}
Topic 0 (vaccine discourse) and Topic 26 (masking in schools) were strong predictors of propagation across all four Twitter affordances. As shown in the kernel density estimates in Figure \ref{fig:kde}, both topics exhibit clear sentiment polarization. Their bimodal coefficients—0.769 for Topic 0 and 0.770 for Topic 26—exceed the 0.555 threshold, indicating a multimodal distribution and the presence of distinct sentiment clusters \citep{math10071042}. Using the top Tag2Vec hashtags and words (see \S\ref{sec:qualitative_analysis}) associated with positive, neutral, and negative sentiment (presented in Table \ref{tab:tag2vec}), we qualitatively analyze the polarized discourse below.

Topic 0 captures the highly polarized discourse surrounding COVID-19 vaccination. On the positive side, hashtags such as \#beatcovid\_19, \#communitymatters, and \#vaccinerollout reflect support for public health initiatives and collective action to control the pandemic. Tweets with these tags typically advocate for vaccination, celebrate being vaccinated, or express solidarity with immunocompromised individuals. 

Neutral content centers around geographic identifiers like \#clarkecounty and \#nmstate which usually included practical and logistical information about where to get vaccinated. They also incorporated more clinical terminology, such as Pfizer COVID-19 vaccine or \#COVID-free, primarily focusing on emerging scientific findings (e.g., vaccine efficacy) and sharing personal updates about their vaccination status. 

In stark contrast, negative sentiment tweets often include misinformation and conspiracy-leaning terms like hydroxychloroquine or \#hoax, and reference controversial identifiers like \#vaccine\_passports which were considered by many as an example of CDC overreach. Discussions about new variants like \#ArcturusRex centered on the waning efficacy of vaccines against new variants with many questioning why vaccines were mandated if they did not protect against new variants. These negative tweets frequently emanate from accounts skeptical of vaccine efficacy or the legitimacy of the pandemic (\#plandemic) itself, reinforcing the polarization of Topic 0. Many of these accounts were of public figures, many with high FFRs, with the majority being verified users (e.g., Dr. Stella Immanuel, Dr. Robert Malone, Dr. Peter McCullough, Dr. Judy Mikovits, and Alex Berenson whom we also discuss in \S\ref{sec:finding_mention}). 

Topic 26 centers on debates about masking policies in schools and reveals a similarly polarized structure. Positive sentiment is expressed through hashtags like \#maskupaz (Arizona), \#protecttheherd, and \#mandatestudentsmask, promoting adherence to CDC guidance and expressing concern for student and community health. These tweets often invoke expert voices (\#medtwitter) and reflect a belief in science-driven public health. 

Neutral tags such as \#vvsdproud and \#schoolboard appear to refer to specific school districts or institutional actors involved in decision-making. These tweets are usually from these institutions and propagated CDC messaging about the vaccine. 

The negative sentiment tags—including \#timetounmask and \#gregabbott—showcase ideological backlash against perceived government overreach. Specifically, they focused on an executive order signed by Greg Abbott, the Texas Governer, which blocked school districts from mandating masking.\footnote{\url{https://www.texastribune.org/2021/08/09/texas-mask-order-schools/}} These tweets often reject mask mandates outright or use them as a vehicle for broader political grievances. While still showing a negative sentiment valence, responses to these Tweets argued against \#trumpsters who were engaged in \#coviddenial, and that \#texasdeservesbetter, specifically, that students with special needs or underlying conditions (e.g., \#Type 1 diabetics) need others to adhere to the \#cdcguidelines to safeguard their lives. This indicates that debates over school health policy served as flashpoints for wider partisan and identity-based conflict.

\begin{table}[ht!]
    \caption{This table presents key hashtags and terms identified through qualitative and Tag2Vec analysis for Topic 0 (vaccines) and Topic 26 (masking in schools), categorized by sentiment. Positive sentiment emphasizes community and scientific support, neutral sentiment conveys logistical or institutional messaging, while negative sentiment includes misinformation and ideologically motivated backlash.}
    \label{tab:tag2vec}
    \centering
    \begin{tabular}{|c p{3.5cm} p{3.5cm} p{3.5cm}|}  
        \hline
        \textbf{Topic} & \textbf{Positive} & \textbf{Neutral} & \textbf{Negative} \\
        \hline
        Topic 0 & \#beatcovid\_19, \#immunodeficient, \#communitymatters, \#gotvaccinated, \#firstresponders, \#vaccinerollout & pfizer covid 19 vaccine, \#clarkecounty, \#richmondcounty, \#painless, \#COVID-free, \#nmstate & hydroxychloroquine, medical professionals sharing anti-vax info, \#ArcturusRex (new COVID variant), \#plandemic, \#hoax, \#vaccine\_passports \\
        \hline
        Topic 26 & \#maskupaz, \#getvaccinated, \#mandatestudentsmask, \#medtwitter, \#protecttheherd, \#CDC & \#vvsdproud (Valley View School District), \#schoolboard, \#pandemicoftheunvaccinated & \#coviddenial, \#texasdeservesbetter, \#trumpsters, \#cdcguidelines, \#timetounmask, \#type\_1\_diabetic, \#gregabbott \\
        \hline
    \end{tabular}
\end{table}

Similar themes were evident in direct responses to the CDC which we focus on next.

\paragraph{Responses to the CDC} 
As presented in Table \ref{tab:reply_to_CDC}, the use of rich content was more predictive of responses to the CDC. Much of the rich content included what users referred to as ``receipts,'' screenshots of earlier CDC messages showing the evolving masking guidelines after vaccinations (Topic 4). Other top topics in responses include the vaccines (Topic 0), the role of the CDC in evictions post-COVID, discussing how COVID is airborne (Topic 36), and comparing the response to COVID with the response to the H1N1 breakout under Obama's administration (Topic 56). 

The framing of these responses went in one of three directions: (1) arguing that the CDC's guidelines are not strict enough; (2) arguing that the CDC is introducing mandates without the right to do so; (3) the CDC is overreacting since the effects of COVID are overblown.

\begin{figure}[ht]
    \centering
    \includegraphics[width=1\linewidth]{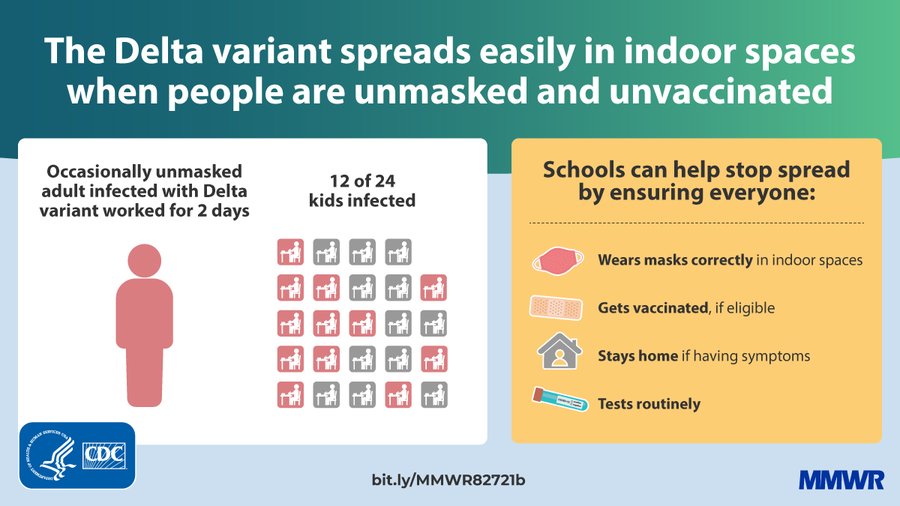}
    \caption{This figure displays a CDC-generated infographic highlighting updated school safety protocols in response to the emergence of the Delta variant. It reflects a shift in guidelines recommending continued masking indoors, even for vaccinated individuals, representing what many saw as a departure from earliery messaging.}
    \label{fig:delta}
\end{figure}

For example, as new variants (e.g., the Delta variant) reduced the efficacy of the vaccine, the CDC introduced new guidelines, especially in school settings. These guidelines included masking in addition to vaccination (see Figure \ref{fig:delta}). 

\begin{figure}[ht]
    \centering
    \includegraphics[width=1\linewidth]{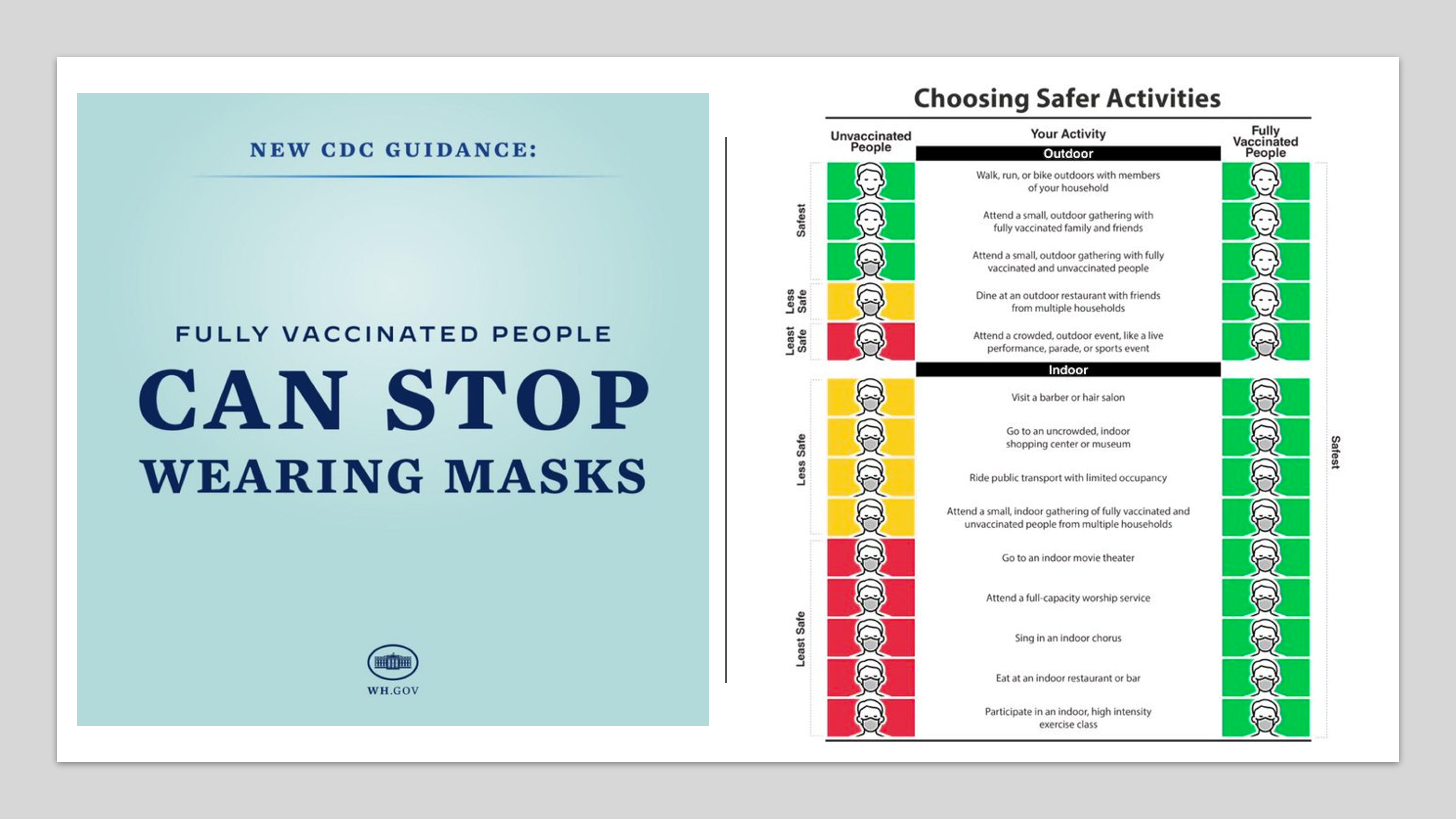}
    \caption{This figure juxtaposes two public-facing CDC messages: one (left) suggesting that vaccinated individuals no longer need to wear masks, and the other(right) provided a more nuanced recommendation masking post-vaccination in different contexts. Both were used as ``receipts'' by critics of the CDC (left) questioning CDC credibility and consistency and defenders of the original guidelines (right) arguing that the evolving COVID situation required nuance, which was present in CDC guidelines.}
    \label{fig:two_vaccines}
\end{figure}

A number of responses to the updated guidelines referenced earlier messaging which was considered contradictory by the CDC (see the image on the right shown in Figure \ref{fig:two_vaccines}). Specifically, the argument was that the message pushed by the CDC, and propagated by the White House specifically said that, after the vaccinating, people would not have the to mask any longer. Now they are arguing that the CDC and associated medical and academic institutions cannot be trusted given that their earlier proclamations proved wrong. In response, many answered with their own ``receipts'' that showed how earlier CDC vaccine communications were more nuanced. They specifically showed the right-side image in Figure \ref{fig:two_vaccines}. The image shows that the CDC still recommended masking for vaccinated individuals, especially with in-doors activities. 

Replies to the CDC were less likely to receive direct replies, likes, retweets, quote-tweets, and usually had lower follower-following ratios.

\paragraph{Direct responses from the CDC}
We found only 58 direct responses from the official CDC  account to 25 accounts based on a question or suggestion. The responses revolved around providing information about the shingles or flu vaccines. A few others were answering questions about the danger of contracting COVID-19 by touching products imported from China. The @CDCgov account answered those accounts by saying there is no danger in handling products from China. 

Now that we analyzed responses to and from the CDC, we will next analyze mention co-occurrence networks along with the @CDCgov account. 

\begin{table}[H]
\centering 
\caption{This Table shows top regression results and the top predictors for Retweets ($\alpha=0.31$, solver=svd) and Likes ($\alpha=0.34$, solver=svd). Table shows significant predictors from regression model. **** p<0.0001; *** p<0.001; ** p<0.01; * p<0.05}
\begin{tabular}{llllll}
\hline
\multicolumn{1}{|c|}{\textbf{Retweets}} & \multicolumn{1}{c|}{\textbf{Ridge}} & \multicolumn{1}{c|}{\textbf{$R^{2}$=0.90}} & \multicolumn{1}{c|}{\textbf{Likes}} & \multicolumn{1}{c|}{\textbf{Ridge}} & \multicolumn{1}{c|}{\textbf{$R^{2}$=0.83}} \\ \hline
\multicolumn{1}{|l|}{\textit{Feature}} & \multicolumn{1}{l|}{\textit{Coefficient}} & \multicolumn{1}{l|}{\textit{p-value}} & \multicolumn{1}{l|}{\textit{Feature}} & \multicolumn{1}{l|}{\textit{Coefficient}} & \multicolumn{1}{l|}{\textit{p-value}} \\ \hline
\multicolumn{1}{|l|}{Topic 0} & \multicolumn{1}{l|}{0.611} & \multicolumn{1}{l|}{****} & \multicolumn{1}{l|}{Topic 0} & \multicolumn{1}{l|}{3.511} & \multicolumn{1}{l|}{****} \\ \hline
\multicolumn{1}{|l|}{Topic 1} & \multicolumn{1}{l|}{0.101} & \multicolumn{1}{l|}{*} & \multicolumn{1}{l|}{Topic 1} & \multicolumn{1}{l|}{-0.614} & \multicolumn{1}{l|}{*} \\ \hline
\multicolumn{1}{|l|}{Topic 38} & \multicolumn{1}{l|}{0.072} & \multicolumn{1}{l|}{**} & \multicolumn{1}{l|}{Topic 38} & \multicolumn{1}{l|}{-0.432} & \multicolumn{1}{l|}{**} \\ \hline
\multicolumn{1}{|l|}{Topic 65} & \multicolumn{1}{l|}{0.045} & \multicolumn{1}{l|}{*} & \multicolumn{1}{l|}{Topic 65} & \multicolumn{1}{l|}{-0.23} & \multicolumn{1}{l|}{*} \\ \hline
\multicolumn{1}{|l|}{Verified} & \multicolumn{1}{l|}{2.329} & \multicolumn{1}{l|}{*} & \multicolumn{1}{l|}{Verified} & \multicolumn{1}{l|}{15.033} & \multicolumn{1}{l|}{****} \\ \hline
\multicolumn{1}{|l|}{Like count} & \multicolumn{1}{l|}{2838.533} & \multicolumn{1}{l|}{****} & \multicolumn{1}{l|}{Quote count} & \multicolumn{1}{l|}{-33146.7} & \multicolumn{1}{l|}{****} \\ \hline
\multicolumn{1}{|l|}{Quote Count} & \multicolumn{1}{l|}{7534.939} & \multicolumn{1}{l|}{****} & \multicolumn{1}{l|}{Reply Count} & \multicolumn{1}{l|}{1189.377} & \multicolumn{1}{l|}{****} \\ \hline
\multicolumn{1}{|l|}{Reply Count} & \multicolumn{1}{l|}{-1084.07} & \multicolumn{1}{l|}{****} & \multicolumn{1}{l|}{Retweet count} & \multicolumn{1}{l|}{47676.71} & \multicolumn{1}{l|}{****} \\ \hline
\multicolumn{1}{|l|}{Count} & \multicolumn{1}{l|}{0.412} & \multicolumn{1}{l|}{****} & \multicolumn{1}{l|}{Count} & \multicolumn{1}{l|}{8.882} & \multicolumn{1}{l|}{****} \\ \hline
\multicolumn{1}{|l|}{} & \multicolumn{1}{l|}{} & \multicolumn{1}{l|}{} & \multicolumn{1}{l|}{FFR} & \multicolumn{1}{l|}{17.385} & \multicolumn{1}{l|}{*} \\ \hline
\end{tabular}
\label{tab:regression_results_1}
\end{table}

 \begin{table}[ht]
\centering 
\caption{This Table shows top regression results and the top predictors for responses ($\alpha$=0.31, solver=svd) and quote-tweets ($\alpha$=0.68, solver=cholesky), Quotes-tweets, and Responses. Table shows significant predictors from regression model. **** p<0.0001; *** p<0.001; ** p<0.01; * p<0.05}
\begin{tabular}{llllll}
\hline
\multicolumn{1}{|c|}{\textbf{Responses}} & \multicolumn{1}{c|}{\textbf{Ridge }} & \multicolumn{1}{c|}{\textbf{$R^{2}$=0.64}} & \multicolumn{1}{c|}{\textbf{Quote-tweets}} & \multicolumn{1}{c|}{\textbf{Ridge}} & \multicolumn{1}{c|}{\textbf{$R^{2}$=0.70}} \\ \hline
\multicolumn{1}{|l|}{\textit{Feature}} & \multicolumn{1}{l|}{\textit{Coefficient}} & \multicolumn{1}{l|}{\textit{p-value}} & \multicolumn{1}{l|}{\textit{Feature}} & \multicolumn{1}{l|}{\textit{Coefficient}} & \multicolumn{1}{l|}{\textit{p-value}} \\ \hline
\multicolumn{1}{|l|}{Topic 0} & \multicolumn{1}{l|}{0.161} & \multicolumn{1}{l|}{****} & \multicolumn{1}{l|}{Topic 0} & \multicolumn{1}{l|}{0.008} & \multicolumn{1}{l|}{****} \\ \hline
\multicolumn{1}{|l|}{Topic 1} & \multicolumn{1}{l|}{-0.06} & \multicolumn{1}{l|}{*} & \multicolumn{1}{l|}{Topic 15} & \multicolumn{1}{l|}{-0.008} & \multicolumn{1}{l|}{**} \\ \hline
\multicolumn{1}{|l|}{Topic 15} & \multicolumn{1}{l|}{0.036} & \multicolumn{1}{l|}{****} & \multicolumn{1}{l|}{Topic 26} & \multicolumn{1}{l|}{0.011} & \multicolumn{1}{l|}{*} \\ \hline
\multicolumn{1}{|l|}{Topic 26} & \multicolumn{1}{l|}{0.004} & \multicolumn{1}{l|}{****} & \multicolumn{1}{l|}{Topic 38} & \multicolumn{1}{l|}{-0.005} & \multicolumn{1}{l|}{*} \\ \hline
\multicolumn{1}{|l|}{Topic 38} & \multicolumn{1}{l|}{-0.001} & \multicolumn{1}{l|}{****} & \multicolumn{1}{l|}{Verified} & \multicolumn{1}{l|}{-0.032} & \multicolumn{1}{l|}{****} \\ \hline
\multicolumn{1}{|l|}{Verified} & \multicolumn{1}{l|}{0.732} & \multicolumn{1}{l|}{****} & \multicolumn{1}{l|}{Reply Count} & \multicolumn{1}{l|}{856.81} & \multicolumn{1}{l|}{****} \\ \hline
\multicolumn{1}{|l|}{Like Count} & \multicolumn{1}{l|}{32.959} & \multicolumn{1}{l|}{****} & \multicolumn{1}{l|}{Retweet Count} & \multicolumn{1}{l|}{962.706} & \multicolumn{1}{l|}{****} \\ \hline
\multicolumn{1}{|l|}{Quote Count} & \multicolumn{1}{l|}{2964.705} & \multicolumn{1}{l|}{****} & \multicolumn{1}{l|}{Count} & \multicolumn{1}{l|}{-1.813} & \multicolumn{1}{l|}{****} \\ \hline
\multicolumn{1}{|l|}{Retweet Count} & \multicolumn{1}{l|}{-504.582} & \multicolumn{1}{l|}{****} & \multicolumn{1}{l|}{FFR} & \multicolumn{1}{l|}{-0.666} & \multicolumn{1}{l|}{****} \\ \hline
\multicolumn{1}{|l|}{Count} & \multicolumn{1}{l|}{10.537} & \multicolumn{1}{l|}{****} & \multicolumn{1}{l|}{} & \multicolumn{1}{l|}{} & \multicolumn{1}{l|}{} \\ \hline
\multicolumn{1}{|l|}{FFR} & \multicolumn{1}{l|}{2.828} & \multicolumn{1}{l|}{****} & \multicolumn{1}{l|}{} & \multicolumn{1}{l|}{} & \multicolumn{1}{l|}{} \\ \hline
\end{tabular}
\label{tab:regression_results_2}
\end{table}

\begin{table}[ht]
\caption{Predictors for direct replies to the CDC. The table shows the average metrics from RandomSearch and the hyperparameters for the top model. They also show the top positive and negative predictors as determined using SHAP.}
\centering
\small
\begin{tabular}{|c||c|}
    \hline
    & \textbf{Reply to CDC} \\
    \hline
    \textbf{Average Metrics} & \\
    Accuracy & 0.645 \\
    Precision & 0.626 \\
    Recall & 0.767 \\
    F1 & 0.684 \\
    AUC & 0.716 \\
    \hline
    \textbf{Top Model (GBC)} &  est=150, max. depth=5, learning rate=0.3\\
    Accuracy & 0.752 \\
    Precision & 0.725 \\
    Recall & 0.811 \\
    F1 & 0.766 \\
    AUC & 0.837 \\
    \hline
    \textbf{Positive Features} & 0, 4, 27, 36,56, Rich \\
    \textbf{Negative Features} & Rep, Like, FFR, RT, 7, 15, 15, 61, QT, 29, 9, 19, 8 \\
    \hline
\end{tabular}
\label{tab:reply_to_CDC}
\end{table}

\begin{table}[ht]
  \caption{Part 1 of Classifiers: Performance metrics for Positive, Neutral, Negative, and Reply to CDC classifiers. The table shows the average metrics from RandomSearch and the hyperparameters for the top model. They also show the top positive and negative predictors as determined using SHAP.}
    \centering
    \small
    \begin{tabular}{|l||p{2.5cm}|p{2.5cm}|p{2.5cm}|}
        \hline
        & \textbf{Positive} & \textbf{Neutral} & \textbf{Negative} \\
        \hline
        \textbf{Average Metrics} & & & \\
        Accuracy & 0.668 & 0.681 & 0.688 \\
        Precision & 0.667 & 0.684 & 0.715 \\
        Recall & 0.717 & 0.713 & 0.593 \\
        F1 & 0.685 & 0.689 & 0.633 \\
        AUC & 0.700 & 0.719 & 0.727 \\
        \hline
        \textbf{Top Model (RF)} & est=50; mins=5, max=None & est=100; mins=5, max=10 & est=100; mins=5, max=10 \\
        Accuracy & 0.984 & 0.983 & 0.985 \\
        Precision & 0.985 & 0.983 & 0.984 \\
        Recall & 0.982 & 0.983 & 0.985 \\
        F1 & 0.984 & 0.983 & 0.985 \\
        AUC & 0.999 & 0.998 & 0.999 \\
        \hline
        \textbf{Positive Features} & RT, 15, QT, FFR, VR, 1 & 5, FFR, 21, 14, 9, 32, 63 & 19, 25 \\
        \textbf{Negative Features} & 25, Rep, 16, 2, 54, 23, 8, 14, 44,29,56  & 70, Rep, QT, Like & FFR, 14, 30, RT, 21 \\
        \hline
    \end{tabular}
\label{tab:classifier_metrics_part1}
\end{table}

\begin{table}[ht]
  \caption{Part 2 of Classifiers: Performance metrics for Photo, Low Quality, and Verified classifiers. The table shows the average metrics from RandomSearch and the hyperparameters for the top model. They also show the top positive and negative predictors as determined using SHAP.}
    \centering
    \small
    \begin{tabular}{|l||p{2.5cm}|p{2.5cm}|p{2.5cm}|}
        \hline
        & \textbf{Rich Media} & \textbf{Low Credibility} & \textbf{Verified} \\
        \hline
        \textbf{Average Metrics} & & & \\
        Accuracy & 0.719 & 0.699 & 0.830 \\
        Precision & 0.820 & 0.708 & 0.838 \\
        Recall & 0.529 & 0.754 & 0.812 \\
        F1 & 0.619 & 0.718 & 0.822 \\
        AUC & 0.756 & 0.789 & 0.905 \\
        \hline
        \textbf{Top Model (RF)} & est=100; mins=5, max=10 & est=100; min=2, max=None & est=100; mins=5, max=10\\
        Accuracy & 0.996 & 0.782 & 0.894 \\
        Precision & 0.998 & 0.803 & 0.876 \\
        Recall & 0.994 & 0.748 & 0.916 \\
        F1 & 0.996 & 0.775 & 0.896 \\
        AUC & 0.995 & 0.872 & 0.964 \\
        \hline
        \textbf{Positive Features} & RT, FFR, QT, Like, Rep, 5 & 19, Rich, 7, VR & FFR, QT, RT, Like, Rep, 14, Rich \\
        \textbf{Negative Features} & VR, 6, 16, 61 & Like, Rep, RT, 23, 32 & Count, 21 \\
        \hline
    \end{tabular}
\label{tab:classifier_metrics_part2}
\end{table}

\subsubsection{What is in a mention?} \label{sec:finding_mention}
\begin{figure}
    \centering
    \includegraphics[width=0.5\linewidth]{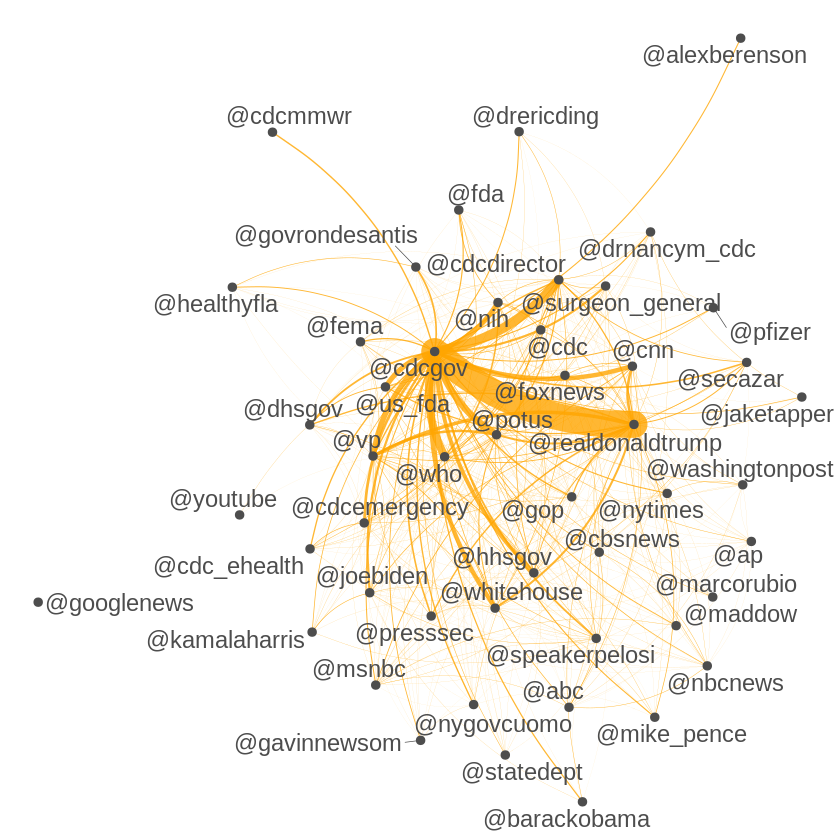}
    \caption{Mention co-occurrence network showing top accounts mentioned along with the CDC account.}
    \label{fig:mention_cooccurence}
\end{figure}

Mentioning accounts can be directing a message to an account expecting a response or increasing attention to the message by mentioning an influential account in the network, in this case, the CDC. The co-occurrence mentions network \citep{Nasim_2017} shown in Figure \ref{fig:mention_cooccurence} where the top accounts mentioned alongside the @CDCGov account are presented in the following groups:
\begin{enumerate}
    \item \textbf{Pharmaceutical companies} like @pfizer (specifically focused on the vaccine production)
    \item \textbf{Public health professionals/academics}: @DrEricDing (see \S\ref{related:verified_acct})
    \item \textbf{Politicians}: @VP, @secazar, @joebiden, @govrondesantis, @mikepense, @presssec, @kamalaharris, @statedepartment, @nycgovcuomo,@joebiden, @gop, @statedept, @cdcdirector
    \item \textbf{News corporations and journalists}: @foxnews, @jaketapper, @cnn, @abc, @cbsnews, @nbcnews @msnbc, @maddow, @alexberenson (this last account is a verified journalist who happens to share low-credibility information sources; more in \S\ref{sec:verified_accts})
    \item \textbf{State and federal agencies}: @fema (Federal Emergency Management Agency), @nih, @us\_fda (US Food and Drug Administration), @dhsgov (Department of Homeland Security), @hhsgov (health and human services), @cdcmmwr (Morbidity and Mortality Weekly Report), @drnancym\_cdc, @HealthyFla (Florida Dept. of Health)
\end{enumerate}

The results of our ERGM model, which analyzes the likelihood of a mention tie, are summarized in Table \ref{tab:ERGM}. Our edges (intercept) parameter is negative and significant. This means that there is a high `cost' to the creation of a tie in the network \cite{keegan_et_al_2012}[p.432]. In our sparse network, we found a positive ``mutual''parameter suggesting a higher likelihood of tie creation associated with reciprocation. In other words, when one user mentions another, they are more likely to mention the original user. 

When analyzing node characteristics, we found that most of the discussion themes identified in \S\ref{sec:topics} decreased the likelihood of creating a mention tie. The only theme that increased the likelihood of the creation of a mention tie was the one focused on non-COVID discourse probably due to the high connectivity between those working with opioid addiction patients, pain activists, and veterans. We found that out-degree tie formation was more likely for users who posted more using less credible sources. In other words, users who relied on less reputable sources were more likely to mention others while they themselves were less likely to be mentioned. Both negative and positive sentiment decreased the likelihood of the creation of mention ties, while neutral sentiment increased the likelihood. The more the use of rich media in Tweets, the less likely the creation of mention ties. Finally, verified users are less likely to have mention ties with other users. 

\begin{table}[ht]
\caption{Exponential family random graph model of mentions network. A smaller value means the likelihood of creating a mention tie is smaller and vice-versa. **** p<0.0001; *** p<0.001; ** p<0.01; * p<0.05}
\begin{tabular}{|l|r|r|r|l|}

\hline
                                & \multicolumn{1}{l|}{Estimate} & \multicolumn{1}{l|}{Std. Error} & \multicolumn{1}{l|}{z value} &     \\ \hline
nodeocov.verified               & -0.062                        & 0.012                           & -4.979                       & *** \\ \hline
nodeicov.verified               & -0.205                        & 0.013                           & -15.236                      & ** \\ \hline
nodematch.top\_theme.Mitigation & -0.147                        & 0.007                           & -20.031                      & *** \\ \hline
nodematch.top\_theme.Non-COVID  & 1.236                         & 0.032                           & 38.792                       & *** \\ \hline
nodematch.top\_theme.Politics   & -0.205                        & 0.011                           & -18.738                      & *** \\ \hline
nodematch.top\_theme.Test       & -0.267                        & 0.022                           & -12.250                      & ** \\ \hline
nodeocov.lowq-credibility             & 0.096                         & 0.024                           & 3.948                        & *** \\ \hline
nodeicov.lowq-credibility             & -0.231                        & 0.029                           & -7.980                       & *** \\ \hline
nodeicov.photo                  & -0.267                        & 0.008                           & -32.107                      & *** \\ \hline
nodematch.sentiment.negative    & -0.237                        & 0.009                           & -25.842                      & *** \\ \hline
nodematch.sentiment.neutral     & 0.381                         & 0.011                           & 33.583                       & *** \\ \hline
nodematch.sentiment.positive    & -0.078                        & 0.008                           & -9.781                       & *** \\ \hline
edges                           & -8.896                        & 0.004                           & -2000.335                    & *** \\ \hline
mutual                          & 3.483                         & 0.063                           & 55.150                       & *** \\ \hline
\end{tabular}
    \label{tab:ERGM}
\end{table}

\subsection{Sentiment, rich media, and low credibility sources} \label{sec:sent_photo_credibility}
In this section, we will analyze the qualities of positive, neutral, and negative labeled Tweets, as well as ones labeled with the use of rich media (e.g., images) and low-credibility sources. 

\subsubsection{Positive Tweets} \label{subsec:positive}
A higher retweet and quote Tweet count, being verified, and having a higher follower-following ratio are all more positively predictive of positive Tweets. Sharing data sources pointing to the seriousness of COVID (Topic 1), but also to good news about dropping rates of infection (Topic 15).

Positive Tweets are less likely to receive direct responses. They are also less likely to discuss COVID and seasonal flu infection and mortality rates (Topics 16 \& 25), the efficacy of masks (Topics 2 and 54), or calling for obedience to the CDC mitigation guidelines in general (Topic 44). They tend to make less mention of political discourse like Trump's CDC (Topic 8), the political/legislative response to COVID (Topic 29), or comparing the COVID response to H1N1 response under Obama's administration (Topic 56).  

\subsubsection{Neutral Tweets} \label{subsec:neutral}
Higher follower-following ratio increases the chance of a neutral Tweet. Neutral tweets tend to receive fewer responses, quote-tweets, and likes. 

Sports players and attendance guidelines (Topic 21) and social distancing guidelines (six feet apart; Topic 63) are also positively associated with neutral Tweets. These are often posts by users with high FFR, usually at a local level, like high school and college athletic departments, coaches, and sports commentators who, while not verified, have large followings. Many of the posts simply stated the current requirements for attending sports events (e.g., in-door masking for intramural sports or social distancing requirements). 

The lack of clarity in CDC guidelines (Topic 9), the attack and defense of the CDC (Topic 14), and the administrative and legislative response to COVID (Topic 32) are positively predictive of neutral Tweets. While defending/attacking the CDC might sound polarizing, the Tweets themselves can have a more neutral sentiment since they can be nuanced. For example, the author of the Tweet could specify that while they support the CDC's efforts, they do not think the organization is doing enough to counter the virus. Specifically, they could focus on the fact that the CDC only publishes data when it can have a more positive context – when the COVID infection numbers are decreasing. For reference, Table \ref{tab:behavior_echo} shows that CDC Tweets are more positive than any other entity, cluster, or theme with an average VADER score of 0.17. 

\subsubsection{Negative Tweets} \label{subsec:negative}
Higher retweets and follower-following ratios are less predictive of negative Tweets. The more a Tweet discusses COVID infections and mortality rates (Topic 25) or the opioid pandemic data (Topic 19), the more they are likely to evince a negative sentiment. 

The first topic can be more negative because it usually focuses on attacking the CDC for reporting mortality numbers that are unexpectedly high or low (note that while COVID infection numbers are heavily discussed in positive tweets which usually reported decreasing trends, positive tweets do not focus on mortality rates). This pattern reflects a political divide: some users downplayed the severity of COVID—labeling it a hoax or `plandemic'—while others used data to underscore the seriousness of the pandemic, citing high infection rates. The conversation often revolved around whether a low mortality rate diminishes the perceived threat of COVID, and whether mortality alone should define the public health response. Some tweets argued that even a low death rate, when multiplied by tens of millions of cases, results in a significant number of deaths. Others emphasized that mortality is not the only outcome of concern, pointing to the risk of long-term health complications following infection

Although Topic 19 included mentions of how social distancing limited access to essential services for opioid addicts, the primary focus was on the ongoing rise in opioid addiction and the shortage of available support services.

\subsubsection{Tweets using rich media} \label{subsec:rich_Tweets}
Tweets using rich media are more likely to be retweeted, liked, quote-tweeted, get a reply, and engage with the CDC through mentions or replies (Topic 5). Tweets using rich media are less likely to be posted from verified accounts, or to discuss lack of up-to-date COVID testing data (Topic 6), seasonal flu mortality rates (Topic 16) or music and stress (Topic 61).

\subsubsection{Tweets using low credibility sources} \label{subsec:low_credibility}
The use of rich media and verified accounts are more predictive of the use of low credibility sources. The focus on discussing the opioid pandemic (Topic 19) and COVID mortality rates (Topic 7) are also positively predictive of low credibility sources. 

Tweets using low credibility sources are less likely to be liked, retweeted, get responses, or engage in discussions regarding legislative and administrative responses to COVID (Topic 32) or Salmonella outbreaks (Topic 23).

\subsection{Verified accounts and accounts with high follower-following ratios} \label{sec:verified_accts}
Verified accounts are more likely to be retweeted, quote-tweeted, liked, get direct responses, use rich media and engage in attacking or defending the CDC (Topic 14). Verified accounts are less likely to tweet (lower total tweet count) and less likely to discuss Sports players and attendance guidelines (Topic 21). While verified accounts engaged with less Tweets than other users, their messages were retweeted at a higher rate which is in-line with the high echo value for verified accounts indicated in Table \ref{tab:behavior_echo}. It is also interesting to note that responses to verified accounts in general, and the CDC in particular were around 5, which means that the user received 5 retweets for an engagement with a verified account or the CDC. 

While the use of low-credibility sources is not predictive of verified accounts, verified accounts are predictive of the use of low credibility sources in a Tweet. In other words, some verified accounts used low credibility sources in their engagement. One example is @alexberenson, a journalist who pushed back against most of the CDC's mitigation strategies. Berenson, with a verified journalist account, used several low-credibility sources and was mentioned along with the @CDCGov account with high frequency (see \S\ref{sec:finding_mention}) which would expand his audience further.

Accounts with high FFR were more likely to receive likes and responses. They were more likely to engage in positive or neutral messaging and use rich media in their Tweets.

\section{Discussion}
We first discuss the lack of crisis message clarity in a continuously evolving pandemic enmeshed in polarized political framing seen through the lens of Twitter affordances in \S\ref{label:theory}. Next, we propose design recommendations based on our findings, specifically, a CDC AI assistant in \S\ref{sec:engage_gen_ai}. Finally, we present policy implications for our findings to support the CDC and other public health organizations in managing online communication for emerging infectious diseases in \S\ref{sec:policy_rec}. 

\subsection{Contributions to current literature} \label{label:theory}
The following three subsections examine how different features of Twitter shape the visibility and impact of public health discourse. First, we analyze how specific affordances—retweets, quote-tweets, replies, and mentions—drive message propagation, revealing patterns of endorsement, contestation, and attention-seeking within echo chambers. Second, we show how sentiment, source credibility, and rich media interact to amplify polarized topics like vaccination and masking, with both low-credibility sources and verified users contributing to virality. Third, we highlight the asymmetry in institutional communication, showing that while CDC tweets achieve high reach, their limited two-way engagement reflects a top-down strategy, underscoring the need for more dialogic, responsive models in public health messaging.

\subsubsection{Affordance-Specific Propagation Patterns}
Consistent with prior work \citep{kim2012role, bonson2019twitter}, we find that retweets and likes are the most common mechanisms through which messages are amplified on Twitter. Regression results (Table~\ref{tab:regression_results_1}) confirm that messages concerning highly salient and polarizing topics, such as vaccine discourse (Topic 0) and masking in schools (Topic 26), are significantly more likely to be retweeted and liked. However, the inverse relationship between direct replies and retweets suggests a divergence in how commitment and endorsement operate, with replies perhaps reflecting contention or engagement rather than support.

Quote-tweets and replies also reflect patterns of polarization. Topics associated with contentious policy or science communication (e.g., masking and vaccine mandates) were highly predictive of quote-tweets and replies, aligning with prior research on contentious engagement \citep{zade_et_al_24, Garimella_et_al_16}. While quote-tweets provide a channel for reinterpretation or reframing \citep{zade_et_al_24}, replies often served to contest the original narrative, either positively or negatively. This supports earlier findings that replies serve a dual role of engagement and pushback \citep{hemsley2018tweeting}.

The mention co-occurrence network (Figure~\ref{fig:mention_cooccurence}) revealed strategic mention practices. Users attempting to extend their audience often co-mentioned influential actors, including politicians, pharmaceutical companies, journalists, and agencies. ERGM results (Table~\ref{tab:ERGM}) showed that verified users were less likely to engage in reciprocal mention ties and that low-credibility content producers were more likely to initiate mentions but less likely to be mentioned in return. These findings extend existing work on attention-seeking behavior on Twitter \citep{mcgregor2019social} and suggest a dynamic of asymmetric influence where peripheral users attempt to leverage high-visibility accounts to gain traction.

Additionally, the ERGM results reveal structural signals consistent with echo chamber dynamics. Specifically, the significant negative association between verified user status and both in-degree and out-degree tie formation suggests a form of stratified engagement where influential users are largely isolated from reciprocal dialogue. Similarly, the positive coefficient for mutual ties indicates that users who engage in mentions tend to do so within tight reciprocal clusters, reinforcing ideologically aligned networks. The negative association with positive and negative sentiment, combined with the positive association with neutral sentiment, suggests that polarized users may not cross ideological lines in their engagements. Instead, they form bounded networks where views are reinforced rather than contested. This aligns with broader literature on echo chambers in social media, where homophily and ideological alignment shape information flows and interaction patterns \citep{keegan_et_al_2012}.

\subsubsection{The Role of Sentiment, Source Credibility, and Rich Media in Information Propagation} 
Building on the concept of imagined audiences \citep{litt2012knock, stewart_et_al_17}, our sentiment and Tag2Vec analyses (Figure~\ref{fig:kde}, Table~\ref{tab:tag2vec}) reveal sharp polarization in widely shared topics. Vaccination discourse (Topic 0) and masking in schools (Topic 26) exhibited bimodal sentiment patterns, with clusters expressing strong support, neutral information-sharing, or firm opposition. Positive sentiment emphasized collective health (e.g., \#beatcovid\_19), while negative sentiment included misinformation and ideological backlash (e.g., \#plandemic, \#timetounmask), reinforcing prior findings that polarization fuels virality \citep{freelon2016beyond}.

While earlier studies often equated negative sentiment with resistance to public health policy (e.g., \cite{unlu2023exploring}), our results show greater nuance. Negative sentiment sometimes reflected frustration that policies were too lenient (e.g., shortened quarantine guidelines; see \S\ref{subsec:negative}). Likewise, neutral posts—frequently overlooked—offered both critical reflections and localized public health messaging (e.g., school event updates; see \S\ref{subsec:neutral}).

Our data also show that rich media significantly boosts message propagation (Table~\ref{tab:behavior_echo}), and tweets from low-credibility sources were more likely to include such media (\S\ref{subsec:low_credibility}). This presents a growing challenge in the context of emerging generative media tools (e.g., text-to-image/video models; \cite{zhao_et_23}). Although tweets citing low-credibility sources were not among the most widely shared (\S\ref{sec:finding_rt}), it remains unclear whether this is due to platform moderation or other factors (cf. \cite{kayla_24,Sharevski_22}). More adaptive moderation tools are needed to identify persuasive misinformation even when shared by more credible or verified users, as further explored in \S\ref{sec:engage_gen_ai}.

Indeed, users relying on low-credibility sources were more active in the CDC public and more likely to mention others, often pairing their posts with persuasive media content \citep{ZHOU2021102679}. While their individual tweets may not achieve high propagation, their influence can be amplified when verified accounts share similar narratives. Notably, we found that verified journalist @alexberenson regularly cited questionable sources while opposing CDC mitigation policies (\S\ref{sec:finding_mention}).

Interestingly, neutral sentiment—often reflecting logistical or institutional communication—was more frequently mentioned and integrated into ongoing discourse. These messages, while less emotionally charged, appear to serve as pragmatic connectors across polarized groups, supporting the value of institutional voices in contested public conversations \citep{papacharissi2016affective}.

\subsubsection{Trusted But Distant: Verified Users and the CDC’s Top-Down Messaging}
Verified users and those with high follower-following ratios (FFRs) were significantly more likely to have their messages liked and retweeted, underscoring the persistent influence of network centrality and credibility cues \citep{Garimella_et_al_16, Suh_et_al_2010}. Given the prominence of such credibility signals, we turn to interactions involving the CDC as a key institutional actor.

While CDC-authored tweets exhibited the highest virality (echo ratio), responses directed at the CDC received comparatively low engagement. This asymmetry echoes critiques of institutional accounts as predominantly broadcast-oriented rather than dialogic \citep{dalrymple2016facts}. Despite over 21,000 tweets addressing the CDC—many incorporating rich media “receipts” referencing previous statements—the CDC issued only 58 direct replies to 25 accounts, mostly centered on vaccine questions and avoiding contentious topics.

This pattern reflects prior findings from crises like H1N1 and Ebola \citep{gesser2014risk, dalrymple2016facts}, where risk communication relied heavily on one-way messaging. However, two-way engagement remains central to the CERC model \citep{reynolds2005crisis}, which emphasizes building trust and incorporating public perspectives into institutional messaging \citep{dozier2013manager}. As discussed in \S\ref{sec:engage_gen_ai}, we offer design recommendations to support more responsive, bottom-up public health communication. While dialogue with the public is vital to fostering scientific understanding \citep{felt2020exploring}, it also opens the door to potential misrepresentation or misinformation—an issue we take up in the next section \citep{allcott2019trends}.

\subsection{Design Recommendation: The CDC AI Assistant} \label{sec:engage_gen_ai}
Building on recent work exploring generative AI in public health communication \citep{sanders2022should, quere2025state, aubin2024llms, schroeder2024large}, we propose the \textit{CDC AI Assistant}, an artificially intelligent communication technology \citep{endacott2022artificial} modeled after systems like Meta AI \citep{facebook_ai_2025} and generative agents with retrieval-augmented architectures \citep{Park_et_al_23}. The assistant operates in collaboration with CDC staff via a human-in-the-loop model \citep{WU2022364}, ensuring organizational control over messaging.The assistant is triggered in two scenarios: (1) replies to @CDCGov, and (2) suspected misinformation. 

In the first case, the AI is fine-tuned on CDC’s prior public communications and functions like a \textit{CoAuthor} assistant \citep{lee_et_al_2022_coauthor}, offering draft suggestions with explanatory chains of thought \citep{zhang_discussion_2023, wei_2022_cot}. These suggestions are reviewed, edited, and archived by CDC staff for future model training.

To address misinformation, the assistant incorporates both community-based moderation strategies \citep{bozarth_et_al_2023,SIMON2015609} and automated tools trained on misinformation datasets \citep{HAYAWI202223}, enabling it to flag and respond to repeated or reworded false claims \citep{sahu2021towards}. This approach responds to distrust in platform-based moderation \citep{Jhaver_et_al_2023} and the evolving nature of health-related misinformation during crises like COVID-19 \citep{wang_et_al_23,hobbs2024low}.

Importantly, our analysis of rich media use and sentiment distributions further supports the need for AI tools that understand the rhetorical and emotional contours of misinformation. Negative sentiment tweets, often paired with rich media and low-credibility sources, were a major vector for polarized discourse. A well-designed assistant would support public health communication professionals to shift public health communication from reactive crisis management toward proactive, participatory dialogue—anchored in transparency, adaptability, and institutional memory.

\subsection{Policy Recommendation: Preparing for the long-now of health crises} \label{sec:policy_rec}
Given the importance of two-way communication—and the lack thereof on Twitter as evidenced in this study—public health institutions like the CDC should invest in more participatory communication channels. Initiatives such as \#CDCChats \citep{dalrymple2016facts} offer a compelling model, allowing the public to directly engage with medical professionals in real time, fostering greater transparency and responsiveness. 

Our design recommendation in Section \S\ref{sec:engage_gen_ai} builds on this model to advocate for long-term engagement infrastructures. By systematically archiving and annotating user interactions, organizations can better contextualize evolving public concerns and trace how changes in guidance or misunderstandings contribute to confusion or polarization. This approach activates the temporal dimension of crisis communication highlighted in \cite{zhang2019social}. These user interactions function as "citizen sensors" \citep{goodchild2007citizens}, where individuals collectively contribute valuable, real-time insights. As \cite{HUANG2022102967} note, such networks facilitate rapid data collection and analysis on emerging issues, providing institutions with a grassroots-level understanding of public sentiment and unmet informational needs.

Moreover, public health agencies should adapt their platform strategies to meet different demographic and cultural communication preferences. For instance, recent findings show that many vaccine-hesitant parents turned to WhatsApp to make sense of COVID-19 information \citep{joy2024if}, highlighting the need for multi-platform engagement that complements formal messaging strategies.

This becomes especially critical when considering the role of earlier CDC communications as flashpoints in polarized discourse. As our analysis showed, Topics 0 (vaccination) and 26 (school masking) were highly propagated and deeply polarized across all Twitter affordances. Neutral messaging often contained institutional updates, while positive and negative clusters revealed affective alignment with or resistance to public health policy. Negative sentiment frequently drew on prior CDC messaging perceived as inconsistent or lacking nuance. Users circulated screenshots—"receipts"—of earlier statements to critique updated guidelines, especially around masking and vaccine efficacy. This was amplified in tweets referencing new variants like \#ArcturusRex and contentious issues like \#vaccine\_passports, sometimes shared by influential accounts known for questioning institutional credibility (e.g., @alexberenson, Dr. Malone). In such contexts, a lack of sustained dialogic engagement allowed misinformation to fester and gain traction.

To counteract this, institutions must not only broadcast accurate information but also contextualize evolving guidance through dialogic engagement, participatory feedback mechanisms, and multimodal outreach. Doing so transforms social media from a unidirectional channel into an adaptive, resilient space for co-producing public health understanding thus producing better framing for public scientific messaging \citep{nisbet_et_al_03}.


\section{limitations and future work}
While our findings show that image use increases the chance of message propagation (see \S\ref{sec:finding_rt}), our work does not analyze the image contents of the tweets over time. Given that images are potent ways to communicate messages, it is important to understand how they factor into public health crisis communication. Future work can use critical discourse analysis methods to cluster and analyze the change in framing in visual content over time \citep{walters_et_al_24}. Additionally, while our study analyzes a substantial amount of data, we have not engaged directly with users to better understand how they viewed the discourse around COVID in the CDC arena. Therefore, future work should build on \cite{Pine_et_al_21}'s analysis by conducting further interviews with users while incorporating not only the way users consumed information in healthcare publics, but how they contributed to the conversation, or are currently transported to new online communities. Our findings have implications for designing interventions to promote credible information and manage polarization. For institutions like the CDC, balancing credibility with dialogic engagement remains a challenge. Amplifying neutral, informative content and engaging strategically with public replies may be key strategies moving forward. Future research should examine affordance usage longitudinally to understand shifts in propagation behavior over time and across evolving crisis stages.

\section{Conclusion}
This study investigates how Twitter functioned as a key arena for public health discourse during the COVID-19 pandemic, particularly in relation to the CDC’s messaging. Analyzing over 275,000 tweets, the research found that the CDC predominantly employed one-way communication, while users actively sought engagement and accountability, often using screenshots of past CDC statements to highlight inconsistencies and question credibility. Public discourse centered on contentious topics like vaccines and school masking, shaped by sentiment, media use, and platform dynamics. Verified users and influential accounts received higher engagement but also circulated more low-credibility content, with rich media amplifying both reliable and unreliable information. These findings highlight the need for public health institutions to adopt more dialogic communication approaches and implement design interventions—such as AI-assisted moderation and engagement tools—to foster trust, responsiveness, and resilience against misinformation in future health crises.


\bibliographystyle{ACM-Reference-Format}
\bibliography{99_refs}


\begin{thebibliography}{163}


\ifx \showCODEN    \undefined \def \showCODEN     #1{\unskip}     \fi
\ifx \showDOI      \undefined \def \showDOI       #1{#1}\fi
\ifx \showISBNx    \undefined \def \showISBNx     #1{\unskip}     \fi
\ifx \showISBNxiii \undefined \def \showISBNxiii  #1{\unskip}     \fi
\ifx \showISSN     \undefined \def \showISSN      #1{\unskip}     \fi
\ifx \showLCCN     \undefined \def \showLCCN      #1{\unskip}     \fi
\ifx \shownote     \undefined \def \shownote      #1{#1}          \fi
\ifx \showarticletitle \undefined \def \showarticletitle #1{#1}   \fi
\ifx \showURL      \undefined \def \showURL       {\relax}        \fi
\providecommand\bibfield[2]{#2}
\providecommand\bibinfo[2]{#2}
\providecommand\natexlab[1]{#1}
\providecommand\showeprint[2][]{arXiv:#2}

\bibitem[Abelson et~al\mbox{.}(2009)]%
        {abelson2009does}
\bibfield{author}{\bibinfo{person}{Julia Abelson}, \bibinfo{person}{Fiona~A Miller}, {and} \bibinfo{person}{Mita Giacomini}.} \bibinfo{year}{2009}\natexlab{}.
\newblock \showarticletitle{What does it mean to trust a health system?: A qualitative study of Canadian health care values}.
\newblock \bibinfo{journal}{\emph{Health policy}} \bibinfo{volume}{91}, \bibinfo{number}{1} (\bibinfo{year}{2009}), \bibinfo{pages}{63--70}.
\newblock


\bibitem[Allcott et~al\mbox{.}(2019)]%
        {allcott2019trends}
\bibfield{author}{\bibinfo{person}{Hunt Allcott}, \bibinfo{person}{Matthew Gentzkow}, {and} \bibinfo{person}{Chuan Yu}.} \bibinfo{year}{2019}\natexlab{}.
\newblock \showarticletitle{Trends in the diffusion of misinformation on social media}.
\newblock \bibinfo{journal}{\emph{Research \& politics}} \bibinfo{volume}{6}, \bibinfo{number}{2} (\bibinfo{year}{2019}), \bibinfo{pages}{2053168019848554}.
\newblock


\bibitem[Ammari et~al\mbox{.}(2019)]%
        {ammari_et_al_19}
\bibfield{author}{\bibinfo{person}{Tawfiq Ammari}, \bibinfo{person}{Sarita Schoenebeck}, {and} \bibinfo{person}{Daniel Romero}.} \bibinfo{year}{2019}\natexlab{}.
\newblock \showarticletitle{Self-declared Throwaway Accounts on Reddit: How Platform Affordances and Shared Norms enable Parenting Disclosure and Support}.
\newblock \bibinfo{journal}{\emph{Proc. ACM Hum.-Comput. Interact.}} \bibinfo{volume}{3}, \bibinfo{number}{CSCW}, Article \bibinfo{articleno}{135} (\bibinfo{date}{nov} \bibinfo{year}{2019}), \bibinfo{numpages}{30}~pages.
\newblock
\urldef\tempurl%
\url{https://doi.org/10.1145/3359237}
\showDOI{\tempurl}


\bibitem[Anger and Kittl(2011)]%
        {anger_kittle_2011}
\bibfield{author}{\bibinfo{person}{Isabel Anger} {and} \bibinfo{person}{Christian Kittl}.} \bibinfo{year}{2011}\natexlab{}.
\newblock \showarticletitle{Measuring influence on Twitter}. In \bibinfo{booktitle}{\emph{Proceedings of the 11th International Conference on Knowledge Management and Knowledge Technologies}} (Graz, Austria) \emph{(\bibinfo{series}{i-KNOW '11})}. \bibinfo{publisher}{Association for Computing Machinery}, \bibinfo{address}{New York, NY, USA}, Article \bibinfo{articleno}{31}, \bibinfo{numpages}{4}~pages.
\newblock
\showISBNx{9781450307321}
\urldef\tempurl%
\url{https://doi.org/10.1145/2024288.2024326}
\showDOI{\tempurl}


\bibitem[Aubin Le~Qu{\'e}r{\'e} et~al\mbox{.}(2024)]%
        {aubin2024llms}
\bibfield{author}{\bibinfo{person}{Marianne Aubin Le~Qu{\'e}r{\'e}}, \bibinfo{person}{Hope Schroeder}, \bibinfo{person}{Casey Randazzo}, \bibinfo{person}{Jie Gao}, \bibinfo{person}{Ziv Epstein}, \bibinfo{person}{Simon~Tangi Perrault}, \bibinfo{person}{David Mimno}, \bibinfo{person}{Louise Barkhuus}, {and} \bibinfo{person}{Hanlin Li}.} \bibinfo{year}{2024}\natexlab{}.
\newblock \showarticletitle{LLMs as research tools: Applications and evaluations in HCI data work}. In \bibinfo{booktitle}{\emph{Extended Abstracts of the CHI Conference on Human Factors in Computing Systems}}. \bibinfo{pages}{1--7}.
\newblock


\bibitem[Axelrod and Hamilton(1981)]%
        {axelrod1981evolution}
\bibfield{author}{\bibinfo{person}{Robert Axelrod} {and} \bibinfo{person}{William~D Hamilton}.} \bibinfo{year}{1981}\natexlab{}.
\newblock \showarticletitle{The evolution of cooperation}.
\newblock \bibinfo{journal}{\emph{science}} \bibinfo{volume}{211}, \bibinfo{number}{4489} (\bibinfo{year}{1981}), \bibinfo{pages}{1390--1396}.
\newblock


\bibitem[Baden et~al\mbox{.}(2020)]%
        {baden2020hybrid}
\bibfield{author}{\bibinfo{person}{Christian Baden}, \bibinfo{person}{Neta Kligler-Vilenchik}, {and} \bibinfo{person}{Moran Yarchi}.} \bibinfo{year}{2020}\natexlab{}.
\newblock \showarticletitle{Hybrid content analysis: Toward a strategy for the theory-driven, computer-assisted classification of large text corpora}.
\newblock \bibinfo{journal}{\emph{Communication Methods and Measures}} \bibinfo{volume}{14}, \bibinfo{number}{3} (\bibinfo{year}{2020}), \bibinfo{pages}{165--183}.
\newblock


\bibitem[Bakshy et~al\mbox{.}(2011)]%
        {bakshy_et_al_2011}
\bibfield{author}{\bibinfo{person}{Eytan Bakshy}, \bibinfo{person}{Jake~M. Hofman}, \bibinfo{person}{Winter~A. Mason}, {and} \bibinfo{person}{Duncan~J. Watts}.} \bibinfo{year}{2011}\natexlab{}.
\newblock \showarticletitle{Everyone's an influencer: quantifying influence on twitter}. In \bibinfo{booktitle}{\emph{Proceedings of the Fourth ACM International Conference on Web Search and Data Mining}} (Hong Kong, China) \emph{(\bibinfo{series}{WSDM '11})}. \bibinfo{publisher}{Association for Computing Machinery}, \bibinfo{address}{New York, NY, USA}, \bibinfo{pages}{65–74}.
\newblock
\showISBNx{9781450304931}
\urldef\tempurl%
\url{https://doi.org/10.1145/1935826.1935845}
\showDOI{\tempurl}


\bibitem[Beigi et~al\mbox{.}(2016)]%
        {beigi2016overview}
\bibfield{author}{\bibinfo{person}{Ghazaleh Beigi}, \bibinfo{person}{Xia Hu}, \bibinfo{person}{Ross Maciejewski}, {and} \bibinfo{person}{Huan Liu}.} \bibinfo{year}{2016}\natexlab{}.
\newblock \showarticletitle{An overview of sentiment analysis in social media and its applications in disaster relief}.
\newblock \bibinfo{journal}{\emph{Sentiment analysis and ontology engineering: An environment of computational intelligence}} (\bibinfo{year}{2016}), \bibinfo{pages}{313--340}.
\newblock


\bibitem[Blei and Lafferty(2006)]%
        {blei_dynamic_2006}
\bibfield{author}{\bibinfo{person}{David~M. Blei} {and} \bibinfo{person}{John~D. Lafferty}.} \bibinfo{year}{2006}\natexlab{}.
\newblock \showarticletitle{Dynamic {Topic} {Models}}. In \bibinfo{booktitle}{\emph{Proceedings of the 23rd {International} {Conference} on {Machine} {Learning}}} \emph{(\bibinfo{series}{{ICML} '06})}. \bibinfo{publisher}{ACM}, \bibinfo{address}{New York, NY, USA}, \bibinfo{pages}{113--120}.
\newblock
\showISBNx{978-1-59593-383-6}
\urldef\tempurl%
\url{https://doi.org/10.1145/1143844.1143859}
\showDOI{\tempurl}


\bibitem[Bons{\'o}n et~al\mbox{.}(2019)]%
        {bonson2019twitter}
\bibfield{author}{\bibinfo{person}{Enrique Bons{\'o}n}, \bibinfo{person}{David Perea}, {and} \bibinfo{person}{Michaela Bedn{\'a}rov{\'a}}.} \bibinfo{year}{2019}\natexlab{}.
\newblock \showarticletitle{Twitter as a tool for citizen engagement: An empirical study of the Andalusian municipalities}.
\newblock \bibinfo{journal}{\emph{Government information quarterly}} \bibinfo{volume}{36}, \bibinfo{number}{3} (\bibinfo{year}{2019}), \bibinfo{pages}{480--489}.
\newblock


\bibitem[Boyd et~al\mbox{.}(2010)]%
        {boyd2010tweet}
\bibfield{author}{\bibinfo{person}{Danah Boyd}, \bibinfo{person}{Scott Golder}, {and} \bibinfo{person}{Gilad Lotan}.} \bibinfo{year}{2010}\natexlab{}.
\newblock \showarticletitle{Tweet, tweet, retweet: Conversational aspects of retweeting on twitter}. In \bibinfo{booktitle}{\emph{2010 43rd Hawaii international conference on system sciences}}. IEEE, \bibinfo{pages}{1--10}.
\newblock


\bibitem[Bozarth et~al\mbox{.}(2023)]%
        {bozarth_et_al_2023}
\bibfield{author}{\bibinfo{person}{Lia Bozarth}, \bibinfo{person}{Jane Im}, \bibinfo{person}{Christopher Quarles}, {and} \bibinfo{person}{Ceren Budak}.} \bibinfo{year}{2023}\natexlab{}.
\newblock \showarticletitle{Wisdom of Two Crowds: Misinformation Moderation on Reddit and How to Improve this Process---A Case Study of COVID-19}.
\newblock \bibinfo{journal}{\emph{Proc. ACM Hum.-Comput. Interact.}} \bibinfo{volume}{7}, \bibinfo{number}{CSCW1}, Article \bibinfo{articleno}{155} (\bibinfo{date}{April} \bibinfo{year}{2023}), \bibinfo{numpages}{33}~pages.
\newblock
\urldef\tempurl%
\url{https://doi.org/10.1145/3579631}
\showDOI{\tempurl}


\bibitem[Bruckman(2002)]%
        {bruckman_studying_2002}
\bibfield{author}{\bibinfo{person}{Amy Bruckman}.} \bibinfo{year}{2002}\natexlab{}.
\newblock \showarticletitle{Studying the amateur artist: {A} perspective on disguising data collected in human subjects research on the {Internet}}.
\newblock \bibinfo{journal}{\emph{Ethics and Information Technology}} \bibinfo{volume}{4}, \bibinfo{number}{3} (\bibinfo{year}{2002}), \bibinfo{pages}{217--231}.
\newblock


\bibitem[Br{\"u}ggemann and Meyer(2023)]%
        {bruggemann2023debates}
\bibfield{author}{\bibinfo{person}{Michael Br{\"u}ggemann} {and} \bibinfo{person}{Hendrik Meyer}.} \bibinfo{year}{2023}\natexlab{}.
\newblock \showarticletitle{When debates break apart: discursive polarization as a multi-dimensional divergence emerging in and through communication}.
\newblock \bibinfo{journal}{\emph{Communication Theory}} \bibinfo{volume}{33}, \bibinfo{number}{2-3} (\bibinfo{year}{2023}), \bibinfo{pages}{132--142}.
\newblock


\bibitem[Campello et~al\mbox{.}(2013)]%
        {campello_density-based_2013}
\bibfield{author}{\bibinfo{person}{Ricardo J. G.~B. Campello}, \bibinfo{person}{Davoud Moulavi}, {and} \bibinfo{person}{Joerg Sander}.} \bibinfo{year}{2013}\natexlab{}.
\newblock \showarticletitle{Density-{Based} {Clustering} {Based} on {Hierarchical} {Density} {Estimates}}. In \bibinfo{booktitle}{\emph{Advances in {Knowledge} {Discovery} and {Data} {Mining}}} \emph{(\bibinfo{series}{Lecture {Notes} in {Computer} {Science}})}, \bibfield{editor}{\bibinfo{person}{Jian Pei}, \bibinfo{person}{Vincent~S. Tseng}, \bibinfo{person}{Longbing Cao}, \bibinfo{person}{Hiroshi Motoda}, {and} \bibinfo{person}{Guandong Xu}} (Eds.). \bibinfo{publisher}{Springer}, \bibinfo{address}{Berlin, Heidelberg}, \bibinfo{pages}{160--172}.
\newblock
\showISBNx{978-3-642-37456-2}
\urldef\tempurl%
\url{https://doi.org/10.1007/978-3-642-37456-2_14}
\showDOI{\tempurl}


\bibitem[Casley(2010)]%
        {christina_2010}
\bibfield{author}{\bibinfo{person}{Christina Casley}.} \bibinfo{year}{2010}\natexlab{}.
\newblock \showarticletitle{{Denying AIDS: Conspiracy theories, pseudoscience and human tragedy, by Seth Kalichman}}.
\newblock \bibinfo{journal}{\emph{African Affairs}} \bibinfo{volume}{109}, \bibinfo{number}{436} (\bibinfo{date}{07} \bibinfo{year}{2010}), \bibinfo{pages}{505--506}.
\newblock
\showISSN{0001-9909}
\urldef\tempurl%
\url{https://doi.org/10.1093/afraf/adq035}
\showDOI{\tempurl}
\showeprint{https://academic.oup.com/afraf/article-pdf/109/436/505/343049/adq035.pdf}


\bibitem[Chang et~al\mbox{.}(2009)]%
        {chang_reading_2009}
\bibfield{author}{\bibinfo{person}{Jonathan Chang}, \bibinfo{person}{Sean Gerrish}, \bibinfo{person}{Chong Wang}, \bibinfo{person}{Jordan~L. Boyd-Graber}, {and} \bibinfo{person}{David~M. Blei}.} \bibinfo{year}{2009}\natexlab{}.
\newblock \showarticletitle{Reading tea leaves: {How} humans interpret topic models}. In \bibinfo{booktitle}{\emph{Advances in neural information processing systems}}. \bibinfo{pages}{288--296}.
\newblock


\bibitem[Chen et~al\mbox{.}(2017)]%
        {chen2017doctag2vec}
\bibfield{author}{\bibinfo{person}{Sheng Chen}, \bibinfo{person}{Akshay Soni}, \bibinfo{person}{Aasish Pappu}, {and} \bibinfo{person}{Yashar Mehdad}.} \bibinfo{year}{2017}\natexlab{}.
\newblock \showarticletitle{Doctag2vec: An embedding based multi-label learning approach for document tagging}.
\newblock \bibinfo{journal}{\emph{arXiv preprint arXiv:1707.04596}} (\bibinfo{year}{2017}).
\newblock


\bibitem[Dalrymple et~al\mbox{.}(2016)]%
        {dalrymple2016facts}
\bibfield{author}{\bibinfo{person}{Kajsa~E Dalrymple}, \bibinfo{person}{Rachel Young}, {and} \bibinfo{person}{Melissa Tully}.} \bibinfo{year}{2016}\natexlab{}.
\newblock \showarticletitle{“Facts, not fear” negotiating uncertainty on social media during the 2014 Ebola crisis}.
\newblock \bibinfo{journal}{\emph{Science communication}} \bibinfo{volume}{38}, \bibinfo{number}{4} (\bibinfo{year}{2016}), \bibinfo{pages}{442--467}.
\newblock


\bibitem[Dennison and Geddes(2019)]%
        {dennison2019rising}
\bibfield{author}{\bibinfo{person}{James Dennison} {and} \bibinfo{person}{Andrew Geddes}.} \bibinfo{year}{2019}\natexlab{}.
\newblock \showarticletitle{A rising tide? The salience of immigration and the rise of anti-immigration political parties in Western Europe}.
\newblock \bibinfo{journal}{\emph{The political quarterly}} \bibinfo{volume}{90}, \bibinfo{number}{1} (\bibinfo{year}{2019}), \bibinfo{pages}{107--116}.
\newblock


\bibitem[{DeVito} et~al\mbox{.}({[n.\,d.]})]%
        {devito_platforms_2017}
\bibfield{author}{\bibinfo{person}{Michael~A. {DeVito}}, \bibinfo{person}{Jeremy Birnholtz}, {and} \bibinfo{person}{Jeffery~T. Hancock}.} \bibinfo{year}{[n.\,d.]}\natexlab{}.
\newblock \showarticletitle{Platforms, people, and perception: using affordances to understand self-presentation on social media}. In \bibinfo{booktitle}{\emph{Proceedings of the 2017 {ACM} Conference on Computer Supported Cooperative Work and Social Computing}} (New York, {NY}, {USA}, 2017-02-25) \emph{(\bibinfo{series}{{CSCW} '17})}. \bibinfo{publisher}{Association for Computing Machinery}, \bibinfo{pages}{740--754}.
\newblock
\showISBNx{978-1-4503-4335-0}
\urldef\tempurl%
\url{https://doi.org/10.1145/2998181.2998192}
\showDOI{\tempurl}


\bibitem[Dou et~al\mbox{.}(2011)]%
        {dou_paralleltopics_2011}
\bibfield{author}{\bibinfo{person}{Wenwen Dou}, \bibinfo{person}{Xiaoyu Wang}, \bibinfo{person}{Remco Chang}, {and} \bibinfo{person}{William Ribarsky}.} \bibinfo{year}{2011}\natexlab{}.
\newblock \showarticletitle{{ParallelTopics}: {A} probabilistic approach to exploring document collections}. In \bibinfo{booktitle}{\emph{2011 {IEEE} {Conference} on {Visual} {Analytics} {Science} and {Technology} ({VAST})}}. \bibinfo{pages}{231--240}.
\newblock
\urldef\tempurl%
\url{https://doi.org/10.1109/VAST.2011.6102461}
\showDOI{\tempurl}


\bibitem[Dozier et~al\mbox{.}(2013)]%
        {dozier2013manager}
\bibfield{author}{\bibinfo{person}{David~M Dozier}, \bibinfo{person}{Larissa~A Grunig}, {and} \bibinfo{person}{James~E Grunig}.} \bibinfo{year}{2013}\natexlab{}.
\newblock \bibinfo{booktitle}{\emph{Manager's guide to excellence in public relations and communication management}}.
\newblock \bibinfo{publisher}{Routledge}.
\newblock


\bibitem[Du et~al\mbox{.}(2022)]%
        {du2022contribution}
\bibfield{author}{\bibinfo{person}{Tingting Du}, \bibinfo{person}{Prasanna Umar}, \bibinfo{person}{Sarah Rajtmajer}, {and} \bibinfo{person}{Anna Squicciarini}.} \bibinfo{year}{2022}\natexlab{}.
\newblock \showarticletitle{The contribution of verified accounts to self-disclosure in COVID-related Twitter conversations}. In \bibinfo{booktitle}{\emph{Proceedings of the international AAAI conference on web and social media}}, Vol.~\bibinfo{volume}{16}. \bibinfo{pages}{1393--1397}.
\newblock


\bibitem[Duskin(2024)]%
        {kayla_24}
\bibfield{author}{\bibinfo{person}{Kayla Duskin}.} \bibinfo{year}{2024}\natexlab{}.
\newblock \showarticletitle{Stemming the Tide of Problematic Information in Online Environments: Assessing Interventions and Identifying Opportunities for Interruption}. In \bibinfo{booktitle}{\emph{Companion Publication of the 16th ACM Web Science Conference}} (Stuttgart, Germany) \emph{(\bibinfo{series}{Websci Companion '24})}. \bibinfo{publisher}{Association for Computing Machinery}, \bibinfo{address}{New York, NY, USA}, \bibinfo{pages}{37–41}.
\newblock
\showISBNx{9798400704536}
\urldef\tempurl%
\url{https://doi.org/10.1145/3630744.3658615}
\showDOI{\tempurl}


\bibitem[Egger and Yu(2022)]%
        {egger2022topic}
\bibfield{author}{\bibinfo{person}{Roman Egger} {and} \bibinfo{person}{Joanne Yu}.} \bibinfo{year}{2022}\natexlab{}.
\newblock \showarticletitle{A topic modeling comparison between lda, nmf, top2vec, and bertopic to demystify twitter posts}.
\newblock \bibinfo{journal}{\emph{Frontiers in sociology}}  \bibinfo{volume}{7} (\bibinfo{year}{2022}), \bibinfo{pages}{886498}.
\newblock


\bibitem[Elbagir and Yang(2019)]%
        {elbagir2019twitter}
\bibfield{author}{\bibinfo{person}{Shihab Elbagir} {and} \bibinfo{person}{Jing Yang}.} \bibinfo{year}{2019}\natexlab{}.
\newblock \showarticletitle{Twitter sentiment analysis using natural language toolkit and VADER sentiment}. In \bibinfo{booktitle}{\emph{Proceedings of the international multiconference of engineers and computer scientists}}, Vol.~\bibinfo{volume}{122}. \bibinfo{pages}{16}.
\newblock


\bibitem[Endacott and Leonardi(2022)]%
        {endacott2022artificial}
\bibfield{author}{\bibinfo{person}{Camille~G Endacott} {and} \bibinfo{person}{Paul~M Leonardi}.} \bibinfo{year}{2022}\natexlab{}.
\newblock \showarticletitle{Artificial intelligence and impression management: Consequences of autonomous conversational agents communicating on one’s behalf}.
\newblock \bibinfo{journal}{\emph{Human Communication Research}} \bibinfo{volume}{48}, \bibinfo{number}{3} (\bibinfo{year}{2022}), \bibinfo{pages}{462--490}.
\newblock


\bibitem[Entman(1993)]%
        {entman1993framing}
\bibfield{author}{\bibinfo{person}{Robert~M Entman}.} \bibinfo{year}{1993}\natexlab{}.
\newblock \showarticletitle{Framing: Toward clarification of a fractured paradigm}.
\newblock \bibinfo{journal}{\emph{Journal of communication}} \bibinfo{volume}{43}, \bibinfo{number}{4} (\bibinfo{year}{1993}), \bibinfo{pages}{51--58}.
\newblock


\bibitem[Ewing and Vu(2021)]%
        {ewing2021navigating}
\bibfield{author}{\bibinfo{person}{Lee-Ann Ewing} {and} \bibinfo{person}{Huy~Quan Vu}.} \bibinfo{year}{2021}\natexlab{}.
\newblock \showarticletitle{Navigating ‘home schooling’during COVID-19: Australian public response on twitter}.
\newblock \bibinfo{journal}{\emph{Media International Australia}} \bibinfo{volume}{178}, \bibinfo{number}{1} (\bibinfo{year}{2021}), \bibinfo{pages}{77--86}.
\newblock


\bibitem[Faraj and Azad({[n.\,d.]})]%
        {faraj_materiality_2013}
\bibfield{author}{\bibinfo{person}{Samer Faraj} {and} \bibinfo{person}{Bijan Azad}.} \bibinfo{year}{[n.\,d.]}\natexlab{}.
\newblock \showarticletitle{The Materiality of Technology: An Affordance Perspective}.
\newblock In \bibinfo{booktitle}{\emph{Materiality and Organizing: Social Interaction in a Technological World}}. \bibinfo{pages}{237--258}.
\newblock
\showISBNx{978-0-19-966405-4}
\urldef\tempurl%
\url{https://doi.org/10.1093/acprof:oso/9780199664054.003.0012}
\showDOI{\tempurl}
\newblock
\shownote{Journal Abbreviation: Materiality and Organizing: Social Interaction in a Technological World}.


\bibitem[Fawcett(2006)]%
        {fawcett2006introduction}
\bibfield{author}{\bibinfo{person}{Tom Fawcett}.} \bibinfo{year}{2006}\natexlab{}.
\newblock \showarticletitle{An introduction to ROC analysis}.
\newblock \bibinfo{journal}{\emph{Pattern recognition letters}} \bibinfo{volume}{27}, \bibinfo{number}{8} (\bibinfo{year}{2006}), \bibinfo{pages}{861--874}.
\newblock


\bibitem[Felt and Davies(2020)]%
        {felt2020exploring}
\bibfield{author}{\bibinfo{person}{Ulrike Felt} {and} \bibinfo{person}{Sarah~R Davies}.} \bibinfo{year}{2020}\natexlab{}.
\newblock \bibinfo{booktitle}{\emph{Exploring science communication: A science and technology studies approach}}.
\newblock \bibinfo{publisher}{Sage}.
\newblock


\bibitem[Fiesler and Proferes(2018)]%
        {fiesler_participant_2018}
\bibfield{author}{\bibinfo{person}{Casey Fiesler} {and} \bibinfo{person}{Nicholas Proferes}.} \bibinfo{year}{2018}\natexlab{}.
\newblock \showarticletitle{“{Participant}” {Perceptions} of {Twitter} {Research} {Ethics}}.
\newblock \bibinfo{journal}{\emph{Social Media + Society}} \bibinfo{volume}{4}, \bibinfo{number}{1} (\bibinfo{date}{Jan.} \bibinfo{year}{2018}), \bibinfo{pages}{2056305118763366}.
\newblock
\showISSN{2056-3051}
\urldef\tempurl%
\url{https://doi.org/10.1177/2056305118763366}
\showDOI{\tempurl}
\newblock
\shownote{Publisher: SAGE Publications Ltd}.


\bibitem[Fox(2012)]%
        {fox2012medical}
\bibfield{author}{\bibinfo{person}{Renee~C Fox}.} \bibinfo{year}{2012}\natexlab{}.
\newblock \showarticletitle{Medical uncertainty revisited}.
\newblock In \bibinfo{booktitle}{\emph{Gender, health and healing}}. \bibinfo{publisher}{Routledge}, \bibinfo{pages}{236--253}.
\newblock


\bibitem[Freelon et~al\mbox{.}(2016)]%
        {freelon2016beyond}
\bibfield{author}{\bibinfo{person}{Deen Freelon}, \bibinfo{person}{Charlton~D McIlwain}, {and} \bibinfo{person}{Meredith Clark}.} \bibinfo{year}{2016}\natexlab{}.
\newblock \showarticletitle{Beyond the hashtags:\# Ferguson,\# Blacklivesmatter, and the online struggle for offline justice}.
\newblock \bibinfo{journal}{\emph{Center for Media \& Social Impact, American University, Forthcoming}} (\bibinfo{year}{2016}).
\newblock


\bibitem[Garimella et~al\mbox{.}(2016)]%
        {Garimella_et_al_16}
\bibfield{author}{\bibinfo{person}{Kiran Garimella}, \bibinfo{person}{Ingmar Weber}, {and} \bibinfo{person}{Munmun De~Choudhury}.} \bibinfo{year}{2016}\natexlab{}.
\newblock \showarticletitle{Quote RTs on Twitter: usage of the new feature for political discourse}. In \bibinfo{booktitle}{\emph{Proceedings of the 8th ACM Conference on Web Science}} (Hannover, Germany) \emph{(\bibinfo{series}{WebSci '16})}. \bibinfo{publisher}{Association for Computing Machinery}, \bibinfo{address}{New York, NY, USA}, \bibinfo{pages}{200–204}.
\newblock
\showISBNx{9781450342087}
\urldef\tempurl%
\url{https://doi.org/10.1145/2908131.2908170}
\showDOI{\tempurl}


\bibitem[Gesser-Edelsburg et~al\mbox{.}(2014)]%
        {gesser2014risk}
\bibfield{author}{\bibinfo{person}{Anat Gesser-Edelsburg}, \bibinfo{person}{Emilio Mordini}, \bibinfo{person}{James~J James}, \bibinfo{person}{Donato Greco}, {and} \bibinfo{person}{Manfred~S Green}.} \bibinfo{year}{2014}\natexlab{}.
\newblock \showarticletitle{Risk communication recommendations and implementation during emerging infectious diseases: a case study of the 2009 H1N1 influenza pandemic}.
\newblock \bibinfo{journal}{\emph{Disaster medicine and public health preparedness}} \bibinfo{volume}{8}, \bibinfo{number}{2} (\bibinfo{year}{2014}), \bibinfo{pages}{158--169}.
\newblock


\bibitem[Gonz{\'a}lez-Bail{\'o}n and De~Domenico(2021)]%
        {gonzalez2021bots}
\bibfield{author}{\bibinfo{person}{Sandra Gonz{\'a}lez-Bail{\'o}n} {and} \bibinfo{person}{Manlio De~Domenico}.} \bibinfo{year}{2021}\natexlab{}.
\newblock \showarticletitle{Bots are less central than verified accounts during contentious political events}.
\newblock \bibinfo{journal}{\emph{Proceedings of the National Academy of Sciences}} \bibinfo{volume}{118}, \bibinfo{number}{11} (\bibinfo{year}{2021}), \bibinfo{pages}{e2013443118}.
\newblock


\bibitem[Goodchild(2007)]%
        {goodchild2007citizens}
\bibfield{author}{\bibinfo{person}{Michael~F Goodchild}.} \bibinfo{year}{2007}\natexlab{}.
\newblock \showarticletitle{Citizens as sensors: the world of volunteered geography}.
\newblock \bibinfo{journal}{\emph{GeoJournal}}  \bibinfo{volume}{69} (\bibinfo{year}{2007}), \bibinfo{pages}{211--221}.
\newblock


\bibitem[Graham(2014)]%
        {Graham_14}
\bibfield{author}{\bibinfo{person}{Melissa~W Graham}.} \bibinfo{year}{2014}\natexlab{}.
\newblock \showarticletitle{Government communication in the digital age: Social media’s effect on local government public relations}.
\newblock \bibinfo{journal}{\emph{Public Relations Inquiry}} \bibinfo{volume}{3}, \bibinfo{number}{3} (\bibinfo{year}{2014}), \bibinfo{pages}{361--376}.
\newblock
\urldef\tempurl%
\url{https://doi.org/10.1177/2046147X14545371}
\showDOI{\tempurl}
\showeprint{https://doi.org/10.1177/2046147X14545371}


\bibitem[Griffiths et~al\mbox{.}(2007)]%
        {griffiths_topics_2007}
\bibfield{author}{\bibinfo{person}{Thomas~L. Griffiths}, \bibinfo{person}{Mark Steyvers}, {and} \bibinfo{person}{Joshua~B. Tenenbaum}.} \bibinfo{year}{2007}\natexlab{}.
\newblock \showarticletitle{Topics in semantic representation}.
\newblock \bibinfo{journal}{\emph{Psychological Review}}  \bibinfo{volume}{114} (\bibinfo{year}{2007}), \bibinfo{pages}{211--244}.
\newblock
\showISSN{1939-1471}
\urldef\tempurl%
\url{https://doi.org/10.1037/0033-295X.114.2.211}
\showDOI{\tempurl}
\newblock
\shownote{Place: US Publisher: American Psychological Association}.


\bibitem[Grootendorst(2022)]%
        {grootendorst2022bertopic}
\bibfield{author}{\bibinfo{person}{Maarten Grootendorst}.} \bibinfo{year}{2022}\natexlab{}.
\newblock \showarticletitle{BERTopic: Neural topic modeling with a class-based TF-IDF procedure}.
\newblock \bibinfo{journal}{\emph{arXiv preprint arXiv:2203.05794}} (\bibinfo{year}{2022}).
\newblock


\bibitem[Grunig and Grunig(2008)]%
        {grunig2008excellence}
\bibfield{author}{\bibinfo{person}{James~E Grunig} {and} \bibinfo{person}{Larissa~A Grunig}.} \bibinfo{year}{2008}\natexlab{}.
\newblock \showarticletitle{Excellence theory in public relations: Past, present, and future}.
\newblock In \bibinfo{booktitle}{\emph{Public relations research: European and international perspectives and innovations}}. \bibinfo{publisher}{Springer}, \bibinfo{pages}{327--347}.
\newblock


\bibitem[Gui et~al\mbox{.}(2017)]%
        {gui_et_al_17}
\bibfield{author}{\bibinfo{person}{Xinning Gui}, \bibinfo{person}{Yubo Kou}, \bibinfo{person}{Kathleen~H. Pine}, {and} \bibinfo{person}{Yunan Chen}.} \bibinfo{year}{2017}\natexlab{}.
\newblock \showarticletitle{Managing Uncertainty: Using Social Media for Risk Assessment during a Public Health Crisis}. In \bibinfo{booktitle}{\emph{Proceedings of the 2017 CHI Conference on Human Factors in Computing Systems}} (Denver, Colorado, USA) \emph{(\bibinfo{series}{CHI '17})}. \bibinfo{publisher}{Association for Computing Machinery}, \bibinfo{address}{New York, NY, USA}, \bibinfo{pages}{4520–4533}.
\newblock
\showISBNx{9781450346559}
\urldef\tempurl%
\url{https://doi.org/10.1145/3025453.3025891}
\showDOI{\tempurl}


\bibitem[Gulati et~al\mbox{.}(2024)]%
        {GULATI2024100495}
\bibfield{author}{\bibinfo{person}{Siddharth Gulati}, \bibinfo{person}{Joe McDonagh}, \bibinfo{person}{Sonia Sousa}, {and} \bibinfo{person}{David Lamas}.} \bibinfo{year}{2024}\natexlab{}.
\newblock \showarticletitle{Trust models and theories in human–computer interaction: A systematic literature review}.
\newblock \bibinfo{journal}{\emph{Computers in Human Behavior Reports}}  \bibinfo{volume}{16} (\bibinfo{year}{2024}), \bibinfo{pages}{100495}.
\newblock
\showISSN{2451-9588}
\urldef\tempurl%
\url{https://doi.org/10.1016/j.chbr.2024.100495}
\showDOI{\tempurl}


\bibitem[Gupta et~al\mbox{.}(2013)]%
        {gupta2013faking}
\bibfield{author}{\bibinfo{person}{Aditi Gupta}, \bibinfo{person}{Hemank Lamba}, \bibinfo{person}{Ponnurangam Kumaraguru}, {and} \bibinfo{person}{Anupam Joshi}.} \bibinfo{year}{2013}\natexlab{}.
\newblock \showarticletitle{Faking sandy: characterizing and identifying fake images on twitter during hurricane sandy}. In \bibinfo{booktitle}{\emph{Proceedings of the 22nd international conference on World Wide Web}}. \bibinfo{pages}{729--736}.
\newblock


\bibitem[Haas({[n.\,d.]})]%
        {haas_regressors}
\bibfield{author}{\bibinfo{person}{Nikhil Haas}.} \bibinfo{year}{[n.\,d.]}\natexlab{}.
\newblock \bibinfo{title}{Regressors: Easy utilities for fitting various regressors, extracting stats, and making relevant plots with scikit-learn models}.
\newblock \bibinfo{howpublished}{\url{https://github.com/nsh87/regressors}}.
\newblock
\newblock
\shownote{Accessed: 2025-03-05}.


\bibitem[Handcock et~al\mbox{.}(2019)]%
        {handcock2019package}
\bibfield{author}{\bibinfo{person}{Mark~S Handcock}, \bibinfo{person}{David~R Hunter}, \bibinfo{person}{Carter~T Butts}, \bibinfo{person}{Steven~M Goodreau}, \bibinfo{person}{Pavel~N Krivitsky}, \bibinfo{person}{Skye Bender-deMoll}, \bibinfo{person}{Martina Morris}, {and} \bibinfo{person}{Maintainer~Martina Morris}.} \bibinfo{year}{2019}\natexlab{}.
\newblock \bibinfo{title}{Package ‘statnet’}.
\newblock
\newblock


\bibitem[Hardy et~al\mbox{.}(2005)]%
        {hardy2005discourse}
\bibfield{author}{\bibinfo{person}{Cynthia Hardy}, \bibinfo{person}{Thomas~B Lawrence}, {and} \bibinfo{person}{David Grant}.} \bibinfo{year}{2005}\natexlab{}.
\newblock \showarticletitle{Discourse and collaboration: The role of conversations and collective identity}.
\newblock \bibinfo{journal}{\emph{Academy of management review}} \bibinfo{volume}{30}, \bibinfo{number}{1} (\bibinfo{year}{2005}), \bibinfo{pages}{58--77}.
\newblock


\bibitem[Harris et~al\mbox{.}(2024)]%
        {harris2024perceived}
\bibfield{author}{\bibinfo{person}{Mallory~J Harris}, \bibinfo{person}{Ryan Murtfeldt}, \bibinfo{person}{Shufan Wang}, \bibinfo{person}{Erin~A Mordecai}, {and} \bibinfo{person}{Jevin~D West}.} \bibinfo{year}{2024}\natexlab{}.
\newblock \showarticletitle{Perceived experts are prevalent and influential within an antivaccine community on Twitter}.
\newblock \bibinfo{journal}{\emph{PNAS nexus}} \bibinfo{volume}{3}, \bibinfo{number}{2} (\bibinfo{year}{2024}), \bibinfo{pages}{pgae007}.
\newblock


\bibitem[Haupt et~al\mbox{.}(2021)]%
        {haupt2021characterizing}
\bibfield{author}{\bibinfo{person}{Michael~Robert Haupt}, \bibinfo{person}{Alex Jinich-Diamant}, \bibinfo{person}{Jiawei Li}, \bibinfo{person}{Matthew Nali}, {and} \bibinfo{person}{Tim~K Mackey}.} \bibinfo{year}{2021}\natexlab{}.
\newblock \showarticletitle{Characterizing twitter user topics and communication network dynamics of the “Liberate” movement during COVID-19 using unsupervised machine learning and social network analysis}.
\newblock \bibinfo{journal}{\emph{Online Social Networks and Media}}  \bibinfo{volume}{21} (\bibinfo{year}{2021}), \bibinfo{pages}{100114}.
\newblock


\bibitem[Hayawi et~al\mbox{.}(2022)]%
        {HAYAWI202223}
\bibfield{author}{\bibinfo{person}{K. Hayawi}, \bibinfo{person}{S. Shahriar}, \bibinfo{person}{M.A. Serhani}, \bibinfo{person}{I. Taleb}, {and} \bibinfo{person}{S.S. Mathew}.} \bibinfo{year}{2022}\natexlab{}.
\newblock \showarticletitle{ANTi-Vax: a novel Twitter dataset for COVID-19 vaccine misinformation detection}.
\newblock \bibinfo{journal}{\emph{Public Health}}  \bibinfo{volume}{203} (\bibinfo{year}{2022}), \bibinfo{pages}{23--30}.
\newblock
\showISSN{0033-3506}
\urldef\tempurl%
\url{https://doi.org/10.1016/j.puhe.2021.11.022}
\showDOI{\tempurl}


\bibitem[Hays(1998)]%
        {hays_cultural_1998}
\bibfield{author}{\bibinfo{person}{Sharon Hays}.} \bibinfo{year}{1998}\natexlab{}.
\newblock \bibinfo{booktitle}{\emph{The {Cultural} {Contradictions} of {Motherhood}}}.
\newblock \bibinfo{publisher}{Yale University Press}.
\newblock
\showISBNx{978-0-300-07652-3}
\newblock
\shownote{Google-Books-ID: 5r5obtRneqoC}.


\bibitem[Hemsley et~al\mbox{.}(2018)]%
        {hemsley2018tweeting}
\bibfield{author}{\bibinfo{person}{Jeff Hemsley}, \bibinfo{person}{Jennifer Stromer-Galley}, \bibinfo{person}{Bryan Semaan}, {and} \bibinfo{person}{Sikana Tanupabrungsun}.} \bibinfo{year}{2018}\natexlab{}.
\newblock \showarticletitle{Tweeting to the target: candidates’ use of strategic messages and@ mentions on Twitter}.
\newblock \bibinfo{journal}{\emph{Journal of Information Technology \& Politics}} \bibinfo{volume}{15}, \bibinfo{number}{1} (\bibinfo{year}{2018}), \bibinfo{pages}{3--18}.
\newblock


\bibitem[Hobbs et~al\mbox{.}(2024)]%
        {hobbs2024low}
\bibfield{author}{\bibinfo{person}{Ali Hobbs}, \bibinfo{person}{Aisha Aldosery}, {and} \bibinfo{person}{Patty Kostkova}.} \bibinfo{year}{2024}\natexlab{}.
\newblock \showarticletitle{Low credibility URL sharing on Twitter during reporting linking rare blood clots with the Oxford/AstraZeneca COVID-19 vaccine}.
\newblock \bibinfo{journal}{\emph{Plos one}} \bibinfo{volume}{19}, \bibinfo{number}{1} (\bibinfo{year}{2024}), \bibinfo{pages}{e0296444}.
\newblock


\bibitem[Howarth(2000)]%
        {howarth2000discourse}
\bibfield{author}{\bibinfo{person}{David Howarth}.} \bibinfo{year}{2000}\natexlab{}.
\newblock \bibinfo{booktitle}{\emph{Discourse}}.
\newblock \bibinfo{publisher}{McGraw-Hill Education (UK)}.
\newblock


\bibitem[Huang and Huang(2023)]%
        {huang2023increasing}
\bibfield{author}{\bibinfo{person}{Alexander~A Huang} {and} \bibinfo{person}{Samuel~Y Huang}.} \bibinfo{year}{2023}\natexlab{}.
\newblock \showarticletitle{Increasing transparency in machine learning through bootstrap simulation and shapely additive explanations}.
\newblock \bibinfo{journal}{\emph{PLoS One}} \bibinfo{volume}{18}, \bibinfo{number}{2} (\bibinfo{year}{2023}), \bibinfo{pages}{e0281922}.
\newblock


\bibitem[Huang et~al\mbox{.}(2010)]%
        {huang_et_al_2010}
\bibfield{author}{\bibinfo{person}{Jeff Huang}, \bibinfo{person}{Katherine~M. Thornton}, {and} \bibinfo{person}{Efthimis~N. Efthimiadis}.} \bibinfo{year}{2010}\natexlab{}.
\newblock \showarticletitle{Conversational tagging in twitter}. In \bibinfo{booktitle}{\emph{Proceedings of the 21st ACM Conference on Hypertext and Hypermedia}} (Toronto, Ontario, Canada) \emph{(\bibinfo{series}{HT '10})}. \bibinfo{publisher}{Association for Computing Machinery}, \bibinfo{address}{New York, NY, USA}, \bibinfo{pages}{173–178}.
\newblock
\showISBNx{9781450300414}
\urldef\tempurl%
\url{https://doi.org/10.1145/1810617.1810647}
\showDOI{\tempurl}


\bibitem[Huang et~al\mbox{.}(2022)]%
        {HUANG2022102967}
\bibfield{author}{\bibinfo{person}{Xiao Huang}, \bibinfo{person}{Siqin Wang}, \bibinfo{person}{Mengxi Zhang}, \bibinfo{person}{Tao Hu}, \bibinfo{person}{Alexander Hohl}, \bibinfo{person}{Bing She}, \bibinfo{person}{Xi Gong}, \bibinfo{person}{Jianxin Li}, \bibinfo{person}{Xiao Liu}, \bibinfo{person}{Oliver Gruebner}, \bibinfo{person}{Regina Liu}, \bibinfo{person}{Xiao Li}, \bibinfo{person}{Zhewei Liu}, \bibinfo{person}{Xinyue Ye}, {and} \bibinfo{person}{Zhenlong Li}.} \bibinfo{year}{2022}\natexlab{}.
\newblock \showarticletitle{Social media mining under the COVID-19 context: Progress, challenges, and opportunities}.
\newblock \bibinfo{journal}{\emph{International Journal of Applied Earth Observation and Geoinformation}}  \bibinfo{volume}{113} (\bibinfo{year}{2022}), \bibinfo{pages}{102967}.
\newblock
\showISSN{1569-8432}
\urldef\tempurl%
\url{https://doi.org/10.1016/j.jag.2022.102967}
\showDOI{\tempurl}


\bibitem[Huang et~al\mbox{.}(2018)]%
        {huang_et_al_18}
\bibfield{author}{\bibinfo{person}{Yo-Ping Huang}, \bibinfo{person}{Nontobeko Hlongwane}, {and} \bibinfo{person}{Li-Jen Kao}.} \bibinfo{year}{2018}\natexlab{}.
\newblock \showarticletitle{Using Sentiment Analysis to Determine Users' Likes on Twitter}. In \bibinfo{booktitle}{\emph{2018 IEEE 16th Intl Conf on Dependable, Autonomic and Secure Computing, 16th Intl Conf on Pervasive Intelligence and Computing, 4th Intl Conf on Big Data Intelligence and Computing and Cyber Science and Technology Congress(DASC/PiCom/DataCom/CyberSciTech)}}. \bibinfo{pages}{1068--1073}.
\newblock
\urldef\tempurl%
\url{https://doi.org/10.1109/DASC/PiCom/DataCom/CyberSciTec.2018.00177}
\showDOI{\tempurl}


\bibitem[Hutto and Gilbert(2014)]%
        {hutto2014vader}
\bibfield{author}{\bibinfo{person}{Clayton Hutto} {and} \bibinfo{person}{Eric Gilbert}.} \bibinfo{year}{2014}\natexlab{}.
\newblock \showarticletitle{Vader: A parsimonious rule-based model for sentiment analysis of social media text}. In \bibinfo{booktitle}{\emph{Proceedings of the international AAAI conference on web and social media}}, Vol.~\bibinfo{volume}{8}. \bibinfo{pages}{216--225}.
\newblock


\bibitem[Javed et~al\mbox{.}(2023)]%
        {javed_et_al_23}
\bibfield{author}{\bibinfo{person}{Muhammad Javed}, \bibinfo{person}{Gerardo~Luis Dimaguila}, \bibinfo{person}{Sedigh~Khademi Habibabadi}, \bibinfo{person}{Chris Palmer}, {and} \bibinfo{person}{Jim Buttery}.} \bibinfo{year}{2023}\natexlab{}.
\newblock \showarticletitle{Learning from Machines? Social Bots Influence on COVID-19 Vaccination-Related Discussions: 2021 in Review}. In \bibinfo{booktitle}{\emph{Proceedings of the 2023 Australasian Computer Science Week}} (Melbourne, VIC, Australia) \emph{(\bibinfo{series}{ACSW '23})}. \bibinfo{publisher}{Association for Computing Machinery}, \bibinfo{address}{New York, NY, USA}, \bibinfo{pages}{190–197}.
\newblock
\showISBNx{9798400700057}
\urldef\tempurl%
\url{https://doi.org/10.1145/3579375.3579400}
\showDOI{\tempurl}


\bibitem[Jenders et~al\mbox{.}(2013)]%
        {jenders_2013}
\bibfield{author}{\bibinfo{person}{Maximilian Jenders}, \bibinfo{person}{Gjergji Kasneci}, {and} \bibinfo{person}{Felix Naumann}.} \bibinfo{year}{2013}\natexlab{}.
\newblock \showarticletitle{Analyzing and predicting viral tweets}. In \bibinfo{booktitle}{\emph{Proceedings of the 22nd International Conference on World Wide Web}} (Rio de Janeiro, Brazil) \emph{(\bibinfo{series}{WWW '13 Companion})}. \bibinfo{publisher}{Association for Computing Machinery}, \bibinfo{address}{New York, NY, USA}, \bibinfo{pages}{657–664}.
\newblock
\showISBNx{9781450320382}
\urldef\tempurl%
\url{https://doi.org/10.1145/2487788.2488017}
\showDOI{\tempurl}


\bibitem[Jhaver et~al\mbox{.}(2023)]%
        {Jhaver_et_al_2023}
\bibfield{author}{\bibinfo{person}{Shagun Jhaver}, \bibinfo{person}{Alice~Qian Zhang}, \bibinfo{person}{Quan~Ze Chen}, \bibinfo{person}{Nikhila Natarajan}, \bibinfo{person}{Ruotong Wang}, {and} \bibinfo{person}{Amy~X. Zhang}.} \bibinfo{year}{2023}\natexlab{}.
\newblock \showarticletitle{Personalizing Content Moderation on Social Media: User Perspectives on Moderation Choices, Interface Design, and Labor}.
\newblock \bibinfo{journal}{\emph{Proc. ACM Hum.-Comput. Interact.}} \bibinfo{volume}{7}, \bibinfo{number}{CSCW2}, Article \bibinfo{articleno}{289} (\bibinfo{date}{Oct.} \bibinfo{year}{2023}), \bibinfo{numpages}{33}~pages.
\newblock
\urldef\tempurl%
\url{https://doi.org/10.1145/3610080}
\showDOI{\tempurl}


\bibitem[Jiang et~al\mbox{.}(2019)]%
        {jiang2019diary}
\bibfield{author}{\bibinfo{person}{Tingting Jiang}, \bibinfo{person}{Qian Guo}, \bibinfo{person}{Yaping Xu}, {and} \bibinfo{person}{Shiting Fu}.} \bibinfo{year}{2019}\natexlab{}.
\newblock \showarticletitle{A diary study of information encountering triggered by visual stimuli on micro-blogging services}.
\newblock \bibinfo{journal}{\emph{Information Processing \& Management}} \bibinfo{volume}{56}, \bibinfo{number}{1} (\bibinfo{year}{2019}), \bibinfo{pages}{29--42}.
\newblock


\bibitem[Jin et~al\mbox{.}(2016)]%
        {jin2016novel}
\bibfield{author}{\bibinfo{person}{Zhiwei Jin}, \bibinfo{person}{Juan Cao}, \bibinfo{person}{Yongdong Zhang}, \bibinfo{person}{Jianshe Zhou}, {and} \bibinfo{person}{Qi Tian}.} \bibinfo{year}{2016}\natexlab{}.
\newblock \showarticletitle{Novel visual and statistical image features for microblogs news verification}.
\newblock \bibinfo{journal}{\emph{IEEE transactions on multimedia}} \bibinfo{volume}{19}, \bibinfo{number}{3} (\bibinfo{year}{2016}), \bibinfo{pages}{598--608}.
\newblock


\bibitem[Joy et~al\mbox{.}(2024)]%
        {joy2024if}
\bibfield{author}{\bibinfo{person}{Karen Joy}, \bibinfo{person}{Michelle Liang}, {and} \bibinfo{person}{Tawfiq Ammari}.} \bibinfo{year}{2024}\natexlab{}.
\newblock \showarticletitle{" If it has an exclamation point, I step away from it, I need facts, not excited feelings": Technologically Mediated Parental COVID Uncertainty}.
\newblock \bibinfo{journal}{\emph{arXiv preprint arXiv:2412.05181}} (\bibinfo{year}{2024}).
\newblock


\bibitem[Kallner(2017)]%
        {kallner2017laboratory}
\bibfield{author}{\bibinfo{person}{Anders Kallner}.} \bibinfo{year}{2017}\natexlab{}.
\newblock \bibinfo{booktitle}{\emph{Laboratory statistics: methods in chemistry and health sciences}}.
\newblock \bibinfo{publisher}{Elsevier}.
\newblock


\bibitem[Keegan et~al\mbox{.}(2012)]%
        {keegan_et_al_2012}
\bibfield{author}{\bibinfo{person}{Brian Keegan}, \bibinfo{person}{Darren Gergle}, {and} \bibinfo{person}{Noshir Contractor}.} \bibinfo{year}{2012}\natexlab{}.
\newblock \showarticletitle{Do Editors or Articles Drive Collaboration? Multilevel Statistical Network Analysis of Wikipedia Coauthorship}. In \bibinfo{booktitle}{\emph{Proceedings of the ACM 2012 Conference on Computer Supported Cooperative Work}} (Seattle, Washington, USA) \emph{(\bibinfo{series}{CSCW '12})}. \bibinfo{publisher}{Association for Computing Machinery}, \bibinfo{address}{New York, NY, USA}, \bibinfo{pages}{427–436}.
\newblock
\showISBNx{9781450310864}
\urldef\tempurl%
\url{https://doi.org/10.1145/2145204.2145271}
\showDOI{\tempurl}


\bibitem[Kim(2025)]%
        {kim2025basic}
\bibfield{author}{\bibinfo{person}{Jaehee Kim}.} \bibinfo{year}{2025}\natexlab{}.
\newblock \showarticletitle{Basic issues and challenges of statistical network analysis.}
\newblock \bibinfo{journal}{\emph{Communications for Statistical Applications \& Methods}} \bibinfo{volume}{32}, \bibinfo{number}{1} (\bibinfo{year}{2025}).
\newblock


\bibitem[Kim and Yoo(2012)]%
        {kim2012role}
\bibfield{author}{\bibinfo{person}{Jihie Kim} {and} \bibinfo{person}{Jaebong Yoo}.} \bibinfo{year}{2012}\natexlab{}.
\newblock \showarticletitle{Role of sentiment in message propagation: Reply vs. retweet behavior in political communication}. In \bibinfo{booktitle}{\emph{2012 international conference on social informatics}}. IEEE, \bibinfo{pages}{131--136}.
\newblock


\bibitem[Kou et~al\mbox{.}(2017)]%
        {kou_at_al_17}
\bibfield{author}{\bibinfo{person}{Yubo Kou}, \bibinfo{person}{Xinning Gui}, \bibinfo{person}{Yunan Chen}, {and} \bibinfo{person}{Kathleen Pine}.} \bibinfo{year}{2017}\natexlab{}.
\newblock \showarticletitle{Conspiracy Talk on Social Media: Collective Sensemaking during a Public Health Crisis}.
\newblock \bibinfo{journal}{\emph{Proc. ACM Hum.-Comput. Interact.}} \bibinfo{volume}{1}, \bibinfo{number}{CSCW}, Article \bibinfo{articleno}{61} (\bibinfo{date}{dec} \bibinfo{year}{2017}), \bibinfo{numpages}{21}~pages.
\newblock
\urldef\tempurl%
\url{https://doi.org/10.1145/3134696}
\showDOI{\tempurl}


\bibitem[Kountouri and Kollias(2023)]%
        {kountouri2023polarizing}
\bibfield{author}{\bibinfo{person}{Fani Kountouri} {and} \bibinfo{person}{Andreas Kollias}.} \bibinfo{year}{2023}\natexlab{}.
\newblock \showarticletitle{Polarizing publics in Twitter through organic targeting tactics of political incivility}.
\newblock \bibinfo{journal}{\emph{Frontiers in Political Science}}  \bibinfo{volume}{5} (\bibinfo{year}{2023}), \bibinfo{pages}{1110953}.
\newblock


\bibitem[Kwok et~al\mbox{.}(2021)]%
        {kwok2021tweet}
\bibfield{author}{\bibinfo{person}{Stephen Wai~Hang Kwok}, \bibinfo{person}{Sai~Kumar Vadde}, {and} \bibinfo{person}{Guanjin Wang}.} \bibinfo{year}{2021}\natexlab{}.
\newblock \showarticletitle{Tweet topics and sentiments relating to COVID-19 vaccination among Australian Twitter users: machine learning analysis}.
\newblock \bibinfo{journal}{\emph{Journal of medical Internet research}} \bibinfo{volume}{23}, \bibinfo{number}{5} (\bibinfo{year}{2021}), \bibinfo{pages}{e26953}.
\newblock


\bibitem[Lachlan et~al\mbox{.}(2016)]%
        {lachlan2016social}
\bibfield{author}{\bibinfo{person}{Kenneth~A Lachlan}, \bibinfo{person}{Patric~R Spence}, \bibinfo{person}{Xialing Lin}, \bibinfo{person}{Kristy Najarian}, {and} \bibinfo{person}{Maria Del~Greco}.} \bibinfo{year}{2016}\natexlab{}.
\newblock \showarticletitle{Social media and crisis management: CERC, search strategies, and Twitter content}.
\newblock \bibinfo{journal}{\emph{Computers in Human Behavior}}  \bibinfo{volume}{54} (\bibinfo{year}{2016}), \bibinfo{pages}{647--652}.
\newblock


\bibitem[Leavitt(2003)]%
        {leavitt2003public}
\bibfield{author}{\bibinfo{person}{Judith~Walzer Leavitt}.} \bibinfo{year}{2003}\natexlab{}.
\newblock \showarticletitle{Public resistance or cooperation? A tale of smallpox in two cities}.
\newblock \bibinfo{journal}{\emph{Biosecurity and Bioterrorism: Biodefense Strategy, Practice, and Science}} \bibinfo{volume}{1}, \bibinfo{number}{3} (\bibinfo{year}{2003}), \bibinfo{pages}{185--192}.
\newblock


\bibitem[Lee et~al\mbox{.}(2022)]%
        {lee_et_al_2022_coauthor}
\bibfield{author}{\bibinfo{person}{Mina Lee}, \bibinfo{person}{Percy Liang}, {and} \bibinfo{person}{Qian Yang}.} \bibinfo{year}{2022}\natexlab{}.
\newblock \showarticletitle{CoAuthor: Designing a Human-AI Collaborative Writing Dataset for Exploring Language Model Capabilities}. In \bibinfo{booktitle}{\emph{Proceedings of the 2022 CHI Conference on Human Factors in Computing Systems}} (New Orleans, LA, USA) \emph{(\bibinfo{series}{CHI '22})}. \bibinfo{publisher}{Association for Computing Machinery}, \bibinfo{address}{New York, NY, USA}, Article \bibinfo{articleno}{388}, \bibinfo{numpages}{19}~pages.
\newblock
\showISBNx{9781450391573}
\urldef\tempurl%
\url{https://doi.org/10.1145/3491102.3502030}
\showDOI{\tempurl}


\bibitem[Lemeshow and Hosmer~Jr(1982)]%
        {lemeshow1982review}
\bibfield{author}{\bibinfo{person}{Stanley Lemeshow} {and} \bibinfo{person}{David~W Hosmer~Jr}.} \bibinfo{year}{1982}\natexlab{}.
\newblock \showarticletitle{A review of goodness of fit statistics for use in the development of logistic regression models}.
\newblock \bibinfo{journal}{\emph{American journal of epidemiology}} \bibinfo{volume}{115}, \bibinfo{number}{1} (\bibinfo{year}{1982}), \bibinfo{pages}{92--106}.
\newblock


\bibitem[Leonardi and Vaast({[n.\,d.]})]%
        {leonardi_social_2017}
\bibfield{author}{\bibinfo{person}{Paul~M. Leonardi} {and} \bibinfo{person}{Emmanuelle Vaast}.} \bibinfo{year}{[n.\,d.]}\natexlab{}.
\newblock \bibinfo{title}{Social Media and Their Affordances for Organizing: A Review and Agenda for Research}.
\newblock
\newblock
\urldef\tempurl%
\url{https://papers.ssrn.com/abstract=2993824}
\showURL{%
\tempurl}


\bibitem[Li et~al\mbox{.}(2022)]%
        {LI2022113752}
\bibfield{author}{\bibinfo{person}{Kai Li}, \bibinfo{person}{Cheng Zhou}, \bibinfo{person}{Xin~(Robert) Luo}, \bibinfo{person}{Jose Benitez}, {and} \bibinfo{person}{Qinyu Liao}.} \bibinfo{year}{2022}\natexlab{}.
\newblock \showarticletitle{Impact of information timeliness and richness on public engagement on social media during COVID-19 pandemic: An empirical investigation based on NLP and machine learning}.
\newblock \bibinfo{journal}{\emph{Decision Support Systems}}  \bibinfo{volume}{162} (\bibinfo{year}{2022}), \bibinfo{pages}{113752}.
\newblock
\showISSN{0167-9236}
\urldef\tempurl%
\url{https://doi.org/10.1016/j.dss.2022.113752}
\showDOI{\tempurl}
\newblock
\shownote{Business and Government Applications of Text Mining \& Natural Language Processing (NLP) for Societal Benefit}.


\bibitem[Liao et~al\mbox{.}(2023)]%
        {liao_23}
\bibfield{author}{\bibinfo{person}{Song Liao}, \bibinfo{person}{Ebuka Okpala}, \bibinfo{person}{Long Cheng}, \bibinfo{person}{Mingqi Li}, \bibinfo{person}{Nishant Vishwamitra}, \bibinfo{person}{Hongxin Hu}, \bibinfo{person}{Feng Luo}, {and} \bibinfo{person}{Matthew Costello}.} \bibinfo{year}{2023}\natexlab{}.
\newblock \showarticletitle{Analysis of COVID-19 Offensive Tweets and Their Targets}. In \bibinfo{booktitle}{\emph{Proceedings of the 29th ACM SIGKDD Conference on Knowledge Discovery and Data Mining}} (Long Beach, CA, USA) \emph{(\bibinfo{series}{KDD '23})}. \bibinfo{publisher}{Association for Computing Machinery}, \bibinfo{address}{New York, NY, USA}, \bibinfo{pages}{4473–4484}.
\newblock
\showISBNx{9798400701030}
\urldef\tempurl%
\url{https://doi.org/10.1145/3580305.3599773}
\showDOI{\tempurl}


\bibitem[Litt(2012)]%
        {litt2012knock}
\bibfield{author}{\bibinfo{person}{Eden Litt}.} \bibinfo{year}{2012}\natexlab{}.
\newblock \showarticletitle{Knock, knock. Who's there? The imagined audience}.
\newblock \bibinfo{journal}{\emph{Journal of broadcasting \& electronic media}} \bibinfo{volume}{56}, \bibinfo{number}{3} (\bibinfo{year}{2012}), \bibinfo{pages}{330--345}.
\newblock


\bibitem[Lundberg and Lee(2017)]%
        {shaplundberg2017unified}
\bibfield{author}{\bibinfo{person}{Scott~M Lundberg} {and} \bibinfo{person}{Su-In Lee}.} \bibinfo{year}{2017}\natexlab{}.
\newblock \showarticletitle{A unified approach to interpreting model predictions}.
\newblock \bibinfo{journal}{\emph{Advances in neural information processing systems}}  \bibinfo{volume}{30} (\bibinfo{year}{2017}).
\newblock


\bibitem[Lunden(2020)]%
        {Lunden2020}
\bibfield{author}{\bibinfo{person}{Ingrid Lunden}.} \bibinfo{year}{2020}\natexlab{}.
\newblock \showarticletitle{Twitter prioritizes blue-check verifications to confirm experts on COVID-19 and the novel coronavirus}.
\newblock \bibinfo{journal}{\emph{TechCrunch}} (\bibinfo{year}{2020}).
\newblock
\urldef\tempurl%
\url{https://techcrunch.com/2020/03/21/twitter-prioritizes-bluecheck-verifications-to-confirm-experts-on-covid-19-and-the-novel-coronavirus/}
\showURL{%
\tempurl}
\newblock
\shownote{Retrieved January 24, 2023}.


\bibitem[Lusher et~al\mbox{.}(2013)]%
        {lusher2013exponential}
\bibfield{author}{\bibinfo{person}{Dean Lusher}, \bibinfo{person}{Johan Koskinen}, {and} \bibinfo{person}{Garry Robins}.} \bibinfo{year}{2013}\natexlab{}.
\newblock \bibinfo{booktitle}{\emph{Exponential random graph models for social networks: Theory, methods, and applications}}.
\newblock \bibinfo{publisher}{Cambridge University Press}.
\newblock


\bibitem[Manguri et~al\mbox{.}(2020)]%
        {manguri2020twitter}
\bibfield{author}{\bibinfo{person}{Kamaran~H Manguri}, \bibinfo{person}{Rebaz~N Ramadhan}, {and} \bibinfo{person}{Pshko R~Mohammed Amin}.} \bibinfo{year}{2020}\natexlab{}.
\newblock \showarticletitle{Twitter sentiment analysis on worldwide COVID-19 outbreaks}.
\newblock \bibinfo{journal}{\emph{Kurdistan Journal of Applied Research}} (\bibinfo{year}{2020}), \bibinfo{pages}{54--65}.
\newblock


\bibitem[Marecos et~al\mbox{.}(2024)]%
        {journalmedia5020046}
\bibfield{author}{\bibinfo{person}{Joao Marecos}, \bibinfo{person}{Duarte Tude~Graça}, \bibinfo{person}{Francisco Goiana-da Silva}, \bibinfo{person}{Hutan Ashrafian}, {and} \bibinfo{person}{Ara Darzi}.} \bibinfo{year}{2024}\natexlab{}.
\newblock \showarticletitle{Source Credibility Labels and Other Nudging Interventions in the Context of Online Health Misinformation: A Systematic Literature Review}.
\newblock \bibinfo{journal}{\emph{Journalism and Media}} \bibinfo{volume}{5}, \bibinfo{number}{2} (\bibinfo{year}{2024}), \bibinfo{pages}{702--717}.
\newblock
\showISSN{2673-5172}
\urldef\tempurl%
\url{https://doi.org/10.3390/journalmedia5020046}
\showDOI{\tempurl}


\bibitem[Matassi and Boczkowski(2023)]%
        {matassi2023know}
\bibfield{author}{\bibinfo{person}{Mora Matassi} {and} \bibinfo{person}{Pablo~J Boczkowski}.} \bibinfo{year}{2023}\natexlab{}.
\newblock \bibinfo{booktitle}{\emph{To know is to compare: Studying social media across nations, media, and Platforms}}.
\newblock \bibinfo{publisher}{MIT Press}.
\newblock


\bibitem[McDonald(2021)]%
        {mcdonald2021please}
\bibfield{author}{\bibinfo{person}{Lucy McDonald}.} \bibinfo{year}{2021}\natexlab{}.
\newblock \showarticletitle{Please like this paper}.
\newblock \bibinfo{journal}{\emph{Philosophy}} \bibinfo{volume}{96}, \bibinfo{number}{3} (\bibinfo{year}{2021}), \bibinfo{pages}{335--358}.
\newblock


\bibitem[McDonald et~al\mbox{.}(2019)]%
        {McDonald_et_al_19}
\bibfield{author}{\bibinfo{person}{Nora McDonald}, \bibinfo{person}{Sarita Schoenebeck}, {and} \bibinfo{person}{Andrea Forte}.} \bibinfo{year}{2019}\natexlab{}.
\newblock \showarticletitle{Reliability and Inter-rater Reliability in Qualitative Research: Norms and Guidelines for CSCW and HCI Practice}.
\newblock \bibinfo{journal}{\emph{Proc. ACM Hum.-Comput. Interact.}} \bibinfo{volume}{3}, \bibinfo{number}{CSCW}, Article \bibinfo{articleno}{72} (\bibinfo{date}{nov} \bibinfo{year}{2019}), \bibinfo{numpages}{23}~pages.
\newblock
\urldef\tempurl%
\url{https://doi.org/10.1145/3359174}
\showDOI{\tempurl}


\bibitem[McGregor(2019)]%
        {mcgregor2019social}
\bibfield{author}{\bibinfo{person}{Shannon~C McGregor}.} \bibinfo{year}{2019}\natexlab{}.
\newblock \showarticletitle{Social media as public opinion: How journalists use social media to represent public opinion}.
\newblock \bibinfo{journal}{\emph{Journalism}} \bibinfo{volume}{20}, \bibinfo{number}{8} (\bibinfo{year}{2019}), \bibinfo{pages}{1070--1086}.
\newblock


\bibitem[{Meta}(2025)]%
        {facebook_ai_2025}
\bibfield{author}{\bibinfo{person}{{Meta}}.} \bibinfo{year}{2025}\natexlab{}.
\newblock \bibinfo{title}{Expand your world with Meta AI}.
\newblock
\newblock
\urldef\tempurl%
\url{https://ai.meta.com/meta-ai/}
\showURL{%
\tempurl}
\newblock
\shownote{Accessed: 2025-03-24}.


\bibitem[Mikolov et~al\mbox{.}(2013)]%
        {mikolov_efficient_2013}
\bibfield{author}{\bibinfo{person}{Tomas Mikolov}, \bibinfo{person}{Kai Chen}, \bibinfo{person}{Greg Corrado}, {and} \bibinfo{person}{Jeffrey Dean}.} \bibinfo{year}{2013}\natexlab{}.
\newblock \showarticletitle{Efficient {Estimation} of {Word} {Representations} in {Vector} {Space}}.
\newblock \bibinfo{journal}{\emph{arXiv:1301.3781 [cs]}} (\bibinfo{date}{Jan.} \bibinfo{year}{2013}).
\newblock
\urldef\tempurl%
\url{http://arxiv.org/abs/1301.3781}
\showURL{%
\tempurl}
\newblock
\shownote{arXiv: 1301.3781}.


\bibitem[Miller et~al\mbox{.}(2021)]%
        {miller2021being}
\bibfield{author}{\bibinfo{person}{Ann~Neville Miller}, \bibinfo{person}{Chad Collins}, \bibinfo{person}{Lindsay Neuberger}, \bibinfo{person}{Andrew Todd}, \bibinfo{person}{Timothy~L Sellnow}, {and} \bibinfo{person}{Laura Boutemen}.} \bibinfo{year}{2021}\natexlab{}.
\newblock \showarticletitle{Being first, being right, and being credible since 2002: A systematic review of Crisis and Emergency Risk Communication (CERC) research}.
\newblock \bibinfo{journal}{\emph{Journal of International Crisis and Risk Communication Research}} \bibinfo{volume}{4}, \bibinfo{number}{1} (\bibinfo{year}{2021}), \bibinfo{pages}{1--27}.
\newblock


\bibitem[Nasim and Rajput(2017)]%
        {Nasim_2017}
\bibfield{author}{\bibinfo{person}{Zarmeen Nasim} {and} \bibinfo{person}{Quratulain Rajput}.} \bibinfo{year}{2017}\natexlab{}.
\newblock \showarticletitle{Understanding role of Twitter in addressing social causes}. In \bibinfo{booktitle}{\emph{2017 International Conference on Innovations in Electrical Engineering and Computational Technologies (ICIEECT)}}. \bibinfo{pages}{1--9}.
\newblock
\urldef\tempurl%
\url{https://doi.org/10.1109/ICIEECT.2017.7916528}
\showDOI{\tempurl}


\bibitem[Nisbet et~al\mbox{.}(2003)]%
        {nisbet_et_al_03}
\bibfield{author}{\bibinfo{person}{Matthew~C. Nisbet}, \bibinfo{person}{Dominique Brossard}, {and} \bibinfo{person}{Adrianne Kroepsch}.} \bibinfo{year}{2003}\natexlab{}.
\newblock \showarticletitle{Framing Science: The Stem Cell Controversy in an Age of Press/Politics}.
\newblock \bibinfo{journal}{\emph{Harvard International Journal of Press/Politics}} \bibinfo{volume}{8}, \bibinfo{number}{2} (\bibinfo{year}{2003}), \bibinfo{pages}{36--70}.
\newblock
\urldef\tempurl%
\url{https://doi.org/10.1177/1081180X02251047}
\showDOI{\tempurl}
\showeprint{https://doi.org/10.1177/1081180X02251047}


\bibitem[Palen and Anderson(2016)]%
        {palen2016crisis}
\bibfield{author}{\bibinfo{person}{Leysia Palen} {and} \bibinfo{person}{Kenneth~M Anderson}.} \bibinfo{year}{2016}\natexlab{}.
\newblock \showarticletitle{Crisis informatics—New data for extraordinary times}.
\newblock \bibinfo{journal}{\emph{Science}} \bibinfo{volume}{353}, \bibinfo{number}{6296} (\bibinfo{year}{2016}), \bibinfo{pages}{224--225}.
\newblock


\bibitem[Papacharissi(2016)]%
        {papacharissi2016affective}
\bibfield{author}{\bibinfo{person}{Zizi Papacharissi}.} \bibinfo{year}{2016}\natexlab{}.
\newblock \showarticletitle{Affective publics and structures of storytelling: Sentiment, events and mediality}.
\newblock \bibinfo{journal}{\emph{Information, communication \& society}} \bibinfo{volume}{19}, \bibinfo{number}{3} (\bibinfo{year}{2016}), \bibinfo{pages}{307--324}.
\newblock


\bibitem[Park et~al\mbox{.}(2023)]%
        {Park_et_al_23}
\bibfield{author}{\bibinfo{person}{Joon~Sung Park}, \bibinfo{person}{Joseph O'Brien}, \bibinfo{person}{Carrie~Jun Cai}, \bibinfo{person}{Meredith~Ringel Morris}, \bibinfo{person}{Percy Liang}, {and} \bibinfo{person}{Michael~S. Bernstein}.} \bibinfo{year}{2023}\natexlab{}.
\newblock \showarticletitle{Generative Agents: Interactive Simulacra of Human Behavior}. In \bibinfo{booktitle}{\emph{Proceedings of the 36th Annual ACM Symposium on User Interface Software and Technology}} (San Francisco, CA, USA) \emph{(\bibinfo{series}{UIST '23})}. \bibinfo{publisher}{Association for Computing Machinery}, \bibinfo{address}{New York, NY, USA}, Article \bibinfo{articleno}{2}, \bibinfo{numpages}{22}~pages.
\newblock
\showISBNx{9798400701320}
\urldef\tempurl%
\url{https://doi.org/10.1145/3586183.3606763}
\showDOI{\tempurl}


\bibitem[Paul et~al\mbox{.}(2019)]%
        {paul2019elites}
\bibfield{author}{\bibinfo{person}{Indraneil Paul}, \bibinfo{person}{Abhinav Khattar}, \bibinfo{person}{Ponnurangam Kumaraguru}, \bibinfo{person}{Manish Gupta}, {and} \bibinfo{person}{Shaan Chopra}.} \bibinfo{year}{2019}\natexlab{}.
\newblock \showarticletitle{Elites tweet? Characterizing the Twitter verified user network}. In \bibinfo{booktitle}{\emph{2019 IEEE 35th International Conference on Data Engineering Workshops (ICDEW)}}. IEEE, \bibinfo{pages}{278--285}.
\newblock


\bibitem[Pavelko and Myrick({[n.\,d.]})]%
        {pavelko_muderinos_2020}
\bibfield{author}{\bibinfo{person}{Rachelle~L. Pavelko} {and} \bibinfo{person}{Jessica~Gall Myrick}.} \bibinfo{year}{[n.\,d.]}\natexlab{}.
\newblock \showarticletitle{Muderinos and media effects: How the my favorite murder podcast and its social media community may promote well-being in audiences with mental illness}.
\newblock  \bibinfo{volume}{27}, \bibinfo{number}{1} (\bibinfo{year}{[n.\,d.]}), \bibinfo{pages}{151--169}.
\newblock
\showISSN{1937-6529}
\urldef\tempurl%
\url{https://doi.org/10.1080/19376529.2019.1638925}
\showDOI{\tempurl}
\newblock
\shownote{Publisher: Routledge \_eprint: https://doi.org/10.1080/19376529.2019.1638925}.


\bibitem[Pedregosa et~al\mbox{.}(2011)]%
        {pedregosa2011scikit}
\bibfield{author}{\bibinfo{person}{Fabian Pedregosa}, \bibinfo{person}{Ga{\"e}l Varoquaux}, \bibinfo{person}{Alexandre Gramfort}, \bibinfo{person}{Vincent Michel}, \bibinfo{person}{Bertrand Thirion}, \bibinfo{person}{Olivier Grisel}, \bibinfo{person}{Mathieu Blondel}, \bibinfo{person}{Peter Prettenhofer}, \bibinfo{person}{Ron Weiss}, \bibinfo{person}{Vincent Dubourg}, {et~al\mbox{.}}} \bibinfo{year}{2011}\natexlab{}.
\newblock \showarticletitle{Scikit-learn: Machine learning in Python}.
\newblock \bibinfo{journal}{\emph{the Journal of machine Learning research}}  \bibinfo{volume}{12} (\bibinfo{year}{2011}), \bibinfo{pages}{2825--2830}.
\newblock


\bibitem[Pierri et~al\mbox{.}(2023)]%
        {pierri_2023}
\bibfield{author}{\bibinfo{person}{Francesco Pierri}, \bibinfo{person}{Matthew~R DeVerna}, \bibinfo{person}{Kai-Cheng Yang}, \bibinfo{person}{David Axelrod}, \bibinfo{person}{John Bryden}, {and} \bibinfo{person}{Filippo Menczer}.} \bibinfo{year}{2023}\natexlab{}.
\newblock \showarticletitle{One Year of COVID-19 Vaccine Misinformation on Twitter: Longitudinal Study}.
\newblock \bibinfo{journal}{\emph{J Med Internet Res}}  \bibinfo{volume}{25} (\bibinfo{date}{24 Feb} \bibinfo{year}{2023}), \bibinfo{pages}{e42227}.
\newblock
\showISSN{1438-8871}
\urldef\tempurl%
\url{https://doi.org/10.2196/42227}
\showDOI{\tempurl}


\bibitem[Pine et~al\mbox{.}(2021)]%
        {Pine_et_al_21}
\bibfield{author}{\bibinfo{person}{Kathleen~H Pine}, \bibinfo{person}{Myeong Lee}, \bibinfo{person}{Samantha~A. Whitman}, \bibinfo{person}{Yunan Chen}, {and} \bibinfo{person}{Kathryn Henne}.} \bibinfo{year}{2021}\natexlab{}.
\newblock \showarticletitle{Making Sense of Risk Information amidst Uncertainty: Individuals’ Perceived Risks Associated with the COVID-19 Pandemic}. In \bibinfo{booktitle}{\emph{Proceedings of the 2021 CHI Conference on Human Factors in Computing Systems}} (Yokohama, Japan) \emph{(\bibinfo{series}{CHI '21})}. \bibinfo{publisher}{Association for Computing Machinery}, \bibinfo{address}{New York, NY, USA}, Article \bibinfo{articleno}{653}, \bibinfo{numpages}{15}~pages.
\newblock
\showISBNx{9781450380966}
\urldef\tempurl%
\url{https://doi.org/10.1145/3411764.3445051}
\showDOI{\tempurl}


\bibitem[Potter and Wetherell(1987)]%
        {potter1987discourse}
\bibfield{author}{\bibinfo{person}{Jonathan Potter} {and} \bibinfo{person}{Margaret Wetherell}.} \bibinfo{year}{1987}\natexlab{}.
\newblock \bibinfo{booktitle}{\emph{Discourse and social psychology: Beyond attitudes and behaviour.}}
\newblock \bibinfo{publisher}{Sage Publications, Inc}.
\newblock


\bibitem[Premat et~al\mbox{.}(2024)]%
        {premat2024introduction}
\bibfield{author}{\bibinfo{person}{Christophe Premat}, \bibinfo{person}{Jean-Michel De~Waele}, {and} \bibinfo{person}{Michel Perottino}.} \bibinfo{year}{2024}\natexlab{}.
\newblock \showarticletitle{Introduction: The legitimacy of experts in the public space during the pandemic}.
\newblock \bibinfo{journal}{\emph{Comparing the place of experts during the first waves of the COVID-19 pandemic}} (\bibinfo{year}{2024}), \bibinfo{pages}{14--30}.
\newblock


\bibitem[Qu{\'e}r{\'e} et~al\mbox{.}(2025)]%
        {quere2025state}
\bibfield{author}{\bibinfo{person}{Marianne Aubin~Le Qu{\'e}r{\'e}}, \bibinfo{person}{Hope Schroeder}, \bibinfo{person}{Casey Randazzo}, {and} \bibinfo{person}{Jie Gao}.} \bibinfo{year}{2025}\natexlab{}.
\newblock \showarticletitle{The State of Large Language Models in HCI Research: Workshop Report}.
\newblock \bibinfo{journal}{\emph{Interactions}} \bibinfo{volume}{32}, \bibinfo{number}{1} (\bibinfo{year}{2025}), \bibinfo{pages}{8--9}.
\newblock


\bibitem[Randazzo and Ammari(2023)]%
        {randazzo_23}
\bibfield{author}{\bibinfo{person}{Casey Randazzo} {and} \bibinfo{person}{Tawfiq Ammari}.} \bibinfo{year}{2023}\natexlab{}.
\newblock \showarticletitle{“If Someone Downvoted My Posts—That’d Be the End of the World”: Designing Safer Online Spaces for Trauma Survivors}. In \bibinfo{booktitle}{\emph{Proceedings of the 2023 CHI Conference on Human Factors in Computing Systems}} (Hamburg, Germany) \emph{(\bibinfo{series}{CHI '23})}. \bibinfo{publisher}{Association for Computing Machinery}, \bibinfo{address}{New York, NY, USA}, Article \bibinfo{articleno}{481}, \bibinfo{numpages}{18}~pages.
\newblock
\showISBNx{9781450394215}
\urldef\tempurl%
\url{https://doi.org/10.1145/3544548.3581453}
\showDOI{\tempurl}


\bibitem[Rao et~al\mbox{.}(2020)]%
        {rao2020retweets}
\bibfield{author}{\bibinfo{person}{H~Raghav Rao}, \bibinfo{person}{Naga Vemprala}, \bibinfo{person}{Patricia Akello}, {and} \bibinfo{person}{Rohit Valecha}.} \bibinfo{year}{2020}\natexlab{}.
\newblock \showarticletitle{Retweets of officials’ alarming vs reassuring messages during the COVID-19 pandemic: Implications for crisis management}.
\newblock \bibinfo{journal}{\emph{International Journal of Information Management}}  \bibinfo{volume}{55} (\bibinfo{year}{2020}), \bibinfo{pages}{102187}.
\newblock


\bibitem[Rathnayake and Suthers(2017)]%
        {Rathnayake_17}
\bibfield{author}{\bibinfo{person}{Chamil Rathnayake} {and} \bibinfo{person}{Daniel~D. Suthers}.} \bibinfo{year}{2017}\natexlab{}.
\newblock \showarticletitle{Twitter Issue Response Hashtags as Affordances for Momentary Connectedness}. In \bibinfo{booktitle}{\emph{Proceedings of the 8th International Conference on Social Media \& Society}} (Toronto, ON, Canada) \emph{(\bibinfo{series}{\#SMSociety17})}. \bibinfo{publisher}{Association for Computing Machinery}, \bibinfo{address}{New York, NY, USA}, Article \bibinfo{articleno}{16}, \bibinfo{numpages}{10}~pages.
\newblock
\showISBNx{9781450348478}
\urldef\tempurl%
\url{https://doi.org/10.1145/3097286.3097302}
\showDOI{\tempurl}


\bibitem[Reimers and Gurevych(2019)]%
        {reimers_sentence-bert_2019}
\bibfield{author}{\bibinfo{person}{Nils Reimers} {and} \bibinfo{person}{Iryna Gurevych}.} \bibinfo{year}{2019}\natexlab{}.
\newblock \showarticletitle{Sentence-{BERT}: {Sentence} {Embeddings} using {Siamese} {BERT}-{Networks}}.
\newblock  (\bibinfo{date}{Aug.} \bibinfo{year}{2019}).
\newblock
\urldef\tempurl%
\url{https://doi.org/10.48550/arXiv.1908.10084}
\showDOI{\tempurl}


\bibitem[Renn(2015)]%
        {renn2015stakeholder}
\bibfield{author}{\bibinfo{person}{Ortwin Renn}.} \bibinfo{year}{2015}\natexlab{}.
\newblock \showarticletitle{Stakeholder and public involvement in risk governance}.
\newblock \bibinfo{journal}{\emph{International Journal of Disaster Risk Science}}  \bibinfo{volume}{6} (\bibinfo{year}{2015}), \bibinfo{pages}{8--20}.
\newblock


\bibitem[Reuter et~al\mbox{.}(2018)]%
        {reuter2018social}
\bibfield{author}{\bibinfo{person}{Christian Reuter}, \bibinfo{person}{Amanda~Lee Hughes}, {and} \bibinfo{person}{Marc-Andr{\'e} Kaufhold}.} \bibinfo{year}{2018}\natexlab{}.
\newblock \showarticletitle{Social media in crisis management: An evaluation and analysis of crisis informatics research}.
\newblock \bibinfo{journal}{\emph{International Journal of Human--Computer Interaction}} \bibinfo{volume}{34}, \bibinfo{number}{4} (\bibinfo{year}{2018}), \bibinfo{pages}{280--294}.
\newblock


\bibitem[Reynolds and Seeger(2005)]%
        {reynolds2005crisis}
\bibfield{author}{\bibinfo{person}{Barbara Reynolds} {and} \bibinfo{person}{Matthew~W Seeger}.} \bibinfo{year}{2005}\natexlab{}.
\newblock \showarticletitle{Crisis and emergency risk communication as an integrative model}.
\newblock \bibinfo{journal}{\emph{Journal of health communication}} \bibinfo{volume}{10}, \bibinfo{number}{1} (\bibinfo{year}{2005}), \bibinfo{pages}{43--55}.
\newblock


\bibitem[Robins and Lusher(2012)]%
        {robins_lusher_2012}
\bibfield{author}{\bibinfo{person}{Garry Robins} {and} \bibinfo{person}{Dean Lusher}.} \bibinfo{year}{2012}\natexlab{}.
\newblock \bibinfo{booktitle}{\emph{What Are Exponential Random Graph Models?}}
\newblock \bibinfo{publisher}{Cambridge University Press}, \bibinfo{pages}{9–15}.
\newblock
\urldef\tempurl%
\url{https://doi.org/10.1017/CBO9780511894701.003}
\showDOI{\tempurl}


\bibitem[Röder et~al\mbox{.}(2015)]%
        {roder_exploring_2015}
\bibfield{author}{\bibinfo{person}{Michael Röder}, \bibinfo{person}{Andreas Both}, {and} \bibinfo{person}{Alexander Hinneburg}.} \bibinfo{year}{2015}\natexlab{}.
\newblock \showarticletitle{Exploring the {Space} of {Topic} {Coherence} {Measures}}. In \bibinfo{booktitle}{\emph{Proceedings of the {Eighth} {ACM} {International} {Conference} on {Web} {Search} and {Data} {Mining}}} \emph{(\bibinfo{series}{{WSDM} '15})}. \bibinfo{publisher}{Association for Computing Machinery}, \bibinfo{address}{New York, NY, USA}, \bibinfo{pages}{399--408}.
\newblock
\showISBNx{978-1-4503-3317-7}
\urldef\tempurl%
\url{https://doi.org/10.1145/2684822.2685324}
\showDOI{\tempurl}


\bibitem[Sahu et~al\mbox{.}(2021)]%
        {sahu2021towards}
\bibfield{author}{\bibinfo{person}{Gaurav Sahu}, \bibinfo{person}{Robin Cohen}, {and} \bibinfo{person}{Olga Vechtomova}.} \bibinfo{year}{2021}\natexlab{}.
\newblock \showarticletitle{Towards a multi-agent system for online hate speech detection}.
\newblock \bibinfo{journal}{\emph{arXiv preprint arXiv:2105.01129}} (\bibinfo{year}{2021}).
\newblock


\bibitem[Sanders et~al\mbox{.}(2022)]%
        {sanders2022should}
\bibfield{author}{\bibinfo{person}{Abraham Sanders}, \bibinfo{person}{Debjani Ray-Majumder}, \bibinfo{person}{John Erickson}, {and} \bibinfo{person}{Kristin Bennett}.} \bibinfo{year}{2022}\natexlab{}.
\newblock \showarticletitle{Should we tweet this? Generative response modeling for predicting reception of public health messaging on Twitter}. In \bibinfo{booktitle}{\emph{Proceedings of the 14th ACM Web Science Conference 2022}}. \bibinfo{pages}{307--318}.
\newblock


\bibitem[Schiffrin(1987)]%
        {schiffrin1987discourse}
\bibfield{author}{\bibinfo{person}{Deborah Schiffrin}.} \bibinfo{year}{1987}\natexlab{}.
\newblock \bibinfo{booktitle}{\emph{Discourse markers}}.
\newblock Number~5. \bibinfo{publisher}{Cambridge University Press}.
\newblock


\bibitem[Schroeder et~al\mbox{.}(2024)]%
        {schroeder2024large}
\bibfield{author}{\bibinfo{person}{Hope Schroeder}, \bibinfo{person}{Marianne Aubin Le~Qu{\'e}r{\'e}}, \bibinfo{person}{Casey Randazzo}, \bibinfo{person}{David Mimno}, {and} \bibinfo{person}{Sarita Schoenebeck}.} \bibinfo{year}{2024}\natexlab{}.
\newblock \showarticletitle{Large language models in qualitative research: Can we do the data justice?}
\newblock \bibinfo{journal}{\emph{arXiv e-prints}} (\bibinfo{year}{2024}), \bibinfo{pages}{arXiv--2410}.
\newblock


\bibitem[Sekimoto et~al\mbox{.}(2020)]%
        {Sekimoto_et_al_2020}
\bibfield{author}{\bibinfo{person}{Kenshin Sekimoto}, \bibinfo{person}{Yoshifumi Seki}, \bibinfo{person}{Mitsuo Yoshida}, {and} \bibinfo{person}{Kyoji Umemura}.} \bibinfo{year}{2020}\natexlab{}.
\newblock \showarticletitle{The metrics of keywords to understand the difference between Retweet and Like in each category}. In \bibinfo{booktitle}{\emph{2020 IEEE/WIC/ACM International Joint Conference on Web Intelligence and Intelligent Agent Technology (WI-IAT)}}. \bibinfo{pages}{560--567}.
\newblock
\urldef\tempurl%
\url{https://doi.org/10.1109/WIIAT50758.2020.00084}
\showDOI{\tempurl}


\bibitem[Shah and Wei(2022)]%
        {shah2022source}
\bibfield{author}{\bibinfo{person}{Zakir Shah} {and} \bibinfo{person}{Lu Wei}.} \bibinfo{year}{2022}\natexlab{}.
\newblock \showarticletitle{Source credibility and the information quality matter in public engagement on social networking sites during the COVID-19 crisis}.
\newblock \bibinfo{journal}{\emph{Frontiers in psychology}}  \bibinfo{volume}{13} (\bibinfo{year}{2022}), \bibinfo{pages}{882705}.
\newblock


\bibitem[Sharevski et~al\mbox{.}(2022)]%
        {Sharevski_22}
\bibfield{author}{\bibinfo{person}{Filipo Sharevski}, \bibinfo{person}{Amy Devine}, \bibinfo{person}{Peter Jachim}, {and} \bibinfo{person}{Emma Pieroni}.} \bibinfo{year}{2022}\natexlab{}.
\newblock \showarticletitle{Meaningful Context, a Red Flag, or Both? Preferences for Enhanced Misinformation Warnings Among US Twitter Users}. In \bibinfo{booktitle}{\emph{Proceedings of the 2022 European Symposium on Usable Security}} (Karlsruhe, Germany) \emph{(\bibinfo{series}{EuroUSEC '22})}. \bibinfo{publisher}{Association for Computing Machinery}, \bibinfo{address}{New York, NY, USA}, \bibinfo{pages}{189–201}.
\newblock
\showISBNx{9781450397001}
\urldef\tempurl%
\url{https://doi.org/10.1145/3549015.3555671}
\showDOI{\tempurl}


\bibitem[Shi et~al\mbox{.}(2018)]%
        {shi2018determinants}
\bibfield{author}{\bibinfo{person}{Juan Shi}, \bibinfo{person}{Ping Hu}, \bibinfo{person}{Kin~Keung Lai}, {and} \bibinfo{person}{Gang Chen}.} \bibinfo{year}{2018}\natexlab{}.
\newblock \showarticletitle{Determinants of users’ information dissemination behavior on social networking sites: An elaboration likelihood model perspective}.
\newblock \bibinfo{journal}{\emph{Internet Research}} \bibinfo{volume}{28}, \bibinfo{number}{2} (\bibinfo{year}{2018}), \bibinfo{pages}{393--418}.
\newblock


\bibitem[Shoemaker et~al\mbox{.}(2023)]%
        {SHOEMAKER202383}
\bibfield{author}{\bibinfo{person}{Eric Shoemaker}, \bibinfo{person}{Haroon Malik}, \bibinfo{person}{Husnu Narman}, {and} \bibinfo{person}{Jamil Chaudri}.} \bibinfo{year}{2023}\natexlab{}.
\newblock \showarticletitle{Explaining the Unseen: Leveraging XAI to Enhance the Trustworthiness of Black-Box Models in Performance Testing}.
\newblock \bibinfo{journal}{\emph{Procedia Computer Science}}  \bibinfo{volume}{224} (\bibinfo{year}{2023}), \bibinfo{pages}{83--90}.
\newblock
\showISSN{1877-0509}
\urldef\tempurl%
\url{https://doi.org/10.1016/j.procs.2023.09.014}
\showDOI{\tempurl}
\newblock
\shownote{18th International Conference on Future Networks and Communications / 20th International Conference on Mobile Systems and Pervasive Computing / 13th International Conference on Sustainable Energy Information Technology}.


\bibitem[Simon et~al\mbox{.}(2015)]%
        {SIMON2015609}
\bibfield{author}{\bibinfo{person}{Tomer Simon}, \bibinfo{person}{Avishay Goldberg}, {and} \bibinfo{person}{Bruria Adini}.} \bibinfo{year}{2015}\natexlab{}.
\newblock \showarticletitle{Socializing in emergencies—A review of the use of social media in emergency situations}.
\newblock \bibinfo{journal}{\emph{International Journal of Information Management}} \bibinfo{volume}{35}, \bibinfo{number}{5} (\bibinfo{year}{2015}), \bibinfo{pages}{609--619}.
\newblock
\showISSN{0268-4012}
\urldef\tempurl%
\url{https://doi.org/10.1016/j.ijinfomgt.2015.07.001}
\showDOI{\tempurl}


\bibitem[Srinivasan and Jiang(2023)]%
        {Srinivasan_Jiang_2023}
\bibfield{author}{\bibinfo{person}{Karthik Srinivasan} {and} \bibinfo{person}{Jinhang Jiang}.} \bibinfo{year}{2023}\natexlab{}.
\newblock \showarticletitle{Examining Disease Multimorbidity in U.S. Hospital Visits Before and During COVID-19 Pandemic: A Graph Analytics Approach}.
\newblock \bibinfo{journal}{\emph{ACM Trans. Manage. Inf. Syst.}} \bibinfo{volume}{14}, \bibinfo{number}{2}, Article \bibinfo{articleno}{17} (\bibinfo{date}{jan} \bibinfo{year}{2023}), \bibinfo{numpages}{17}~pages.
\newblock
\showISSN{2158-656X}
\urldef\tempurl%
\url{https://doi.org/10.1145/3564274}
\showDOI{\tempurl}


\bibitem[Stephens et~al\mbox{.}(2005)]%
        {keri_et_al_2005}
\bibfield{author}{\bibinfo{person}{Keri~K. Stephens}, \bibinfo{person}{Patty~Callish Malone}, {and} \bibinfo{person}{Christine~M. Bailey}.} \bibinfo{year}{2005}\natexlab{}.
\newblock \showarticletitle{Communicating with stakeholders During a Crisis: Evaluating Message Strategies}.
\newblock \bibinfo{journal}{\emph{The Journal of Business Communication (1973)}} \bibinfo{volume}{42}, \bibinfo{number}{4} (\bibinfo{year}{2005}), \bibinfo{pages}{390--419}.
\newblock
\urldef\tempurl%
\url{https://doi.org/10.1177/0021943605279057}
\showDOI{\tempurl}
\showeprint{https://journals.sagepub.com/doi/pdf/10.1177/0021943605279057}


\bibitem[Stewart et~al\mbox{.}(2017)]%
        {stewart_et_al_17}
\bibfield{author}{\bibinfo{person}{Leo~Graiden Stewart}, \bibinfo{person}{Ahmer Arif}, \bibinfo{person}{A.~Conrad Nied}, \bibinfo{person}{Emma~S. Spiro}, {and} \bibinfo{person}{Kate Starbird}.} \bibinfo{year}{2017}\natexlab{}.
\newblock \showarticletitle{Drawing the Lines of Contention: Networked Frame Contests Within \#BlackLivesMatter Discourse}.
\newblock \bibinfo{journal}{\emph{Proc. ACM Hum.-Comput. Interact.}} \bibinfo{volume}{1}, \bibinfo{number}{CSCW}, Article \bibinfo{articleno}{96} (\bibinfo{date}{dec} \bibinfo{year}{2017}), \bibinfo{numpages}{23}~pages.
\newblock
\urldef\tempurl%
\url{https://doi.org/10.1145/3134920}
\showDOI{\tempurl}


\bibitem[Suh et~al\mbox{.}(2010)]%
        {Suh_et_al_2010}
\bibfield{author}{\bibinfo{person}{Bongwon Suh}, \bibinfo{person}{Lichan Hong}, \bibinfo{person}{Peter Pirolli}, {and} \bibinfo{person}{Ed~H. Chi}.} \bibinfo{year}{2010}\natexlab{}.
\newblock \showarticletitle{Want to be Retweeted? Large Scale Analytics on Factors Impacting Retweet in Twitter Network}. In \bibinfo{booktitle}{\emph{Proceedings of the 2010 IEEE Second International Conference on Social Computing}} \emph{(\bibinfo{series}{SOCIALCOM '10})}. \bibinfo{publisher}{IEEE Computer Society}, \bibinfo{address}{USA}, \bibinfo{pages}{177–184}.
\newblock
\showISBNx{9780769542119}
\urldef\tempurl%
\url{https://doi.org/10.1109/SocialCom.2010.33}
\showDOI{\tempurl}


\bibitem[Support(2020)]%
        {TwitterSupport2020}
\bibfield{author}{\bibinfo{person}{Twitter Support}.} \bibinfo{year}{2020}\natexlab{}.
\newblock \bibinfo{title}{PSA about what we’re doing to Verify Twitter accounts that are providing credible updates around \#COVID19}.
\newblock \bibinfo{howpublished}{Twitter}.
\newblock
\urldef\tempurl%
\url{https://twitter.com/TwitterSupport/status/1241155701822476288}
\showURL{%
\tempurl}
\newblock
\shownote{Retrieved January 24, 2023}.


\bibitem[Swets(2014)]%
        {swets2014signal}
\bibfield{author}{\bibinfo{person}{John~A Swets}.} \bibinfo{year}{2014}\natexlab{}.
\newblock \bibinfo{booktitle}{\emph{Signal detection theory and ROC analysis in psychology and diagnostics: Collected papers}}.
\newblock \bibinfo{publisher}{Psychology Press}.
\newblock


\bibitem[Tarbă et~al\mbox{.}(2022)]%
        {math10071042}
\bibfield{author}{\bibinfo{person}{Nicolae Tarbă}, \bibinfo{person}{Mihai-Lucian Voncilă}, {and} \bibinfo{person}{Costin-Anton Boiangiu}.} \bibinfo{year}{2022}\natexlab{}.
\newblock \showarticletitle{On Generalizing Sarle’s Bimodality Coefficient as a Path towards a Newly Composite Bimodality Coefficient}.
\newblock \bibinfo{journal}{\emph{Mathematics}} \bibinfo{volume}{10}, \bibinfo{number}{7} (\bibinfo{year}{2022}).
\newblock
\showISSN{2227-7390}
\urldef\tempurl%
\url{https://doi.org/10.3390/math10071042}
\showDOI{\tempurl}


\bibitem[Tong et~al\mbox{.}(2022)]%
        {tong_et_al_22}
\bibfield{author}{\bibinfo{person}{Xin Tong}, \bibinfo{person}{Yixuan Li}, \bibinfo{person}{Jiayi Li}, \bibinfo{person}{Rongqi Bei}, {and} \bibinfo{person}{Luyao Zhang}.} \bibinfo{year}{2022}\natexlab{}.
\newblock \showarticletitle{What are People Talking about in \#BackLivesMatter and \#StopAsianHate? Exploring and Categorizing Twitter Topics Emerged in Online Social Movements through the Latent Dirichlet Allocation Model}. In \bibinfo{booktitle}{\emph{Proceedings of the 2022 AAAI/ACM Conference on AI, Ethics, and Society}} (Oxford, United Kingdom) \emph{(\bibinfo{series}{AIES '22})}. \bibinfo{publisher}{Association for Computing Machinery}, \bibinfo{address}{New York, NY, USA}, \bibinfo{pages}{723–738}.
\newblock
\showISBNx{9781450392471}
\urldef\tempurl%
\url{https://doi.org/10.1145/3514094.3534202}
\showDOI{\tempurl}


\bibitem[Unlu et~al\mbox{.}(2023)]%
        {unlu2023exploring}
\bibfield{author}{\bibinfo{person}{Ali Unlu}, \bibinfo{person}{Sophie Truong}, \bibinfo{person}{Tuukka Tammi}, {and} \bibinfo{person}{Anna-Leena Lohiniva}.} \bibinfo{year}{2023}\natexlab{}.
\newblock \showarticletitle{Exploring political mistrust in pandemic risk communication: mixed-method study using social media data analysis}.
\newblock \bibinfo{journal}{\emph{Journal of Medical Internet Research}}  \bibinfo{volume}{25} (\bibinfo{year}{2023}), \bibinfo{pages}{e50199}.
\newblock


\bibitem[Valdez et~al\mbox{.}(2020)]%
        {valdez2020social}
\bibfield{author}{\bibinfo{person}{Danny Valdez}, \bibinfo{person}{Marijn Ten~Thij}, \bibinfo{person}{Krishna Bathina}, \bibinfo{person}{Lauren~A Rutter}, {and} \bibinfo{person}{Johan Bollen}.} \bibinfo{year}{2020}\natexlab{}.
\newblock \showarticletitle{Social media insights into US mental health during the COVID-19 pandemic: longitudinal analysis of Twitter data}.
\newblock \bibinfo{journal}{\emph{Journal of medical Internet research}} \bibinfo{volume}{22}, \bibinfo{number}{12} (\bibinfo{year}{2020}), \bibinfo{pages}{e21418}.
\newblock


\bibitem[Valiavska and Smith-Frigerio(2022)]%
        {valiavska2022politics}
\bibfield{author}{\bibinfo{person}{Anna Valiavska} {and} \bibinfo{person}{Sarah Smith-Frigerio}.} \bibinfo{year}{2022}\natexlab{}.
\newblock \showarticletitle{Politics over public health: Analysis of Twitter and Reddit posts concerning the role of politics in the public health response to COVID-19}.
\newblock \bibinfo{journal}{\emph{Health Communication}} (\bibinfo{year}{2022}), \bibinfo{pages}{1--10}.
\newblock


\bibitem[{van der Meer} and Verhoeven(2014)]%
        {VANDERMEER2014526}
\bibfield{author}{\bibinfo{person}{Toni~G.L.A. {van der Meer}} {and} \bibinfo{person}{Joost~W.M. Verhoeven}.} \bibinfo{year}{2014}\natexlab{}.
\newblock \showarticletitle{Emotional crisis communication}.
\newblock \bibinfo{journal}{\emph{Public Relations Review}} \bibinfo{volume}{40}, \bibinfo{number}{3} (\bibinfo{year}{2014}), \bibinfo{pages}{526--536}.
\newblock
\showISSN{0363-8111}
\urldef\tempurl%
\url{https://doi.org/10.1016/j.pubrev.2014.03.004}
\showDOI{\tempurl}


\bibitem[{van Doorn} and Koster(2019)]%
        {VANDOORN201974}
\bibfield{author}{\bibinfo{person}{Janne {van Doorn}} {and} \bibinfo{person}{Nathalie~N. Koster}.} \bibinfo{year}{2019}\natexlab{}.
\newblock \showarticletitle{Emotional victims and the impact on credibility: A systematic review}.
\newblock \bibinfo{journal}{\emph{Aggression and Violent Behavior}}  \bibinfo{volume}{47} (\bibinfo{year}{2019}), \bibinfo{pages}{74--89}.
\newblock
\showISSN{1359-1789}
\urldef\tempurl%
\url{https://doi.org/10.1016/j.avb.2019.03.007}
\showDOI{\tempurl}


\bibitem[Vanessa~Murphree and Blevens(2009)]%
        {Murphree29052009}
\bibfield{author}{\bibinfo{person}{Bryan H.~Reber Vanessa~Murphree} {and} \bibinfo{person}{Frederick Blevens}.} \bibinfo{year}{2009}\natexlab{}.
\newblock \showarticletitle{Superhero, Instructor, Optimist: FEMA and the Frames of Disaster in Hurricanes Katrina and Rita}.
\newblock \bibinfo{journal}{\emph{Journal of Public Relations Research}} \bibinfo{volume}{21}, \bibinfo{number}{3} (\bibinfo{year}{2009}), \bibinfo{pages}{273--294}.
\newblock
\urldef\tempurl%
\url{https://doi.org/10.1080/10627260802640732}
\showDOI{\tempurl}
\showeprint{https://doi.org/10.1080/10627260802640732}


\bibitem[Vishnu et~al\mbox{.}(2023)]%
        {vishnu2023recurrent}
\bibfield{author}{\bibinfo{person}{MK Vishnu}, \bibinfo{person}{VR~Vishal Rupak}, \bibinfo{person}{S Vedhapriyaa}, \bibinfo{person}{M Sangeetha}, \bibinfo{person}{R Manjuladevi}, {and} \bibinfo{person}{C Sagana}.} \bibinfo{year}{2023}\natexlab{}.
\newblock \showarticletitle{Recurrent gastric cancer prediction using randomized search cv optimizer}. In \bibinfo{booktitle}{\emph{2023 International Conference on Computer Communication and Informatics (ICCCI)}}. IEEE, \bibinfo{pages}{1--5}.
\newblock


\bibitem[Vishwamitra et~al\mbox{.}(2024)]%
        {Vishwamitra_24}
\bibfield{author}{\bibinfo{person}{Nishant Vishwamitra}, \bibinfo{person}{Keyan Guo}, \bibinfo{person}{Song Liao}, \bibinfo{person}{Jaden Mu}, \bibinfo{person}{Zheyuan Ma}, \bibinfo{person}{Long Cheng}, \bibinfo{person}{Ziming Zhao}, {and} \bibinfo{person}{Hongxin Hu}.} \bibinfo{year}{2024}\natexlab{}.
\newblock \showarticletitle{Understanding and Analyzing COVID-19-related Online Hate Propagation Through Hateful Memes Shared on Twitter}. In \bibinfo{booktitle}{\emph{Proceedings of the 2023 IEEE/ACM International Conference on Advances in Social Networks Analysis and Mining}} (Kusadasi, Turkiye) \emph{(\bibinfo{series}{ASONAM '23})}. \bibinfo{publisher}{Association for Computing Machinery}, \bibinfo{address}{New York, NY, USA}, \bibinfo{pages}{103–107}.
\newblock
\showISBNx{9798400704093}
\urldef\tempurl%
\url{https://doi.org/10.1145/3625007.3630111}
\showDOI{\tempurl}


\bibitem[Walters et~al\mbox{.}(0)]%
        {walters_et_al_24}
\bibfield{author}{\bibinfo{person}{Alyvia Walters}, \bibinfo{person}{Tawfiq Ammari}, \bibinfo{person}{Kiran Garimella}, {and} \bibinfo{person}{Shagun Jhaver}.} \bibinfo{year}{0}\natexlab{}.
\newblock \showarticletitle{Online knowledge production in polarized political memes: The case of critical race theory}.
\newblock \bibinfo{journal}{\emph{New Media \& Society}} \bibinfo{volume}{0}, \bibinfo{number}{0} (\bibinfo{year}{0}), \bibinfo{pages}{14614448241252591}.
\newblock
\urldef\tempurl%
\url{https://doi.org/10.1177/14614448241252591}
\showDOI{\tempurl}
\showeprint{https://doi.org/10.1177/14614448241252591}


\bibitem[Wang et~al\mbox{.}(2011)]%
        {wang2011random}
\bibfield{author}{\bibinfo{person}{Sijian Wang}, \bibinfo{person}{Bin Nan}, \bibinfo{person}{Saharon Rosset}, {and} \bibinfo{person}{Ji Zhu}.} \bibinfo{year}{2011}\natexlab{}.
\newblock \showarticletitle{Random lasso}.
\newblock \bibinfo{journal}{\emph{The annals of applied statistics}} \bibinfo{volume}{5}, \bibinfo{number}{1} (\bibinfo{year}{2011}), \bibinfo{pages}{468}.
\newblock


\bibitem[Wang et~al\mbox{.}(2023)]%
        {wang_et_al_23}
\bibfield{author}{\bibinfo{person}{Yuping Wang}, \bibinfo{person}{Chen Ling}, {and} \bibinfo{person}{Gianluca Stringhini}.} \bibinfo{year}{2023}\natexlab{}.
\newblock \showarticletitle{Understanding the Use of Images to Spread COVID-19 Misinformation on Twitter}.
\newblock \bibinfo{journal}{\emph{Proc. ACM Hum.-Comput. Interact.}} \bibinfo{volume}{7}, \bibinfo{number}{CSCW1}, Article \bibinfo{articleno}{108} (\bibinfo{date}{apr} \bibinfo{year}{2023}), \bibinfo{numpages}{32}~pages.
\newblock
\urldef\tempurl%
\url{https://doi.org/10.1145/3579542}
\showDOI{\tempurl}


\bibitem[Wang et~al\mbox{.}(2016)]%
        {wang2016catching}
\bibfield{author}{\bibinfo{person}{Yu Wang}, \bibinfo{person}{Jiebo Luo}, \bibinfo{person}{Richard Niemi}, \bibinfo{person}{Yuncheng Li}, {and} \bibinfo{person}{Tianran Hu}.} \bibinfo{year}{2016}\natexlab{}.
\newblock \showarticletitle{Catching fire via" likes": Inferring topic preferences of trump followers on twitter}. In \bibinfo{booktitle}{\emph{Proceedings of the International AAAI Conference on Web and Social Media}}, Vol.~\bibinfo{volume}{10}. \bibinfo{pages}{719--722}.
\newblock


\bibitem[Waters and Jamal(2011)]%
        {waters2011tweet}
\bibfield{author}{\bibinfo{person}{Richard~D Waters} {and} \bibinfo{person}{Jia~Y Jamal}.} \bibinfo{year}{2011}\natexlab{}.
\newblock \showarticletitle{Tweet, tweet, tweet: A content analysis of nonprofit organizations’ Twitter updates}.
\newblock \bibinfo{journal}{\emph{Public relations review}} \bibinfo{volume}{37}, \bibinfo{number}{3} (\bibinfo{year}{2011}), \bibinfo{pages}{321--324}.
\newblock


\bibitem[Wei et~al\mbox{.}(2022)]%
        {wei_2022_cot}
\bibfield{author}{\bibinfo{person}{Jason Wei}, \bibinfo{person}{Xuezhi Wang}, \bibinfo{person}{Dale Schuurmans}, \bibinfo{person}{Maarten Bosma}, \bibinfo{person}{Brian Ichter}, \bibinfo{person}{Fei Xia}, \bibinfo{person}{Ed~H. Chi}, \bibinfo{person}{Quoc~V. Le}, {and} \bibinfo{person}{Denny Zhou}.} \bibinfo{year}{2022}\natexlab{}.
\newblock \showarticletitle{Chain-of-thought prompting elicits reasoning in large language models}. In \bibinfo{booktitle}{\emph{Proceedings of the 36th International Conference on Neural Information Processing Systems}} (New Orleans, LA, USA) \emph{(\bibinfo{series}{NIPS '22})}. \bibinfo{publisher}{Curran Associates Inc.}, \bibinfo{address}{Red Hook, NY, USA}, Article \bibinfo{articleno}{1800}, \bibinfo{numpages}{14}~pages.
\newblock
\showISBNx{9781713871088}


\bibitem[Wu et~al\mbox{.}(2022)]%
        {WU2022364}
\bibfield{author}{\bibinfo{person}{Xingjiao Wu}, \bibinfo{person}{Luwei Xiao}, \bibinfo{person}{Yixuan Sun}, \bibinfo{person}{Junhang Zhang}, \bibinfo{person}{Tianlong Ma}, {and} \bibinfo{person}{Liang He}.} \bibinfo{year}{2022}\natexlab{}.
\newblock \showarticletitle{A survey of human-in-the-loop for machine learning}.
\newblock \bibinfo{journal}{\emph{Future Generation Computer Systems}}  \bibinfo{volume}{135} (\bibinfo{year}{2022}), \bibinfo{pages}{364--381}.
\newblock
\showISSN{0167-739X}
\urldef\tempurl%
\url{https://doi.org/10.1016/j.future.2022.05.014}
\showDOI{\tempurl}


\bibitem[Yarchi et~al\mbox{.}(2021)]%
        {yarchi2021political}
\bibfield{author}{\bibinfo{person}{Moran Yarchi}, \bibinfo{person}{Christian Baden}, {and} \bibinfo{person}{Neta Kligler-Vilenchik}.} \bibinfo{year}{2021}\natexlab{}.
\newblock \showarticletitle{Political polarization on the digital sphere: A cross-platform, over-time analysis of interactional, positional, and affective polarization on social media}.
\newblock \bibinfo{journal}{\emph{Political Communication}} \bibinfo{volume}{38}, \bibinfo{number}{1-2} (\bibinfo{year}{2021}), \bibinfo{pages}{98--139}.
\newblock


\bibitem[Yin and Zhang(2020)]%
        {yin2020incorporating}
\bibfield{author}{\bibinfo{person}{Chunxiao Yin} {and} \bibinfo{person}{Xiaofei Zhang}.} \bibinfo{year}{2020}\natexlab{}.
\newblock \showarticletitle{Incorporating message format into user evaluation of microblog information credibility: A nonlinear perspective}.
\newblock \bibinfo{journal}{\emph{Information Processing \& Management}} \bibinfo{volume}{57}, \bibinfo{number}{6} (\bibinfo{year}{2020}), \bibinfo{pages}{102345}.
\newblock


\bibitem[Zade et~al\mbox{.}(2024)]%
        {zade_et_al_24}
\bibfield{author}{\bibinfo{person}{Himanshu Zade}, \bibinfo{person}{Spencer Williams}, \bibinfo{person}{Theresa~T. Tran}, \bibinfo{person}{Christina Smith}, \bibinfo{person}{Sukrit Venkatagiri}, \bibinfo{person}{Gary Hsieh}, {and} \bibinfo{person}{Kate Starbird}.} \bibinfo{year}{2024}\natexlab{}.
\newblock \showarticletitle{To Reply or to Quote: Comparing Conversational Framing Strategies on Twitter}.
\newblock \bibinfo{journal}{\emph{ACM J. Comput. Sustain. Soc.}} \bibinfo{volume}{2}, \bibinfo{number}{1}, Article \bibinfo{articleno}{9} (\bibinfo{date}{Jan.} \bibinfo{year}{2024}), \bibinfo{numpages}{27}~pages.
\newblock
\urldef\tempurl%
\url{https://doi.org/10.1145/3625680}
\showDOI{\tempurl}


\bibitem[Zahry et~al\mbox{.}(2023)]%
        {zahry2023risk}
\bibfield{author}{\bibinfo{person}{Nagwan~R Zahry}, \bibinfo{person}{Michael McCluskey}, {and} \bibinfo{person}{Jiying Ling}.} \bibinfo{year}{2023}\natexlab{}.
\newblock \showarticletitle{Risk governance during the COVID-19 pandemic: A quantitative content analysis of governors' narratives on twitter}.
\newblock \bibinfo{journal}{\emph{Journal of Contingencies and Crisis Management}} \bibinfo{volume}{31}, \bibinfo{number}{1} (\bibinfo{year}{2023}), \bibinfo{pages}{77--91}.
\newblock


\bibitem[Zappavigna and Martin(2018)]%
        {zappavigna2018communing}
\bibfield{author}{\bibinfo{person}{Michele Zappavigna} {and} \bibinfo{person}{James~R Martin}.} \bibinfo{year}{2018}\natexlab{}.
\newblock \showarticletitle{\# Communing affiliation: Social tagging as a resource for aligning around values in social media}.
\newblock \bibinfo{journal}{\emph{Discourse, context \& media}}  \bibinfo{volume}{22} (\bibinfo{year}{2018}), \bibinfo{pages}{4--12}.
\newblock


\bibitem[Zhang et~al\mbox{.}(2019)]%
        {zhang2019social}
\bibfield{author}{\bibinfo{person}{Cheng Zhang}, \bibinfo{person}{Chao Fan}, \bibinfo{person}{Wenlin Yao}, \bibinfo{person}{Xia Hu}, {and} \bibinfo{person}{Ali Mostafavi}.} \bibinfo{year}{2019}\natexlab{}.
\newblock \showarticletitle{Social media for intelligent public information and warning in disasters: An interdisciplinary review}.
\newblock \bibinfo{journal}{\emph{International Journal of Information Management}}  \bibinfo{volume}{49} (\bibinfo{year}{2019}), \bibinfo{pages}{190--207}.
\newblock


\bibitem[Zhang et~al\mbox{.}(2023)]%
        {zhang_discussion_2023}
\bibfield{author}{\bibinfo{person}{Zheng Zhang}, \bibinfo{person}{Jie Gao}, \bibinfo{person}{Ranjodh~Singh Dhaliwal}, {and} \bibinfo{person}{Toby Jia-Jun Li}.} \bibinfo{year}{2023}\natexlab{}.
\newblock \showarticletitle{VISAR: A Human-AI Argumentative Writing Assistant with Visual Programming and Rapid Draft Prototyping}. In \bibinfo{booktitle}{\emph{Proceedings of the 36th Annual ACM Symposium on User Interface Software and Technology}} (San Francisco, CA, USA) \emph{(\bibinfo{series}{UIST '23})}. \bibinfo{publisher}{Association for Computing Machinery}, \bibinfo{address}{New York, NY, USA}, Article \bibinfo{articleno}{5}, \bibinfo{numpages}{30}~pages.
\newblock
\showISBNx{9798400701320}
\urldef\tempurl%
\url{https://doi.org/10.1145/3586183.3606800}
\showDOI{\tempurl}


\bibitem[Zhao et~al\mbox{.}(2023)]%
        {zhao_et_23}
\bibfield{author}{\bibinfo{person}{Wanqing Zhao}, \bibinfo{person}{Yuta Nakashima}, \bibinfo{person}{Haiyuan Chen}, {and} \bibinfo{person}{Noboru Babaguchi}.} \bibinfo{year}{2023}\natexlab{}.
\newblock \showarticletitle{Enhancing Fake News Detection in Social Media via Label Propagation on Cross-modal Tweet Graph}. In \bibinfo{booktitle}{\emph{Proceedings of the 31st ACM International Conference on Multimedia}} (Ottawa ON, Canada) \emph{(\bibinfo{series}{MM '23})}. \bibinfo{publisher}{Association for Computing Machinery}, \bibinfo{address}{New York, NY, USA}, \bibinfo{pages}{2400–2408}.
\newblock
\showISBNx{9798400701085}
\urldef\tempurl%
\url{https://doi.org/10.1145/3581783.3612086}
\showDOI{\tempurl}


\bibitem[Zhou et~al\mbox{.}(2021a)]%
        {ZHOU2021102679}
\bibfield{author}{\bibinfo{person}{Cheng Zhou}, \bibinfo{person}{Kai Li}, {and} \bibinfo{person}{Yanhong Lu}.} \bibinfo{year}{2021}\natexlab{a}.
\newblock \showarticletitle{Linguistic characteristics and the dissemination of misinformation in social media: The moderating effect of information richness}.
\newblock \bibinfo{journal}{\emph{Information Processing \& Management}} \bibinfo{volume}{58}, \bibinfo{number}{6} (\bibinfo{year}{2021}), \bibinfo{pages}{102679}.
\newblock
\showISSN{0306-4573}
\urldef\tempurl%
\url{https://doi.org/10.1016/j.ipm.2021.102679}
\showDOI{\tempurl}


\bibitem[Zhou et~al\mbox{.}(2021b)]%
        {zhou2021characterizing}
\bibfield{author}{\bibinfo{person}{Cheng Zhou}, \bibinfo{person}{Haoxin Xiu}, \bibinfo{person}{Yuqiu Wang}, {and} \bibinfo{person}{Xinyao Yu}.} \bibinfo{year}{2021}\natexlab{b}.
\newblock \showarticletitle{Characterizing the dissemination of misinformation on social media in health emergencies: An empirical study based on COVID-19}.
\newblock \bibinfo{journal}{\emph{Information Processing \& Management}} \bibinfo{volume}{58}, \bibinfo{number}{4} (\bibinfo{year}{2021}), \bibinfo{pages}{102554}.
\newblock


\bibitem[Zhou et~al\mbox{.}(2023)]%
        {zhou_et_al_23}
\bibfield{author}{\bibinfo{person}{Kaitlyn Zhou}, \bibinfo{person}{Tom Wilson}, \bibinfo{person}{Kate Starbird}, {and} \bibinfo{person}{Emma~S. Spiro}.} \bibinfo{year}{2023}\natexlab{}.
\newblock \showarticletitle{Spotlight Tweets: A Lens for Exploring Attention Dynamics within Online Sensemaking During Crisis Events}.
\newblock \bibinfo{journal}{\emph{Trans. Soc. Comput.}} \bibinfo{volume}{6}, \bibinfo{number}{1–2}, Article \bibinfo{articleno}{2} (\bibinfo{date}{June} \bibinfo{year}{2023}), \bibinfo{numpages}{33}~pages.
\newblock
\urldef\tempurl%
\url{https://doi.org/10.1145/3577213}
\showDOI{\tempurl}


\bibitem[Řehůřek(nd)]%
        {rehurek_coherencemodel}
\bibfield{author}{\bibinfo{person}{Radim Řehůřek}.} \bibinfo{year}{n.d.}\natexlab{}.
\newblock \bibinfo{title}{CoherenceModel – Gensim}.
\newblock
\newblock
\urldef\tempurl%
\url{https://radimrehurek.com/gensim/models/coherencemodel.html}
\showURL{%
\tempurl}
\newblock
\shownote{Accessed: 2025-03-11}.


\end{thebibliography}

\end{document}